\documentclass[journal]{IEEEtran}

\ifCLASSINFOpdf
  \usepackage[pdftex]{graphicx}
  \graphicspath{{../pdf/}{../jpeg/}}
  \DeclareGraphicsExtensions{.pdf,.jpeg,.png}
\else
\fi

\usepackage{subfigure}
\usepackage{graphicx}
\usepackage[cmex10]{amsmath}
\usepackage{amssymb}
\usepackage{authblk}
\usepackage{stfloats}
\usepackage{url}
\usepackage{color}
\usepackage{pifont}
\usepackage{booktabs}
\usepackage{multirow}
\usepackage[utf8]{inputenc}
\usepackage{hyperref}
\newcommand{\LVendo}{LV\textsubscript{Endo}}
\newcommand{\LVepi}{LV\textsubscript{Epi}}
\newcommand{\LVef}{LV\textsubscript{EF}}
\newcommand{\LVedv}{LV\textsubscript{EDV}}
\newcommand{\LVesv}{LV\textsubscript{ESV}}

\newcommand{\ie}{\textit{i.e.}}
\newcommand{\eg}{\textit{e.g.}}
\newcommand{\etal}{\textit{et al.}}

\definecolor{pink}{rgb}{0.858, 0.188, 0.478}
\definecolor{orange}{rgb}{1.0, 0.5, 0.0}

\definecolor{cyan}{rgb}{0.113, 0.776, 0.686}
\definecolor{brown}{rgb}{0.741, 0.552, 0.180}
\definecolor{mygreen}{rgb}{0.0, 0.5, 0.0}

\newcommand \modification[1]{\textcolor{black}{#1}}
\newcommand \correction[1]{\textcolor{black}{#1}}

\def\mathcolor#1#{\@mathcolor{#1}}
\def\@mathcolor#1#2#3{
\protect\leavevmode
\begingroup
\color#1{#2}#3
\endgroup
}

\begin{document}



\title{Deep Learning for Segmentation using an Open Large-Scale Dataset in 2D Echocardiography}


\author{Sarah~Leclerc, 
		Erik~Smistad,
        Jo\~ao~Pedrosa,
        Andreas~{\O}stvik,
        Frederic Cervenansky,
        Florian~Espinosa,
        Torvald~Espeland,
        Erik~Andreas~Rye~Berg,
        Pierre-Marc~Jodoin,
        Thomas~Grenier,
        Carole~Lartizien,
        Jan~D'hooge,
        Lasse~Lovstakken, 
        and Olivier~Bernard
\thanks{S.~Leclerc, T.~Grenier, C.~Lartizien, F.~Cervenansky and O.~Bernard are with the University of Lyon, CREATIS, CNRS UMR5220, Inserm U1044, INSA-Lyon, University of Lyon 1, Villeurbanne, France. E-mail: olivier.bernard@creatis.insa-lyon.fr.}
\thanks{E.~Smistad, A.~Ostvik and L.~Lovstakken are with the Center of Innovative Ultrasound Solutions (CIUS), Department of Circulation and Medical Imaging, Norwegian University of Science and Technology (NTNU), Trondheim, Norway}
\thanks{J.~Pedrosa and J.~D'hooge are with the Dept. of Cardiovascular Sciences, KU Leuven, Leuven, Belgium}
\thanks{F.~Espinosa is with the Cardiovascular department Centre Hospitalier de Saint-Etienne
Saint-Etienne, France}
\thanks{T.~Espeland and E.A.~Rye~Berg are with the Center of Innovative Ultrasound Solutions (CIUS) and the Clinic of cardiology, St. Olavs Hospital, Trondheim, Norway}
\thanks{P.-M.~Jodoin is with the Computer Science Department, University of Sherbrooke, Sherbrooke, Canada.}
\vspace{-1cm}}

\markboth{Draft version for IEEE TMI journal}%
{Shell \MakeLowercase{\textit{et al.}}: Bare Demo of IEEEtran.cls for IEEE Journals}

\maketitle

\begin{abstract}
Delineation of the cardiac structures from 2D echocardiographic images is a common clinical task to establish a diagnosis. Over the past decades, the automation of this task has been the subject of intense research. In this paper, we evaluate how far the state-of-the-art \correction{encoder-decoder} deep convolutional neural network methods can go at assessing \mbox{2D} echocardiographic images, \ie~segmenting cardiac structures as well as estimating clinical indices, on a dataset especially designed to answer this objective. We therefore introduce the Cardiac Acquisitions for Multi-structure Ultrasound Segmentation (CAMUS) dataset, the largest publicly-available and fully-annotated dataset for the purpose of echocardiographic assessment. The dataset contains two and four-chamber acquisitions from 500 patients with reference measurements from one cardiologist on the full dataset and from three cardiologists on a fold of 50 patients. Results show that encoder-decoder based architectures outperform state-of-the-art non-deep learning methods and faithfully reproduce the expert analysis for the end-diastolic and end-systolic left ventricular volumes, with a mean correlation of 0.95 and an absolute mean error of 9.5 ml. Concerning the ejection fraction of the left ventricle, results are more contrasted with a mean correlation coefficient of 0.80 and an absolute mean error of 5.6\%. Although these results are below the inter-observer scores, they remain slightly worse than the intra-observer's ones. Based on this observation, areas for improvement are defined, which open the door for accurate and fully-automatic analysis  \modification{of} 2D echocardiographic  \modification{images}.
\end{abstract}

\begin{IEEEkeywords}
Cardiac segmentation and diagnosis, deep learning, ultrasound, left ventricle, myocardium, left atrium.
\end{IEEEkeywords}

\IEEEpeerreviewmaketitle

\section{Introduction}
\label{sec:introduction}
Analysis of 2D echocardiographic images plays a crucial role in clinical routine to measure the cardiac morphology and function and to reach a diagnosis. Such analysis is based on the interpretation of clinical indices which are extracted from low-level image processing such as segmentation and tracking. For instance, the extraction of the ejection fraction (EF) of the left ventricle (LV) requires accurate delineation of the left ventricular endocardium in both end diastole (ED) and end systole (ES). In clinical routine, semi-automatic or manual annotation is still daily work due to the lack of accuracy and reproducibility of fully-automatic cardiac segmentation methods. This leads to time consuming tasks prone to intra- and inter-observer variability \cite{Armstrong2015}. The inherent difficulties for segmenting echocardiographic images are well documented: \emph{i)} poor contrast between the myocardium and the blood pool; \emph{ii)} brightness inhomogeneities; \emph{iii)} variation in the speckle pattern along the myocardium due to the orientation of the cardiac probe with respect to tissue; \emph{iv)} presence of trabeculae and papillary muscles with intensities similar to the myocardium; \emph{v)} significant tissue echogenicity variability within the population; \emph{vi)} shape, intensity and motion variability of the heart structures across patients and pathologies.

The lack of \modification{large} and publicly-available dataset has prevented a thorough evaluation of the potential of deep learning methods to estimate clinical indices, while these techniques are actively applied with great success for other modalities~\cite{Bernard2018}. Indeed, while the number of medical imaging challenges comparing deep learning methods has exploded this last decade, only one focused on cardiac ultrasound image segmentation \cite{Bernard2016}. \correction{Unfortunately, since the challenge was held in 2014, none of the participant used convolutional neural networks (CNNs) because these methods had not yet gained popularity in medical imaging.} The lack of well-annotated echocardiographic datasets can be explained by the difficulty of exporting data from clinical ultrasound equipments and getting a large amount of images carefully annotated by cardiologists due to the very nature of echocardiography as mentioned above. In this context, the purpose of this paper is to provide answers to the following four questions:
\begin{enumerate}
\item How well do CNNs perform compared to non-deep learning state-of-the-art techniques ?
\item How many patients are needed to train a CNN to get highly accurate results in 2D echocardiographic image segmentation ?
\item How accurate can the volumes and ejection fraction be estimated from the segmentation of CNNs compared to the inter/intra-expert variability ?
\item What improvement can be brought by sophisticated architectures compared to simpler CNN models for 2D echocardiographic segmentation ?
\end{enumerate}
For that purpose, we present a new public dataset called CAMUS (Cardiac Acquisitions for Multi-structure Ultrasound Segmentation). It contains $2$D echocardiographic sequences \modification{with two and four-chamber views} of $500$ patients that were acquired with the same equipment in the same medical center. The size of this dataset and its tight connection to every-day clinical issues give the possibility to train deep learning methods to automatically analyze echocardiographic data. In addition, CAMUS includes manual expert annotations for the left ventricle endocardium (\LVendo), the myocardium (epicardium contour more specifically, named \LVepi) and the left atrium (LA).

\section{Previous work}
\label{sec:previous_work}

\noindent\textit{Previous cardiac ultrasound datasets}

\vspace*{0.1cm}

To date, only one echocardiographic dataset has been broadly validated. This dataset was released in conjunction with the Challenge on Endocardial Three-dimensional Ultrasound Segmentation (CETUS) which took place during the MICCAI 2014 conference\footnote{https://www.creatis.insa-lyon.fr/Challenge/CETUS/}. The CETUS dataset is composed of $45$ $3$D echocardiographic sequences ($15$ for training, $30$ for testing) equally distributed among three different subgroups: healthy subjects, patients with previous myocardial infarction examined at least 3 months after the event and patients with dilated cardiomyopathy. The data is provided with two reference meshes of the \LVendo~per patient (one at ED and one at ES), each reference corresponding to the mean shape computed from the annotations of three different experienced cardiologists. Five fully-automatic (deformable models, Hough random forest, Kalman filter, active appearance model) and four semi-automatic methods (graph-cut method, structured random forest, multi-atlas and level-set approaches) were evaluated through this challenge. No challenger implemented a deep neural network. The outcome of the challenge revealed that the overall best scores were obtained by the B-spline explicit active surface, a fully-automatic method proposed by Barbosa \etal~\cite{Barbosa2014}.  This method was \modification{later on} improved by Pedrosa \etal~\cite{Pedrosa2017}~thanks to the integration of a shape prior derived from a conventional principal component analysis scheme. By doing so, the authors obtained the following scores for the segmentation of the $3$D \LVendo: \emph{i)} average Dice values of $0.909$ (ED) and $0.875$ (ES); \emph{ii)} average Hausdorff distances of $6.3$ mm (ED) and $6.9$ mm (ES) and \emph{iii)} average mean absolute distances of $1.8$ mm (ED) and $2.0$ mm (ES). 

\vspace*{0.2cm}

\noindent\textit{Non-deep learning methods}

\vspace*{0.1cm}

Several surveys of echocardiographic segmentation methods have been proposed, both in 2D \cite{Noble2006,Carneiro2012} and 3D \cite{Bernard2016,Leung2010}. Most of the reported methods focused on the segmentation of the \LVendo~border. Among those reviews, only the one by Bernard \etal~published in 2016 benchmarked different techniques on the same dataset, leading to a fair comparison \cite{Bernard2016}. In this study, the authors listed the results obtained by nine different techniques. The reported methods can be divided in two main categories: those with a weak prior and those with a strong prior. The first group involves weak assumptions such as spatial, intensity, motion or anatomical information. It includes image-based techniques (multi-scale quadrature filter) \cite{Wang2014}, a motion-based method (Kalman filter) \cite{Smistad2014}, deformable models (BEAS, level-set) \cite{Barbosa2014,Wang2014} and a graph-based approach (graph-cut) \cite{Bernier2014}. The second group uses approaches with strong priors including a shape-prior (Hough forest) \cite{Milletari2014}, an active appearance model \cite{Vanstralen2014}, an atlas-based method \cite{Oktay2014} and a machine learning algorithm (random forest) \cite{Milletari2014,Keraudren2014,Domingos2014}, each requiring a manually-annotated training set.

\vspace*{0.2cm}

\noindent\textit{Deep learning methods}

\vspace*{0.1cm}

Deep-learning methods have been successfully applied to the segmentation of the \LVendo~in echocardiography. In 2012, Carneiro \etal~developed a two-stage deep learning method for the segmentation of the \LVendo~for 2D echocardiographic images restricted to four-chamber view acquisitions~\cite{Carneiro2012}. Based on a maximum {\em a posteriori} framework, the authors formulated the LV segmentation problem according to two successive steps: \emph{i)} the automatic selection of several regions in the tested image where the \LVendo~is fully present; \emph{ii)} the automatic extraction of the \LVendo~contour from the previously selected regions. These two steps involved a deep belief network. Their method was trained on 400 images from $12$ different patient sequences with various pathologies and tested on 50 images from $2$ healthy subject sequences. They obtained an average Hausdorff distance of $\sim\!\!18$ mm and an average mean absolute distance of $\sim\!\!8$ mm for the \LVendo.

In 2017, Smistad \etal~\cite{Smistad2017} showed that the U-Net CNN method~\cite{Ronneberger2015} could be trained to successfully segment the left ventricle in 2D ultrasound images. However, due to lack of training data, the network was trained with the output of a state-of-the-art deformable model segmentation method \cite{Smistad2014}. On a manually segmented test set, the results showed that the network and the deformable model obtained the same accuracy with a Dice score of 0.87.

Recently, Oktay \etal~\cite{Oktay2018}~used CNNs to segment the 3D \LVendo~structure using an approach named anatomically constrained neural network (ACNN). The core of their neural network is based on an architecture similar to the 3D U-Net~\cite{Cicek2016}, whose segmentation output is constrained to fit a non-linear compact representation of the underlying anatomy derived from an auto-encoder network. The performance of their method was assessed on the CETUS dataset. They obtained the following scores for the segmentation of the 3D \LVendo~structure: \emph{i)} average Dice values of $0.912$ (ED) and $0.873$ (ES); \emph{ii)} average Hausdorff distances of $7.0$ mm (ED) and $7.7$ mm (ES) and \emph{iii)} average mean absolute distances of $1.9$ mm (ED) and $2.1$ mm (ES)~\cite{Oktay2018}. Interestingly, these results are quite close to those obtained by Pedrosa \mbox{\etal}~\cite{Pedrosa2017}. Additionally, the use of only $15$ patients during the training phase illustrates the strong potential of deep learning techniques to analyze echocardiographic images.

\section{CAMUS dataset}
\label{sec:camus_dataset}

\subsection{Dataset}
\label{sec:dataset}
As described in the previous section, only one echocardiographic dataset, composed of 45 3D sequences, \modification{has been  broadly validated by the community}. This dataset is not appropriate to study the behavior of deep learning approaches in the particular case of 2D sequences. In this context, we introduce the largest publicly-available and fully annotated dataset for the purpose of 2D echocardiographic assessment.

\subsubsection{Patient selection}

The proposed dataset consists of clinical exams from 500 patients, acquired at the University Hospital of St Etienne (France) and included in this study within the regulation set by the local ethical committee of the hospital. The acquisitions were optimized to perform \LVef~measurements. In order to enforce clinical realism, neither prerequisite nor data selection have been performed. Consequently, \emph{i)} some cases were difficult to trace; \emph{ii)} the dataset involves a wide variability of acquisition settings; \emph{iii)} for some patients, parts of the wall were not visible in the images; \emph{iv)} for some cases, the probe orientation recommendation to acquire a rigorous four-chambers view was simply impossible to follow and a five-chambers view was acquired instead. This produced a highly heterogeneous dataset, both in terms of image quality and pathological cases, which is typical of daily clinical practice data. Table \ref{tab:dataset} provides the main information which characterizes the collected dataset. From this table, one can see that half of the dataset population has a \LVef~lower than $45$\%, thus being considered at pathological risk (beyond the uncertainty of the measurement). Also, $19$\% of the images have a poor quality (based on the opinion of one expert \emph{O\textsubscript{1a}}), indicating that for this subgroup the localization of the \LVendo~ and \LVepi~ as well as the estimation of clinical indices are not considered clinically accurate and workable. In classical analysis, poor quality images are usually removed from the dataset because of their clinical uselessness. Therefore, those data were not involved in this project during the computation of the different metrics but were used to study their influence as part of the training and validation sets for deep learning techniques. 

\begin{table}[tbp]
\renewcommand{\arraystretch}{1.}
  \caption{The main characteristics of the CAMUS echocardiographic dataset collected from $500$ patients}
  \centering
  \begin{tabular}{*{7}{c}}
 \toprule 
 \multicolumn{1}{c}{\multirow{3}{*}{\bf \small Dataset}} & \multicolumn{3}{c}{\bf \small Image Quality} & \multicolumn{3}{c}{\bf \small \LVef} \\
 & \multicolumn{3}{c}{\small \emph{(in percentage)}} & \multicolumn{3}{c}{\small \emph{(in percentage)}} \\
 \cmidrule(r){2-4} \cmidrule(r){5-7} 
 & Good & Medium & Poor & $\leq 45$\% & $\geq 55$\% & else \\
 \midrule
 \multicolumn{1}{l}{Full} & 35 & 46 & 19 & 49 & 19 & 32 \\
 \cmidrule(r){1-1}
 \multicolumn{1}{l}{\emph{fold 1}} & 34 & 48 & 18 & 48 & 20 & 32 \\
 \multicolumn{1}{l}{\emph{fold 2}} & 34 & 46 & 20 & 50 & 18 & 32 \\
 \multicolumn{1}{l}{\emph{fold 3}} & 34 & 46 & 20 & 48 & 20 & 32 \\ 
 \multicolumn{1}{l}{\emph{fold 4}} & 34 & 46 & 20 & 50 & 20 & 30 \\  
 \multicolumn{1}{l}{\emph{fold 5}} & 34 & 46 & 20 & 48 & 20 & 32 \\ 
 \multicolumn{1}{l}{\emph{fold 6}} & 36 & 46 & 18 & 50 & 20 & 30 \\ 
 \multicolumn{1}{l}{\emph{fold 7}} & 36 & 46 & 18 & 50 & 20 & 30 \\ 
 \multicolumn{1}{l}{\emph{fold 8}} & 36 & 46 & 18 & 50 & 18 & 32 \\ 
 \multicolumn{1}{l}{\emph{fold 9}} & 36 & 46 & 18 & 48 & 20 & 32 \\ 
 \multicolumn{1}{l}{\emph{fold 10}} & 36 & 46 & 18 & 50 & 18 & 32 \\ 
 \midrule
 \vspace{-0.7cm}
\label{tab:dataset}
  \end{tabular}
\end{table}

The dataset was divided into 10 folds to perform standard cross-validation for the machine learning methods. Each fold contains $50$ patients with the same distributions in terms of image quality and \LVef~as the full dataset (see table~\ref{tab:dataset}). For each of the 10 test sets, the remaining $450$ patients ($9$ folds) were used during the training/validation phases of the machine learning techniques. In particular, $8$ folds ($400$ patients) were used for training and $1$ ($50$ patients) for validation, \ie~parameters optimization. The full dataset is available for download at \href{https://camus.creatis.insa-lyon.fr/challenge/}{https://camus.creatis.insa-lyon.fr/challenge/},.

\vspace*{0.1cm}

\subsubsection{Acquisition protocol}

The full dataset was acquired from GE Vivid E95 ultrasound scanners (GE Vingmed Ultrasound, Horten Norway), with a GE M5S probe (GE Healthcare, US). No additional protocol than the one used in clinical routine was put in place. For each patient, $2$D apical four-chamber and two-chamber view sequences were exported from EchoPAC analysis software (GE Vingmed Ultrasound, Horten, Norway). These standard cardiac views were chosen for this study to enable the estimation of \LVef~values based on the Simpson's biplane method of discs \cite{Folland1979}. Each exported sequence corresponds to a set of B-mode images expressed in polar coordinates. The same interpolation procedure was used to express all sequences in Cartesian coordinates with a unique grid resolution, \ie~$\lambda/2=0.3$~mm along the x-axis (axis parallel to the probe) and $\lambda/4=0.15$~mm along the z-axis (axis perpendicular to the probe), where $\lambda$ corresponds to the wavelength of the ultrasound probe. At least one full cardiac cycle was acquired for each patient in each view, allowing manual annotation of cardiac structures at ED and ES.

\subsection{Reference segmentation and contouring protocol}

Establishing a well-defined ground-truth segmentation was of utmost importance for this work. The main difficulty when delineating 2D echocardiographic images comes from poor contrast in some regions along with the presence of well-known artifacts (\eg~reverberation, clutter, acoustic shadowing). One direct consequence is that embedded fully-automatic ultrasound cardiac segmentation softwares do not perform well. During the clinical exam, the clinicians delineate the different contours using semi-automatic tools under time constraints. In this context, the use of manual annotations extracted from clinical exams is not optimal to design a reference dataset for machine learning where the coherence and accuracy in the manual contouring play an important role during the learning phase. To illustrate this point, it is interesting to note that the only existing shared echocardiographic dataset with reference annotations was realized off-line for the purpose of the CETUS MICCAI challenge in 2014. In particular, it required more than 10 months for 3 different cardiologists to manually contour the 3D endocardium surfaces of $45$ patients including consensus revisions \cite{Bernard2016}. This illustrates the extreme difficulty in designing such a high-quality dataset. 

\vspace*{0.1cm}

\subsubsection{Cardiologists involvement}

Three cardiologists (\modification{\emph{O\textsubscript{1}}, \emph{O\textsubscript{2}} and \emph{O\textsubscript{3}}}) participated in the annotation of the dataset. Considerable effort was spent to define a consistent manual segmentation protocol. This protocol was designed with the help of \emph{O\textsubscript{1}} and was then strictly followed \modification{\emph{O\textsubscript{2}} and \emph{O\textsubscript{3}}}. In particular, we asked \emph{O\textsubscript{1}} to perform the manual annotation and to determine ED and ES of the full dataset, while the two others contoured the test set of fold $5$ ($50$ patients). \emph{O\textsubscript{1}} also annotated twice fold $5$ seven months apart (we call those annotations \emph{O\textsubscript{1a}} and \emph{O\textsubscript{1b}}). This fold was therefore used to measure both the inter- and intra-observer variability.

\vspace*{0.1cm}

\subsubsection{Contouring protocol}

According to the recommendation of the American Society of Echocardiography and the European Association of Cardiovascular Imaging~\cite{Lang2015}, ED is preferably defined as the first frame after mitral valve closure or the frame in the cardiac cycle in which the respective LV dimension or volume measurement is the largest. ES is best defined as the frame after aortic valve closure (\eg~using an apical long axis view image) or the frame in which the cardiac dimension or volume is the smallest. In this work, ED and ES were selected as the frames where the LV dimension was at its largest or smallest, which is not the most accurate way, especially in the presence of abnormalities. This simpler approach was used due to the lack of reliable ECG. Thus the clinical indices, ED/ES volume and EF, reported in this work have to be interpreted with this in mind. While only the extraction of the \LVendo~contour is necessary to estimate \LVef~values, we also asked the cardiologists to manually outline the \LVepi~and the LA for all patients. This was done to study the influence of contextualization (segmentation of several structures at once) on the performance of the \LVendo~segmentation using deep learning techniques. The following protocol was set up.

\begin{itemize}
\item \LVendo: Convention was used for the LV wall, mitral valve plane, trabeculations, papillary muscles and apex~\cite{Lang2015}. Basic points were to \emph{i)} include trabeculae and papillary muscles in the LV cavity; \emph{ii)} keep tissue consistency between ED and ES instants; \emph{iii)} terminate the contours in the mitral valve plane on the ventricular side of the bright ridge, at the points where the valve leaflets are hinging; \emph{iv)} partially exclude left ventricular outflow tract from the cavity by drawing from septal mitral valve hinge point to the septal wall to create a smooth shape.
\item \LVepi: There is no recommendation for delineating the epicardium. We thus outlined the epicardium as the interface between the pericardium and the myocardium for the anterior, anterolateral and inferior segments and the frontier between the right ventricle cavity and the septum for the inferoseptal segments.
\item LA: There are recommendations for LA segmentation to assess the full LA area from dedicated LA recordings. However, since we have used acquisitions focusing on the LV, part of the dataset does not cover the full LA surface and is thus not suited to perform such measurement. Having this in mind, we used the following contouring protocol: \emph{i)} start the LA contour from the extremities of the \LVendo~contour, at the points where the valve leaflets are hinging; \emph{ii)} have the contour pass by the LA inner border.
\end{itemize}

\vspace*{0.2cm}

\begin{figure}[tp]
\centering
\subfigure[Good image quality]{\includegraphics[width=0.48\textwidth]{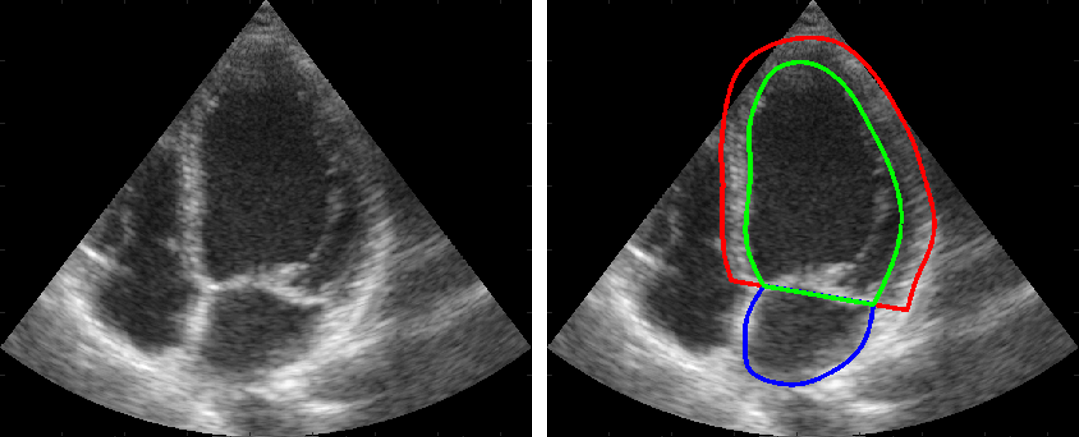}}%
\vfil
\subfigure[Medium image quality]{\includegraphics[width=0.48\textwidth]{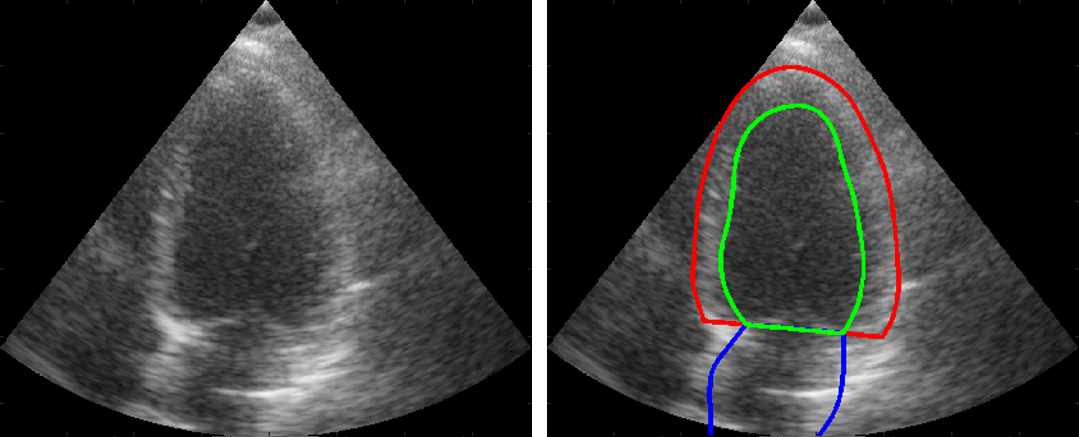}}%
\vfil
\subfigure[Poor image quality]{\includegraphics[width=0.48\textwidth]{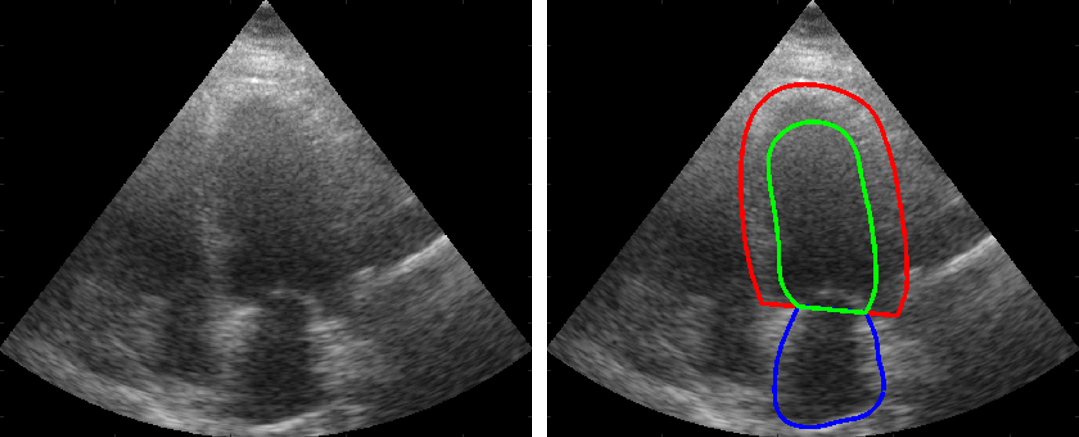}}%
\caption{Typical images extracted from the proposed dataset. \correction{Endocardium and epicardium of the left ventricle and left atrium wall are shown respectively in green, red and blue.} [Left] input images; [Right] corresponding manual annotations.}
\label{fig:manual_annotations}
\end{figure}

Fig.~\ref{fig:manual_annotations} illustrates our manual contouring protocol for a good, a medium, and a poor-quality image.

\section{Evaluated methods}
\label{sec:evaluated_methods}

\subsection{CNN techniques based on an encoder-decoder architecture}

The goal of this study is to assess how far CNNs can go at segmenting 2D echocardiographic images. As such, we chose to focus on the well-known encoder-decoder networks (EDNs) which have been the cornerstone of a wide variety of CNNs that were successfully applied in medical imaging~\cite{Litjens2017}. EDNs are based on a two-stage convolutional network architecture well suited for segmentation. The first part, known as the encoder, consists of a series of convolutions and downsampling operations. These operations extract features from the images while spatially compressing them, thus enabling extraction of high-level features. The second part is the decoder, which uses features from the encoder and applies a set of convolutions and upsampling operations to gradually transform feature maps into a final segmentation.

Among the existing EDNs, one of the most popular architectures used in medical imaging corresponds the U-Net model proposed by Ronneberger \etal~in 2015 \cite{Ronneberger2015}. This network integrates \modification{skip} connections between the encoder and decoder parts with the goal of helping in retrieving details that were potentially lost during the downsampling while also stabilizing gradients. The original \mbox{U-Net} follows a specific scheme of convolutions, where each downsampling and upsampling step is proceeded by two 3x3 convolutional layers, while the number of features is doubled per downsampling and reduced in half per upsampling. \mbox{U-Net} has been successfully applied to a wide range of medical applications~\cite{Litjens2017}, but for each application, the network design has usually been adapted and optimized to get the best segmentation performance on each application. The main U-Net design choices can be classified in three categories: \emph{i)} layer choices: convolutional layer size, activation functions, normalization layers, down- and upsampling strategies (\eg~max pooling, striding, deconvolution and repeat); \emph{ii)} the optimization process (gradient descent strategy, weight initialization, loss function, batch size, regularization constraints, stopping criteria, deep supervision, dropout); \emph{iii)} data handling (pre-processing, augmentation, sampling). Since the seminal paper in 2015, several studies based on the EDN structure have been carried out with the goal of outperforming the \mbox{U-Net}. Among those methods, two types of approaches have been proposed: those based on U-Net architecture but with extensions such as shape regularization~\cite{Oktay2018} and those with more sophisticated architectures~\cite{newell2016,zhou2018}. In this context, we decided to benchmark the following EDNs for the purpose of segmenting 2D echocardiographic images:

\subsubsection{U-Net} Taking into account the wide range of possible U-Net designs, we decided to compare the performance of two independent implementations, \ie~\mbox{U-Net 1} optimized for speed, and \mbox{U-Net 2} optimized for accuracy. This leads to two different architectures (which both differ from the original one proposed by Ronneberger \etal) with their own hyperparameters settings, as shown in table \ref{tab:unets}. The “number of feature maps” column given in table \ref{tab:unets} corresponds to the number of convolutions in the convolution layers. For each U-Net implementation, we successively indicate the values for the first, the bottom (where the spatial information is the most compressed), and the last convolution layers. \mbox{U-Net 1 \& 2} enable to investigate the impact of hyperparameters choices on the quality of the results. \modification{The total number of parameters for \mbox{U-Net 1 and 2} are 2.0M and 17.5M, respectively.}
\subsubsection{ACNN} Starting from a given segmentation architecture, this method integrates an auxiliary loss to constrain the segmentation output to fit a non-linear compact representation of the underlying anatomy derived from an auto-encoder network ~\cite{Oktay2018}. For comparison purposes, we used the U-Net 1 architecture described in table~\ref{tab:unets} as the segmentation module in our ACNN implementation. Moreover, the following choices were made to obtain the best results on our dataset: \emph{i)} a code of $32$ coefficients was set for the auto-encoder network (which allows an average reconstruction accuracy of $97\%$); \emph{ii)} the hyperparameter balancing the segmentation and shape regularization losses was set so that the two losses had close initialization values. \modification{The ACNN models in our study have 2.2M parameters.}
\subsubsection{SHG} Stacked Hourglasses (SHG) method integrates three successive encoder-decoder networks (usually three times the same architecture) where the first two are used as residual blocks~\cite{newell2016}. Each output of the encoder-decoder networks is associated with an intermediate segmentation loss. This strategy is \modification{called} deep supervision. The output of the third network is used as the final segmentation result. For comparison purposes, we also used the U-Net 1 architecture as the key encoder-decoder network in our SHG implementation. \modification{The number of parameters of the SHG method is 4.5M (not 6M in order to keep the same batch size as U-Net 1).}
\subsubsection{U-Net++} This method is also based on deep supervision technique but with the integration of additional convolution layers in the form of dense skip connections~\cite{zhou2018}. Starting from the official online version of the code, we adapted the corresponding architecture to obtain the best results on our dataset. The following changes were made: \emph{i)} dropout was removed; \emph{ii)}  averaging of the last feature maps of the intermediate outputs was removed; \emph{iii)} the original design of layers was adapted according to the choices we made to optimized U-Net 1 architecture; \emph{iv)} the batch size was set to 20. \modification{The \mbox{U-Net++} method comprises 1.1M parameters. The original version had 9M parameters but we adapted it, in particular the number of feature maps, so to get the best possible results on the CAMUS dataset.}

\modification{Please note that the same data pre- and post-processing strategies (\ie~connected component analysis keeping the largest region and removing holes) were applied for each of the evaluated method.} For completeness' sake, implementation details of the two \mbox{U-Net} architectures as well as additional tests on ACNN and \mbox{U-Net++} can be found in the supplementary materials. Supplementary materials are available in the supplementary files /multimedia tab.

\begin{table*}[tbp]
\renewcommand{\arraystretch}{1.0}
  \caption{Main characteristics of U-Net 1 and U-Net 2. More details are provided in the supplementary materials /multimedia tab.}
  \centering
  \begin{tabular}{{c}*{8}{c}}

\toprule 

\multicolumn{1}{c}{\multirow{2}{*}{Architectures}} & Number of & Lowest & Upsampling & Normalization & Batch & Learning & Loss & \# Trainable\\
 & \correction{feature maps} & Resolution & Scheme & Scheme & Size & Rate & Function & Parameters \\

\midrule
\addlinespace[+1.5ex]

\multicolumn{1}{c}{\multirow{1}{*}{U-Net 1}} & $32 \downarrow 128 \uparrow 16$  & $8*8$  &  $2*2$ repeats  & None & 32 & 1e-3 & \correction{Multi-class Dice} & 2M \\
\addlinespace[+1.5ex]
\multicolumn{1}{c}{\multirow{2}{*}{U-Net 2}} & \multicolumn{1}{c}{\multirow{2}{*}{$32 \downarrow 512 \uparrow 32$}}  & \multicolumn{1}{c}{\multirow{2}{*}{$16*16$}} & \multicolumn{1}{c}{\multirow{2}{*}{Deconvolutions}} & \multicolumn{1}{c}{\multirow{2}{*}{BatchNorm}} & \multicolumn{1}{c}{\multirow{2}{*}{10}} & \multicolumn{1}{c}{\multirow{2}{*}{1e-4}}  & \correction{Cross-entropy} & \multicolumn{1}{c}{\multirow{2}{*}{18M}}  \\  
 & & & & & & & \correction{+ weight decay} & \\  

\addlinespace[+0.5ex]
\bottomrule
\label{tab:unets}
  \end{tabular}
\end{table*}

\subsection{Non-deep learning state-of-the-art techniques}

To compare the performance of the EDN methods described above, we implemented the following non-deep learning \mbox{state-of-the-art} methods which obtained among the best results during the CETUS challenge~\cite{Bernard2016} and which were recently improved \cite{Pedrosa2017} and applied in 2D~\cite{Leclerc2017}.

\subsubsection{SRF} Structured Random Forests (SRF) refer to an ensemble learning method for classification or regression. It operates at training time by building a set of decision trees that assign a label patch to each input image patch, computed as the mean prediction of the individual trees~\cite{dollar2015}. During the training phase, each tree individually learns a set of split functions from a random subset of the training dataset and input features. Those functions are intended to group patches sharing close image intensities and segmentation patterns. During the testing phase, the image to segment is fragmented into different overlapping patches. Each image patch goes through the splitting functions of each tree so that the mean label patch computed from the reached leaves forms its segmentation. Detailed description of the SRF algorithm implemented in this project can be found in~\cite{Leclerc2017}. Compared to our previous study, data was not split between ES and ED nor between 4 chambers and 2 chambers views but processed in one indistinctive pool of images. Since CAMUS has a larger number of patients than the dataset used in~\cite{Leclerc2017}, we trained $12$ individual trees for each subset of $100$ patients.

\subsubsection{BEASM} The key concept of the B-Spline Explicit Active Surface Model (BEASM) framework is to consider the boundary of a deformable interface as an explicit function, where one of the coordinates of the points within the surface is given explicitly as a function $\psi$ of the remaining coordinates. In this framework, $\psi$ is defined as a linear combination of B-spline basis functions whose controlled knots are located on a regular rectangular grid defined on the chosen coordinate system (polar space in our case). Based on a standard variational approach, the evolution of the deformable surface is then governed by the minimization of an energy function according to the B-spline coefficients~\cite{Barbosa2012}. This framework has been successfully applied in~\cite{Pedrosa2016} for the coupled segmentation of the \LVendo~and \LVepi~structures in echocardiography and further extended by the integration of a shape prior directly into the B-spline space in ~\cite{Pedrosa2017}, named as BEASM in the rest of the paper. Because BEASM amounts to a deformable-based model, the initialization of the contour plays a crucial role on the quality of the results. We thus decided to implement two different strategies: \emph{i)} one named BEASM-fully where the evolving contour is automatically initialized from a method inspired by the work proposed in~\cite{Barbosa2013}; \emph{ii)} another named BEASM-semi where the evolving contour is initialized from three points (two at the base and one at the apex of the \LVendo~structure) extracted from the reference contours. By doing so, we gave the possibility to quantify the influence of the initialization procedure for BEASM on an heterogeneous ultrasound dataset.

\vspace*{-0.4cm}
\section{Results}
\label{sec:results}

As stated in Section \ref{sec:dataset}, the 19\% of poor quality images were not used to compute the different metrics provided in this part. Moreover, to avoid the use of different models according to the acquisition settings, we trained only one model for each machine learning method on both apical four-chamber and two-chamber views regardless of the time instant in the cardiac sequence. For a detailed analysis of the results, a set of complementary information are provided in the supplementary materials (\ie~figures of segmentation results for all the methods, Bland-Altman plots, limitations examples). Supplementary materials are available in the supplementary files /multimedia tab.

\subsection{Evaluation metrics}
\label{sec:evaluation_metrics}

\subsubsection{Geometrical metrics}
To measure the accuracy of the segmentation output (\LVendo, \LVepi~or LA) of a given method, the Dice metric, the mean absolute distance ($d_{m}$) and the 2D Hausdorff distance ($d_H$) were used. The Dice similarity index is defined as \mbox{$D=2\left(\left|S_{user}\cap S_{ref}\right|\right)/\left(\left|S_{user}\right|+\left|S_{ref}\right|\right)$} and is a measure of overlap between the segmented surface $S_{user}$ extracted from a method and the corresponding reference surface $S_{ref}$.  The Dice index gives a value between 0 (no overlap) and 1 (full overlap). $d_{m}$ corresponds to the average distance between $S_{user}$ and $S_{ref}$ while $d_H$ measures the local maximum distance between the two surfaces.

\subsubsection{Clinical metrics} 
We gauge the methods' performance with $3$ clinical indices: \emph{i)} the ED volume (\LVedv~in $ml$); \emph{ii)} the ES volume (\LVesv~in $ml$); \emph{iii)} the ejection fraction (\LVef~as a percentage) for which we computed four metrics: the correlation ($corr$), the bias and the standard deviation ($std$) values (computed from conventional definitions) and the mean absolute error ($mae$). The combination of the bias and standard deviation also provides useful information on the corresponding limit of agreement values.

\subsection{Empirical results} 
\begin{table*}[tbp]
\renewcommand{\arraystretch}{1.0}
  \caption{Segmentation accuracy (\LVendo~and \LVepi) of the $8$ evaluated methods on the ten test folds of table \ref{tab:dataset} restricted to patients having good \& medium image quality ($406$ patients in total). All the metrics were computed using the annotations of expert \emph{O\textsubscript{1a}}. \modification{All the scores obtained with the \mbox{U-Net 2}, which is the overall best performing method, were statistically different with a p-value $< 0.05$ (computed with a Wilcoxon signed-rank test) compared to all the tested methods, except \mbox{U-Net 1} and SHG for the \emph{d\textsubscript{H}} metric.}}
 \centering
  \begin{tabular}{c*{6}{c}|*{6}{c}}

 \toprule 
 \multicolumn{1}{c}{\multirow{6}{*}{\bf \small Methods *}} &  \multicolumn{6}{c}{\bf \small ED} & \multicolumn{6}{c}{\bf \small ES} \\
 \cmidrule(r){2-7} \cmidrule(r){8-13}
 &  \multicolumn{3}{c}{\bf$\boldsymbol{\LVendo}$} & \multicolumn{3}{c}{\bf $\boldsymbol{\LVepi}$} & 
 \multicolumn{3}{c}{\bf $\boldsymbol{\LVendo}$} & \multicolumn{3}{c}{\bf $\boldsymbol{\LVepi}$} \\
 \cmidrule(r){2-4} \cmidrule(r){5-7} \cmidrule(r){8-10} \cmidrule(r){11-13}
 & \multicolumn{1}{c}{\bf \emph{D}} & \multicolumn{1}{c}{\bf \emph{d\textsubscript{m}}} & \multicolumn{1}{c}{\bf \bf \emph{d\textsubscript{H}}} & \multicolumn{1}{c}{\bf \emph{D}} & \multicolumn{1}{c}{\bf \emph{d\textsubscript{m}}} & \multicolumn{1}{c}{\bf \emph{d\textsubscript{H}}} & \multicolumn{1}{c}{\bf \emph{D}} & \multicolumn{1}{c}{\bf \emph{d\textsubscript{m}}} & \multicolumn{1}{c}{\bf \emph{d\textsubscript{H}}} & \multicolumn{1}{c}{\bf \emph{D}} & \multicolumn{1}{c}{\bf \emph{d\textsubscript{m}}} & \multicolumn{1}{c}{\bf \emph{d\textsubscript{H}}} \\
\cmidrule(r){2-2} \cmidrule(r){3-3} \cmidrule(r){4-4} \cmidrule(r){5-5} \cmidrule(r){6-6} \cmidrule(r){7-7} \cmidrule(r){8-8} \cmidrule(r){9-9} \cmidrule(r){10-10} \cmidrule(r){11-11} \cmidrule(r){12-12} \cmidrule(r){13-13}
 & val. & mm & mm & val.& mm & mm & val. & mm & mm & val. & mm & mm \\
 
\midrule

\multicolumn{1}{l}{\emph{O\textsubscript{1a}} vs \emph{O\textsubscript{2}}} & 0.919 & 2.2 & 6.0 & 0.913
& 3.5 & 8.0 & 0.873 & 2.7 & 6.6 & 0.890
& 3.9 & 8.6 \\

\multicolumn{1}{l}{(inter-obs)} & \scriptsize{$\pm$0.033} & \scriptsize{$\pm$0.9}
& \scriptsize{$\pm$2.0} & \scriptsize{$\pm$0.037}
& \scriptsize{$\pm$1.7} & \scriptsize{$\pm$2.9}
& \scriptsize{$\pm$0.060} & \scriptsize{$\pm$1.2}
& \scriptsize{$\pm$2.4} & \scriptsize{$\pm$0.047}
& \scriptsize{$\pm$1.8} & \scriptsize{$\pm$3.3} \\

\addlinespace[+0.7ex]

\multicolumn{1}{l}{\emph{O\textsubscript{1a}} vs \emph{O\textsubscript{3}}} & 0.886 & 3.3 & 8.2 & 0.943
& 2.3 & 6.5 & 0.823 & 4.0 & 8.8 & 0.931
& 2.4 & 6.4 \\

\multicolumn{1}{l}{(inter-obs)} & \scriptsize{$\pm$0.050} & \scriptsize{$\pm$1.5}
& \scriptsize{$\pm$2.5} & \scriptsize{$\pm$0.018}
& \scriptsize{$\pm$0.8} & \scriptsize{$\pm$2.6}
& \scriptsize{$\pm$0.091} & \scriptsize{$\pm$2.0}
& \scriptsize{$\pm$3.5} & \scriptsize{$\pm$0.025}
& \scriptsize{$\pm$1.0} & \scriptsize{$\pm$2.4} \\

\addlinespace[+0.7ex]

\multicolumn{1}{l}{\emph{O\textsubscript{2}} vs \emph{O\textsubscript{3}}} & 0.921 & 2.3 & 6.3 & 0.922
& 3.0 & 7.4 & 0.888 & 2.6 & 6.9 & 0.885
& 3.9 & 8.4 \\
\multicolumn{1}{l}{(inter-obs)} & \scriptsize{$\pm$0.037} & \scriptsize{$\pm$1.2}
& \scriptsize{$\pm$2.5} & \scriptsize{$\pm$0.036}
& \scriptsize{$\pm$1.5} & \scriptsize{$\pm$3.0}
& \scriptsize{$\pm$0.058} & \scriptsize{$\pm$1.3}
& \scriptsize{$\pm$2.9} & \scriptsize{$\pm$0.054}
& \scriptsize{$\pm$1.9} & \scriptsize{$\pm$2.8} \\

\addlinespace[+0.7ex] 

\multicolumn{1}{l}{\emph{O\textsubscript{1a}} vs \emph{O\textsubscript{1b}}} & 0.945 & 1.4 & 4.6 & 0.957
& 1.7 & 5.0 & 0.930 & 1.3 & 4.5 & 0.951
& 1.7 & 5.0 \\

\multicolumn{1}{l}{(intra-obs)} & \scriptsize{$\pm$0.019} & \scriptsize{$\pm$0.5}
& \scriptsize{$\pm$1.8} & \scriptsize{$\pm$0.019}
& \scriptsize{$\pm$0.9} & \scriptsize{$\pm$2.3}
& \scriptsize{$\pm$0.031} & \scriptsize{$\pm$0.5}
& \scriptsize{$\pm$1.8} & \scriptsize{$\pm$0.021}
& \scriptsize{$\pm$0.8} & \scriptsize{$\pm$2.1} \\
 
\addlinespace[+0.7ex]  
 
\midrule

\addlinespace[+0.7ex]

\multicolumn{1}{l}{\multirow{2}{*}{SRF}} & 0.895 & 2.8 & 11.2 & 0.914 & 3.2 & 13.0 & 0.848 & 3.6 & 11.6 & 0.901 & 3.5 & 13.0 \\

 & \scriptsize{$\pm$0.074} & \scriptsize{$\pm$3.6}
& \scriptsize{$\pm$10.2} & \scriptsize{$\pm$0.057}
& \scriptsize{$\pm$2.0} & \scriptsize{$\pm$9.1}
& \scriptsize{$\pm$0.137} & \scriptsize{$\pm$7.8}
& \scriptsize{$\pm$13.6} & \scriptsize{$\pm$0.078}
& \scriptsize{$\pm$4.7} & \scriptsize{$\pm$11.1} \\

\addlinespace[+0.7ex]

\multicolumn{1}{l}{\multirow{2}{*}{BEASM-fully}} & 0.879 & 3.3 & 9.2 & 0.895 & 3.9 & 10.6 & 0.826 & 3.8 & 9.9 & 0.880 & 4.2 & 11.2 \\

 & \scriptsize{$\pm$0.065} & \scriptsize{$\pm$1.8}
& \scriptsize{$\pm$4.9} & \scriptsize{$\pm$0.051}
& \scriptsize{$\pm$2.1} & \scriptsize{$\pm$5.1}
& \scriptsize{$\pm$0.092} & \scriptsize{$\pm$2.1}
& \scriptsize{$\pm$5.1} & \scriptsize{$\pm$0.054}
& \scriptsize{$\pm$2.0} & \scriptsize{$\pm$5.1} \\

\addlinespace[+0.7ex]

\multicolumn{1}{l}{\multirow{2}{*}{BEASM-semi}} & 0.920 & 2.2 & 6.0 & 0.917 & 3.2 & 8.2 & 0.861 & 3.1 & 7.7 & 0.900 & 3.5 & 9.2 \\

 & \scriptsize{$\pm$0.039} & \scriptsize{$\pm$1.2}
& \scriptsize{$\pm$2.4} & \scriptsize{$\pm$0.038}
& \scriptsize{$\pm$1.6} & \scriptsize{$\pm$3.0}
& \scriptsize{$\pm$0.070} & \scriptsize{$\pm$1.6}
& \scriptsize{$\pm$3.2} & \scriptsize{$\pm$0.042}
& \scriptsize{$\pm$1.7} & \scriptsize{$\pm$3.4} \\

\addlinespace[+0.7ex]

\midrule

\addlinespace[+0.7ex]

\multicolumn{1}{l}{\multirow{2}{*}{U-Net~1}} &  0.934 & 1.7 & 5.5 & 0.951
& 1.9 & 5.9 & 0.905 & 1.8 & 5.7 & 0.943 & 2.0 & 6.1 \\

& \scriptsize{$\pm$0.042} & \scriptsize{$\pm$1.0}
& \scriptsize{$\pm$2.9} & \scriptsize{$\pm$0.024}
& \scriptsize{$\pm$0.9} & \scriptsize{$\pm$3.4}
& \scriptsize{$\pm$0.063} & \scriptsize{$\pm$1.3}
& \scriptsize{$\pm$3.7} & \scriptsize{$\pm$0.035}
& \scriptsize{$\pm$1.2} & \scriptsize{$\pm$4.1} \\

\addlinespace[+0.7ex]

\multicolumn{1}{l}{\multirow{2}{*}{U-Net~2}} & \bf 0.939 & \bf 1.6 & \bf 5.3 & \bf 0.954 & \bf 1.7 & 6.0 & \bf 0.916 & \bf 1.6 & \bf 5.5 & \bf 0.945
& \bf 1.9 & 6.1 \\

& \scriptsize{$\pm$0.043} & \scriptsize{$\pm$1.3}
& \scriptsize{$\pm$3.6} & \scriptsize{$\pm$0.023}
& \scriptsize{$\pm$0.9} & \scriptsize{$\pm$3.4}
& \scriptsize{$\pm$0.061} & \scriptsize{$\pm$1.6}
& \scriptsize{$\pm$3.8} & \scriptsize{$\pm$0.039}
& \scriptsize{$\pm$1.2} & \scriptsize{$\pm$4.6} \\

\addlinespace[+0.7ex]

\multicolumn{1}{l}{\multirow{2}{*}{ACNN}} & 0.932 & 1.7 & 5.8 & 0.950 & 1.9 & 6.4 & 0.903 & 1.9 & 6.0 & 0.942 & 2.0 & 6.3 \\
 & \scriptsize{$\pm 0.034$} & \scriptsize{$\pm 0.9$}
& \scriptsize{$\pm 3.1$} & \scriptsize{$\pm 0.026$}
& \scriptsize{$\pm 1.1$} & \scriptsize{$\pm 4.1$}
& \scriptsize{$\pm 0.059$} & \scriptsize{$\pm 1.1$}
& \scriptsize{$\pm 3.9$} & \scriptsize{$\pm 0.034$}
& \scriptsize{$\pm 1.2$} & \scriptsize{$\pm 4.2$} \\

\addlinespace[+0.7ex]

\multicolumn{1}{l}{\multirow{2}{*}{SHG}} & 0.934 & 1.7 & 5.6 & 0.951 & 1.9 & \bf 5.7 & 0.906 & 1.8 & 5.8 & 0.944 & 2.0 & \bf 6.0\\

 & \scriptsize{$\pm 0.034 $} & \scriptsize{$\pm 0.9$}
& \scriptsize{$\pm 2.8 $} & \scriptsize{$\pm 0.023$}
& \scriptsize{$\pm 1.0 $} & \scriptsize{$\pm 3.3$}
& \scriptsize{$\pm 0.057$} & \scriptsize{$\pm 1.1$}
& \scriptsize{$\pm 3.8$} & \scriptsize{$\pm 0.034$}
& \scriptsize{$\pm 1.2$} & \scriptsize{$\pm 4.3$} \\

\addlinespace[+0.7ex]

\multicolumn{1}{l}{\multirow{2}{*}{U-Net~++}} & 0.927 & 1.8 & 6.5 & 0.945 & 2.1 & 7.2 & 0.904 & 1.8 & 6.3 & 0.939 & 2.1 & 7.1 \\

 & \scriptsize{$\pm0.046$} & \scriptsize{$\pm 1.1$} 
 & \scriptsize{$\pm 3.9$} & \scriptsize{$\pm 0.026$} 
 & \scriptsize{$\pm 1.0$} & \scriptsize{$\pm 4.5$} 
 & \scriptsize{$\pm 0.060 $} & \scriptsize{$\pm 1.0$}
& \scriptsize{$\pm4.2$} & \scriptsize{$\pm 0.034$}
& \scriptsize{$\pm 1.1$} & \scriptsize{$\pm 5.1$} \\


%
%

\addlinespace[+0.7ex]

\bottomrule

\multicolumn{13}{l}{} \\
\multicolumn{13}{l}{*~\emph{\LVendo: Endocardial contour of the left ventricle; \LVepi: Epicardial contour of the left ventricle; ED: End diastole}} \\
\multicolumn{13}{l}{~~\emph{ES: End systole; D: Dice index; d\textsubscript{m}: mean absolute distance; d\textsubscript{H}: Hausdorff distance; mae: mean absolute error}} \\
\multicolumn{13}{l}{~~\emph{The values in bold refer to the best performance for each measure.}}\\
\multicolumn{13}{l}{~~\emph{The inter and intra-observer measurements were computed from fold 5 restricted to patients having good \& medium}}\\
\multicolumn{13}{l}{~~\emph{image quality (40 patients)}\vspace{-0.4cm}}
  \label{tab:comparison_segmentation_dl_vs_others}
  \end{tabular}
\end{table*}

\subsubsection{Geometrical scores}

Table \ref{tab:comparison_segmentation_dl_vs_others} shows the segmentation testing accuracy computed from patients having good and medium image quality ($406$ patients) for the 8 algorithms described in section \ref{sec:evaluated_methods}. Mean and standard deviation values for each metric were obtained from cross-validating on the $10$ folds of the dataset. The values in bold correspond to the best scores for each metric. From these results, one can see that the EDN implementations get the overall best segmentation scores on all metrics, for both ED and ES. Interestingly, while the EDN methods are fully-automatic, they still get better segmentation results than the semi-automatic BEASM algorithm.

The two U-Nets achieve equivalent results for all the metrics compared to the ones obtained by the more sophisticated encoder-decoder architectures. This hints to the idea that a plateau has been reached, which classical tuning, shape regularization techniques and more sophisticated architectures have difficulties to overcome. This also suggests that a \mbox{U-Net} implementation, which requires less parameters than SHG and U-Net++ methods and less training time than ACNN, offers the best compromise between the network size and performance for the particular task of 2D echocardiographic image segmentation.

To assess the influence of the layer design in the performance between \mbox{U-Net 1 and 2}, statistical significance of their respective results was analyzed by performing the Wilcoxon signed-rank test for each metric. \modification{Results showed that \mbox{U-Net 1 and 2} produced scores that are statistically different (\mbox{p-value $<0.05$}) for most metrics at ED and ES, apart for the \LVepi~Hausdorff distance.} However, this must be nuanced by the fact that \emph{i)} the \mbox{U-Net} geometrical scores are very close (mean \emph{d\textsubscript{m}} and \emph{d\textsubscript{H}} difference of $0.1$~mm and $0.1$~mm, respectively), producing distributions with high degree of overlap as shown in the supplementary materials /multimedia tab; \emph{ii)} the \mbox{U-Net} geometrical results lie between the inter-observer and intra-observer scores for all metrics, proving the robustness of this method in obtaining accurate segmentation results. As a complement to the above, we investigated in the supplementary materials /multimedia tab the influence of each of the \mbox{U-Net} hyperparameters presented in table~\ref{tab:unets}. Results highlight the importance of the choice of the normalization scheme.

As for the fully automatic non-deep learning state-of-the-art methods, BEASM-auto obtained on average better Hausdorff distances (mean $d_H$ of $9.9$~mm at ED and $10.5$~mm at ES) while the SRF got better Dice and $d_m$ scores (mean $d_m$ of $3.0$~mm at ED and $3.5$~mm at ES). However, the large standard deviation values for the SRF illustrate the difficulties of this method in obtaining consistent segmentations over the entire dataset. As for the BEASM-semi, one can see that the manual initialization has a strong impact on the quality of the results, with a mean improvement of $0.8$~mm and $2.4$~mm for the $d_m$ and $d_H$ metrics, respectively. Moreover, it is well known that the left ventricle shape is more difficult to segment at ES, leading to slightly worse performance for classical algorithms on this time instant. This property is also confirmed in our study since all the evaluated methods produced better results at ED on every metric. 

As complement, we provided in the supplementary materials /multimedia tab the geometrical scores obtained on the poor quality images ($94$ patients) for the 8 evaluated algorithms. For this part of the dataset, the EDNs also obtained the best segmentation results on all metrics. Interestingly, while EDN scores on poor quality images are slightly worse than those computed on good and medium quality, they remain very competitive compared to the scores given in table~\ref{tab:comparison_segmentation_dl_vs_others} (mean \LVendo~\emph{d\textsubscript{m}} and \emph{d\textsubscript{H}} of \correction{$2.2$~mm and $7.0$~mm and mean \LVepi~\emph{d\textsubscript{m}} and \emph{d\textsubscript{H}} of $2.3$~mm and $7.6$~mm}).

\begin{table*}[tbp]
\renewcommand{\arraystretch}{1.0}
  \caption{Clinical metrics of the $8$ evaluated methods on the ten test folds of table \ref{tab:dataset} restricted to patients having good \& medium image quality ($406$ patients in total). All the metrics were computed using the annotations of expert \emph{O\textsubscript{1a}}. \modification{Volumes and ejection fraction obtained with \mbox{U-Net 2} were statistically different with p-values $<0.05$ (computed with the Wilcoxon signed-rank test) compared to all the tested methods, except SHG for the \LVesv indice.}}
  \centering
  \begin{tabular}{c*{3}{c}|*{3}{c}|*{3}{c}}

 \toprule 
 \multicolumn{1}{c}{\multirow{5}{*}{\bf \small Methods *}} &  \multicolumn{3}{c}{\bf \small \emph{LV\textsubscript{EDV}}} & \multicolumn{3}{c}{\bf \small \emph{LV\textsubscript{ESV}}} & \multicolumn{3}{c}{\bf \small \emph{LV\textsubscript{EF}}}\\
 \cmidrule(r){2-4} \cmidrule(r){5-7} \cmidrule(r){8-10}
 & \multicolumn{1}{c}{\bf \emph{corr}} & \multicolumn{1}{c}{\bf \emph{bias$\pm\sigma$}} & \multicolumn{1}{c}{\bf \emph{mae}} & \multicolumn{1}{c}{\bf \emph{corr}} & \multicolumn{1}{c}{\bf \emph{bias$\pm\sigma$}} & \multicolumn{1}{c}{\bf \emph{mae}} & \multicolumn{1}{c}{\bf \emph{corr}} & \multicolumn{1}{c}{\bf \emph{bias$\pm\sigma$}} & \multicolumn{1}{c}{\bf \emph{mae}} \\
\cmidrule(r){2-2} \cmidrule(r){3-3} \cmidrule(r){4-4} \cmidrule(r){5-5} \cmidrule(r){6-6} \cmidrule(r){7-7} \cmidrule(r){8-8} \cmidrule(r){9-9} \cmidrule(r){10-10}
 & val. & ml & ml & val.& ml & ml & val. & \% & \% \\
 
\midrule

\multicolumn{1}{l}{\emph{O\textsubscript{1a}} vs \emph{O\textsubscript{2}} (inter-obs)} & 0.940 & 18.7$\pm$12.9& 18.7 & 0.956 & 18.9$\pm$9.3 & 18.9 & 0.801 & -9.1$\pm$8.1 & 10.0 \\

\multicolumn{1}{l}{\emph{O\textsubscript{1a}} vs \emph{O\textsubscript{3}} (inter-obs)} & 0.895 & 39.0$\pm$18.8& 39.0 & 0.860 & 35.9$\pm$17.1 & 35.9 & 0.646 & -12.6$\pm$10.0 & 13.4 \\

\multicolumn{1}{l}{\emph{O\textsubscript{2}} vs \emph{O\textsubscript{3}} (inter-obs)} & 0.926 & -20.3$\pm$15.6& 21.0 & 0.916 & -17.0$\pm$13.5 & 17.7 & 0.569 & 3.5$\pm$11.0 & 8.5 \\

\multicolumn{1}{l}{\emph{O\textsubscript{1a}} vs \emph{O\textsubscript{1b}} (intra-obs)} & 0.978 & -2.8$\pm$7.1& 6.2 & 0.981 & -0.1$\pm$5.8 & 4.5 & 0.896 & -2.3$\pm$5.7 & 0.9 \\

\midrule

\multicolumn{1}{l}{SRF} & 0.755 & -0.2$\pm$25.7 & 17.4 & 0.827 & 9.3$\pm$18.0 & 14.8 & 0.465 & -11.5$\pm$15.4 & 12.8 \\

\multicolumn{1}{l}{BEASM-fully} & 0.704 & 13.4$\pm$30.6 & 22.9 & 0.713 & 18.0$\pm$25.8 & 22.5 & 0.731 & -9.8$\pm$8.3 & 10.7 \\

\multicolumn{1}{l}{BEASM-semi} & 0.886 & 14.6$\pm$19.2 & 17.8 & 0.880 & 18.3$\pm$16.9 & 19.5 & 0.790 & -9.4$\pm$7.2 & 10.0 \\

\midrule

\multicolumn{1}{l}{U-Net~1} & 0.947 & -8.3$\pm$12.6 & 10.9 & 0.955 & -4.9$\pm$9.9 & 8.2 & 0.791 & -0.5$\pm$7.7 & 5.6 \\

\multicolumn{1}{l}{U-Net~2} & \bf 0.954 & -6.9 $\pm$11.8 & \bf 9.8 & \bf 0.964 & -3.7$\pm$ 9.0 & \bf 6.8  & \bf 0.823 & -1.0$\pm$7.1 & \bf 5.3\\

\multicolumn{1}{l}{ACNN} & 0.945 &-6.7$\pm $ 12.9 & 10.8 & 0.947 & -4.0$\pm $ 10.8 & 8.3 & 0.799& -0.8$\pm$7.5 & 5.7\\

\multicolumn{1}{l}{SHG} & 0.943 & \-6.4$\pm$12.8 & 10.5 & 0.938 & -3.2$\pm$11.3 & 8.2 &  0.770 & -1.4$\pm$7.8 & 5.7\\

\multicolumn{1}{l}{U-Net~++} & 0.946 & -11.4$\pm$ 12.9& 13.2 &  0.952& -5.7$\pm$ 10.7 & 8.6& 0.789& -1.8$\pm$ 7.7& 5.6 \\

%

\bottomrule
 
 \multicolumn{10}{l}{} \\

 \multicolumn{10}{l}{*~\emph{corr: Pearson correlation coefficient; mae: mean absolute error.}} \\
 \multicolumn{10}{l}{~~\emph{The values in bold refer to the best performance for each measure.}}\\
 \multicolumn{10}{l}{~~\emph{The inter and intra-observer measurements were computed from fold 5 restricted to patients having good \& medium}}\\
 \multicolumn{10}{l}{~~\emph{image quality (40 patients.)}\vspace{-0.4cm}}
 
\label{tab:comparison_clinical_indices_dl_vs_others_full}
  \end{tabular}
\end{table*}


\subsubsection{Clinical scores}

Table \ref{tab:comparison_clinical_indices_dl_vs_others_full} contains the clinical metrics for the $8$ methods. Those indices were computed with the Simpson's rule \cite{Folland1979} from the segmentation results of each algorithm. The values in bold represent the best scores for the corresponding index.
As for segmentation, the EDNs obtained the best clinical scores on all the tested metrics (bias was not taken into account since the lowest bias value in itself does not necessarily mean the best performing method). Regarding the estimation of the \LVedv~and \LVesv, the EDNs obtained high correlation scores (all above $0.94$) and reasonably small biases (at most $11.4$ ml), standard deviations (less than $12.9$ ml) and mean absolute errors (at most $13.2$ ml). Results are more contrasted for the estimation of the \LVef. For this metric, the EDNs got lower correlation scores (at most $0.82$) but smaller biases (less than $1.8$ \%), standard deviations (at most $7.8$ \%) and mean absolute errors (less than $5.7$ \%). It is worth pointing out that average EDN scores are all below the inter-observer scores. This proves the clinical interest of such approaches but also reveals the needs for improvement as discussed in Section~\ref{sec:discussions}. Here again, even if the U-Net methods involved simpler architecture, they obtained similar results compared to the more sophisticated EDNs. \modification{Finally, using the Wilcoxon signed-rank test, \mbox{U-Net 1} and \mbox{U-Net 2} produced \LVedv,~\LVesv~and \LVef~results whose difference is statistically significant (p-values $<0.05$), although their measurements are very close.}

\subsection{U-Net behavior}
\label{sec:unet_behavior}

\begin{figure*}[htbp]
\centering
\includegraphics[height=0.28\textheight]{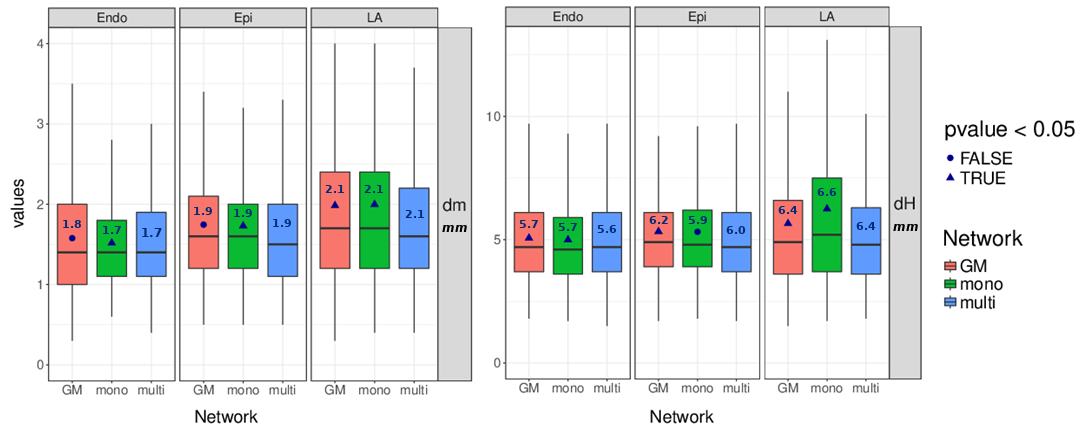}
\caption{Tukey box plots computed from the geometrical results of the U-Net 1 architecture for three different schemes (\emph{GM} for learning to simultaneously segment all three structures from good \& medium image quality, \emph{mono} for learning to segment one structure from all image quality, \emph{multi} for learning to  simultaneously segment all three structures from all image quality). Blue numbers correspond to mean values computed from each set of measurements. \modification{All p-values are based on the Wilcoxon signed-rank test computed with the \emph{multi} strategy as reference.}}
\label{fig:boxplots}
\end{figure*}

From the results given in table \ref{tab:comparison_segmentation_dl_vs_others} and \ref{tab:comparison_clinical_indices_dl_vs_others_full}, it appears that the \mbox{U-Net} method has the most effective architecture among the tested EDN models in terms of trade-off between the number of parameters and the achieved performance for the particular task of 2D echocardiographic image analysis. To better analyze the behavior of this model, we set up several additional experiments whose results are provided in Fig.~\ref{fig:boxplots} and Fig.~\ref{fig:inc_size}. For all these experiments, even if the acquisitions were optimized to perform \LVef~measurements (meaning that part of the LA may or may not be fully visible depending on the acquisitions), we also investigated the capacity of \mbox{U-Net} to segment the LA in addition to the $LV_{endo}$ and $LV_{epi}$.
Moreover, since the two tested \mbox{U-Nets} produced overall close geometrical and clinical scores, we only used in this part the \mbox{U-Net 1} model since it requires considerably less parameters to learn. Finally, all the given metrics were computed from both four and two-chamber views and at ED and ES time instants to facilitate the interpretation of the results.

\vspace*{0.1cm}

\subsubsection{Mono versus multi-structures approaches}
\label{sec: dlb1}
We assessed the influence of learning strategies on the quality of the segmentation of the \LVendo, \LVepi~and LA. In particular, we trained 4 models with the same \mbox{U-Net 1} architecture but with different training sets including all image quality, \ie~one network trained on predicting only the \LVendo, one the \LVepi, one the LA, and one all structures. Results on the full dataset are plotted in green and blue in Fig.~\ref{fig:boxplots} and are referred to as {\em mono} and {\em multi}.

From the derived box plots, one can see that, unrelated to the structure, the mono and multi-structures approaches produced very close results even if the corresponding differences are statistically different. These results show that, with the proposed implementations, learning the segmentation of one structure (\eg~\LVendo) in the context of the others (\eg~\LVepi~\&~LA) does not improve significantly the results compared to learning the segmentation of the structure alone. This hints at designing dedicated architectures and/or loss functions to better exploit the contextual information provided in the segmentation masks. \correction{Furthermore, even if the segmentation of the LA structure is challenging compared to ~\LVepi~\ and \LVendo~\ due to acquisition conditions, the \mbox{U-Net 1} manages to get close results both in terms of mean absolute distance (mean $d_m$ equals to $1.7$, $1.9$ and $2.1$~mm for the \LVendo, \LVepi~and LA respectively) and average Hausdorff distance (mean $d_H$ equals to $5.6$, $6.0$ and $6.4$~mm for the \LVendo, \LVepi~and LA respectively).}

\vspace*{0.2cm}

\subsubsection{The effect of poor quality images}
\label{sec: dlb2}
We investigated in Fig.~\ref{fig:boxplots} the influence of involving images of poor quality during the training phase. Based on a multi-structures scheme, we trained two \mbox{U-Net 1} models with the same architecture, one using the full training dataset not caring for image quality (plotted in blue and referred as \emph{multi} in Fig.~\ref{fig:boxplots}) and one using the training dataset restricted to patients having good and medium image quality (plotted in red and referred as \emph{GM} in Fig.~\ref{fig:boxplots}). From the obtained box plots, one can observe that the two different strategies produced very close results \correction{even if the corresponding differences are mostly statistically significant (apart for the $d_m$ metric for the \LVendo~and \LVepi).} These results suggest that the $19$\% ($94$ patients) of poor quality images \emph{i)} do not bring additional information (supporting that the remaining deep learning issues are weakly linked to image quality); \emph{ii)} do not decrease performance compared to a model trained on the $406$ patients with good and medium image quality. This result suggests that poor image quality, in itself, does not complicate the segmentation task as much as could be expected and that encoder-decoder based techniques are able to cope with the variability in image quality found in echocardiography.

\vspace*{0.2cm}

\subsubsection{Influence of the size of the training dataset}
\begin{figure}[tbp]
\centering\includegraphics[height=0.26\textheight]{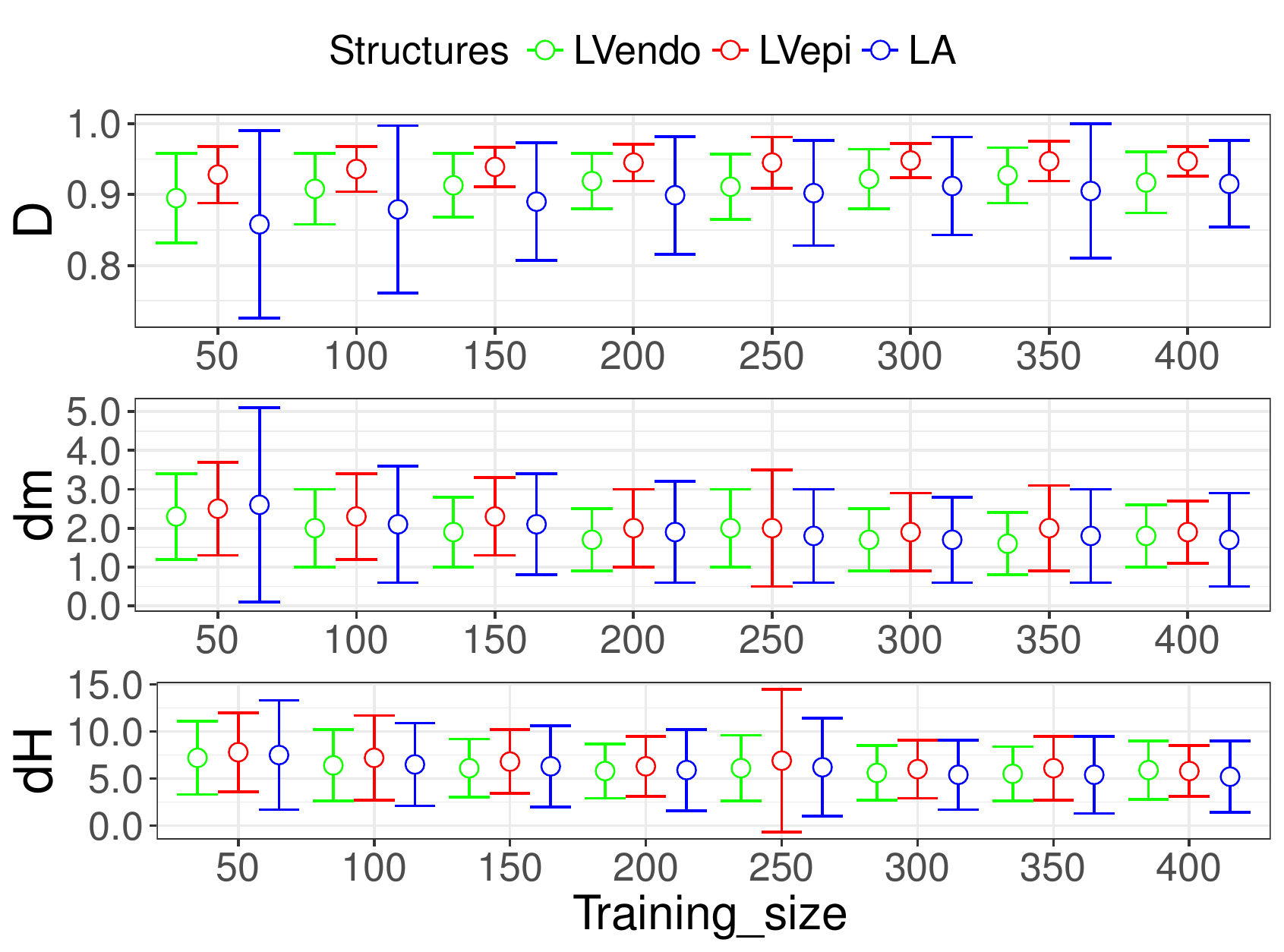}
\caption{Evolution of the segmentation scores (computed from fold 5 and from both ED and ES time instants) derived from the U-Net 1 architecture according to an incremental increase of the number of patients involved in the training dataset. Dots correspond to mean values while bars correspond to standard deviations.}
\label{fig:inc_size}
\end{figure}

We studied in Fig.~\ref{fig:inc_size} the influence of the size of the training dataset on the quality of the segmentation of the \LVendo, \LVepi~and LA structures. To this aim, we set up $8$ different experiments, where the same fold $5$ and $6$ were respectively used as test and validation sets. As for the training set, starting from 50 patients, we added for each new experiment 50 additional patients until 400 patients was reached for the last trial. In each experiment, the same \mbox{U-Net 1} architecture was used and optimized in the same way to derive the best performing parameters from the validation set. Moreover, the number of training epochs was proportionally lowered to ensure that each network went through the same number of iterations. 
 
From this figure, one can first observe an overall improvement of all metrics for the three cardiac structures with the increasing number of patients in the training set. Interestingly, while the improvement between $50$ to $200$ patients is quite pronounced (\eg~a decrease of $d_H$ for the \LVendo~from $7.2$~mm to $5.8$~mm), one can observe a change in the evolution of the performance of the \mbox{U-Net 1} method from $250$ patients. Indeed, for this particular value, results worsen a bit, which may be explained by the bias brought by the validation and test data as we are not doing cross-validation in this experiment. Moreover, from this value, the $d_m$ scores seem to stabilize around $1.8$~mm for the \LVendo, $1.9$~mm for the \LVepi~and $1.8$~mm for the LA structure. The same conclusions can be made for the Dice metric, with a convergence value around $0.920$ for the \LVendo, $0.947$ for the \LVepi~and $0.909$ for the LA structure. As for the $d_H$ metric, while some improvement can still be observed from $250$ to $400$ patients for the \LVepi~and LA structures ($1.1$~mm and $1.0$~mm for the \LVepi~and the LA, respectively), it is not obvious to draw the same conclusion for the \LVendo~structure since the decrease of its corresponding value is less pronounced ($0.2$~mm). In the light of these results, the \mbox{U-Net 1} implementation performs better that the state-of-the-art non-deep learning methods after training with only $50$ patients. Moreover, this method needs at least $250$ patients during the training phase to reach highly competitive results, which can be slightly improved with a larger training set.
 
\vspace*{0.2cm} 

\subsubsection{Influence of the expert annotations}
\label{sec:annotator}
\begin{figure}[htbp]
	\centering
	{\includegraphics[height=0.24\textheight]{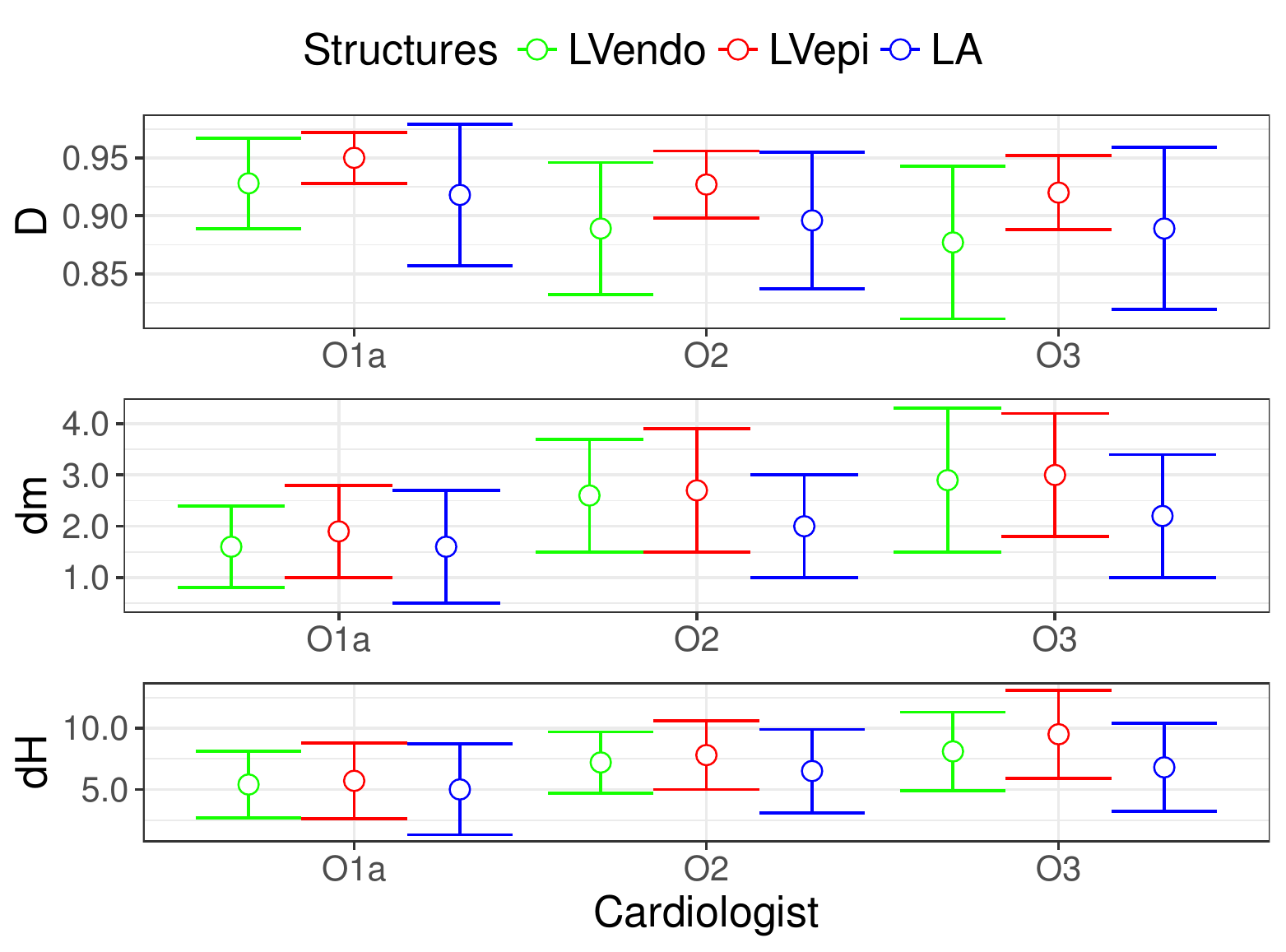}}
	\caption{Geometric scores of the three cardiologist-specific models on 400 patients (1600 images)}%
	\label{fig:bias}
\end{figure}

We investigated in Fig.~\ref{fig:bias} the influence of the expert annotations during the training phase. To this aim, we trained three models on fold 5 from the same \mbox{U-Net 1} architecture based each time on the manual contouring from a different annotator. The validation fold was kept the same for each experiment to avoid any bias error. The models were then evaluated on the remaining 400 patients annotated by cardiologist \emph{O\textsubscript{1a}}.

From this figure, one can observe that the best scores for the three structures are obtained for the model trained on the annotations of cardiologist $O_{1a}$, who performed the manual contouring on the test and validation sets. This observation is consistent with the inter-variability results provided in tables \ref{tab:comparison_segmentation_dl_vs_others} and \ref{tab:comparison_clinical_indices_dl_vs_others_full}. It confirms that cardiologists have consistent differences in their way of contouring images and that an EDN has the capacity to learn a specific way of segmenting.

\vspace*{0.2cm}

\subsubsection{Runtime performance}

The two U-Nets were implemented in Python with the same version of the TensorFlow and Keras libraries and an Nvidia Tesla M60 GPUs (8 Go RAM). Because of the larger number of trainable parameters involved in the \mbox{U-Net 2} solution (see table \ref{tab:unets}), the running time of the two networks is different. For the training phase, the time required to train on $400$ patients is $24\pm5$ \emph{min} and $73\pm~1$ \emph{min} for the \mbox{U-Net 1} and \mbox{U-Net 2}, respectively. At test time, the segmentation of a single image takes $0.09\pm~0.03$ \emph{s} and $0.14\pm~0.06$ \emph{s} for the \mbox{U-Net 1} and \mbox{U-Net 2}, respectively.

\subsection{Discussion}
\label{sec:discussions}

\subsubsection{Statistical differences between U-Net 1 and 2 results}

From Tables \ref{tab:comparison_segmentation_dl_vs_others} and \ref{tab:comparison_clinical_indices_dl_vs_others_full} it has been observed that although \mbox{U-Net} 1 and \mbox{U-Net} 2 have very similar performances, their results were judged most of the time as being statistically different by the Wilcoxon Signed Rank Test (p-value $< 0.05$). This can be explained by the fact that when the number of samples is quite high, any slight but consistent deviation between the two distributions will make the difference statistically significant. In our study, since we worked on a large scale dataset, most of the statistical tests were performed on a large amount of samples (for instance Tables \ref{tab:comparison_segmentation_dl_vs_others} and \ref{tab:comparison_clinical_indices_dl_vs_others_full} involve more than 800 paired observations for each statistical test), encouraging situations where the differences between results produced by two methods are recognized as statistically significant (even if the evaluated distributions are very close).

\vspace*{0.2cm}

\subsubsection{Inter and intra-observer variability}

To further assess the quality of the EDN segmentation results, we added in table~\ref{tab:comparison_segmentation_dl_vs_others} the inter and intra-observer variability measurements computed from fold $5$ (restricted to $40$ patients with good and medium image quality). Concerning the inter-observer variability, the corresponding Dice scores vary between $0.82$ and $0.93$, the $d_m$ between $2.2$~mm and $4.0$~mm and the $d_H$ between $6.0$~mm and $8.8$~mm. The \LVepi~is the most difficult structure to annotate while both \LVendo~and \LVepi~are harder to contour at ES than at ED. One should also note the large $d_m$ value of $4.0$~mm between observer 1a and 3 for the \LVendo~structure at ES. This illustrates \emph{i)} the difficulty in getting coherent manual annotations between experts from daily clinical practice data; \emph{ii)} the difficulty for the experts to use unfamiliar software for the analysis; \emph{iii)} the needs to provide interactively the volumetric results to the experts for instant comparison (this was not done during the manual annotations); \emph{iv)} the difficulty in contouring some data acquired with non-standard views. Concerning the \mbox{intra-observer} variability, one can observe that the results obtained on all the segmentation scores are better than the inter-observer ones, with a mean difference of $1.5$~mm for the $d_m$ metric and $2.6$~mm for the Hausdorff distance. This illustrates the high consistency of manual contouring from experienced cardiologists, even on challenging data. Those results also provide important information on the limits to reach in order to consider that a machine learning algorithm faithfully reproduces the expertise of one cardiologist.

In table~\ref{tab:comparison_clinical_indices_dl_vs_others_full}, we also reported the inter and intra-observer variability measurements computed from fold $5$ (restricted to $40$ patients with good and medium image quality) for the \LVedv, \LVesv~and \LVef~metrics. From these results, one can observe that the experts reached good agreements for the estimation of the \LVedv~and \LVesv~with mean correlation scores of $0.92$ and $0.91$, respectively. However, the \LVef~results are worse with a mean correlation value of $0.67$. This reveals the extreme difficulty in getting fully  consistent manual annotations from ED to ES and between clinicians. It also illustrates the need for semi- or fully-automatic solutions to get higher temporal coherency, as illustrated by the higher \LVef~scores obtained by the semi-automatic BEASM method ($0.79$) and the EDN approaches ($0.79$ on average). Concerning the intra-observer variability, results are much more consistent with mean correlation scores of $0.98$, $0.98$ and $0.90$ of the \LVedv, \LVesv~and \LVef~metrics, respectively.

\vspace*{0.2cm}

\subsubsection{\mbox{U-Net} versus more sophisticated encoder-decoder architectures}

Tables \ref{tab:comparison_segmentation_dl_vs_others} and \ref{tab:comparison_clinical_indices_dl_vs_others_full} underlines that U-Net results are very close to those obtained by more sophisticated architectures. This is surprising as one might expect that more complex deep learning designs would improve results, at least marginally. As for ACNN, similar scores may be explained by the simple shapes encountered in 2D echocardiography. Indeed, the reference contours drawn by the experts involve truncated ellipse-like shapes whose information seems to be easily learned by the different EDNs. As a result, the anatomical constraint of the ACNN does not bring any additional value during the segmentation process, leading to similar or even slightly lower performance due to the regularization effect (which can lead to simpler shapes than expected). Further results which support this hypothesis are provided in the supplementary materials /multimedia tab. Concerning SHG and \mbox{U-Net++}, the similar scores may be explained by the results in Fig.~\ref{fig:inc_size}. From this figure, we observed that U-Net reaches a plateau in terms of its performances when training on more than 250 patients. This suggests that the capacity of a U-Net is sufficient to generalize well on CAMUS dataset. \modification{Thus more complex architectures like SHG and U-Net++ did not bring any improvement in our case.}

\vspace*{0.2cm}

\subsubsection{Effect of the stochasticity during the training phase}

\modification{The training of all models in this work is stochastic, thus training a network will not converge to the exact same model each time. To estimate its effect on performance, we provide in table III of the supplementary materials /multimedia tab the results produced by \mbox{U-Net 1} and \mbox{U-Net 2} for two different trainings, showing that at worst the \emph{d\textsubscript{m}} and \emph{d\textsubscript{H}} respectively varied of 0.1 and 0.2 mm. The scores obtained with the two \mbox{U-Net 2} models were consistently better than the ones produced by the \mbox{U-Net 1} models. This indicates that the effect of the stochastic nature of the training process is limited in our case.}

\vspace*{0.2cm}

\subsubsection{Accuracy of EDNs at delineating the \LVendo~, \LVepi~and LA structures}

Segmentation results given in table~\ref{tab:comparison_segmentation_dl_vs_others} show that the five EDN implementations clearly outperform the state-of-the-art fully and semi-automatic non-deep-learning methods. In particular, while also learning from annotated data, SRF does not perform as consistently as the EDNs. Concerning the deformable model-based BEASM, even if it integrates the annotated information through a shape prior, this method produces overall significantly less accurate segmentation results. It thus appears from this study that a well-designed EDN can reach impressive segmentation scores in echocardiography. \modification{In addition, results given in Section \ref{sec:unet_behavior} show that U-Net provides the same performances whether low quality data is included or not, and whether the model learns to segment all structures simultaneously or separately (involving three models instead of a single one). \mbox{U-Net 1} obtained very close results compared to \mbox{U-Net 2}, but with 9 times less parameters. The number of parameters directly influences runtime performance, training time, storage and memory consumption. Since ultrasound is a real-time imaging modality, it would be a huge advantage if a single compact EDN could accurately segment multiple cardiac structures. In this regard, this study indicates that \mbox{U-Net 1} would be the best candidate to embed into clinical equipment.}

Interestingly, the EDN results are better than the inter-observer scores, on all structures and metrics. Although further investigations shall be made to validate this assertion, the obtained results tend to show that, when properly trained, deep learning techniques are able to reproduce manual annotations with high fidelity. The results presented in this pilot study should thus stimulate the community to set up public multi-centric and multi-vendor datasets in echocardiography with annotations from cardiologists having passed high level consensus criteria. It is also interesting to note that EDN results are slightly worse than the intra-observer scores, on all structures and metrics (apart for the $d_m$ metric for the \LVepi~at ED). This reveals that even if EDNs produce remarkable results, there still exists room for improvement to faithfully reproduce the manual annotations of one expert, taking into account its variability due to the ultrasound image quality. 

In complement, we counted the number of cases for which the EDNs produced results outside the inter-observer variability, \ie~a $d_m$ value higher than $3.5$~mm and $4.0$ at ED and ES, respectively and a $d_H$ value higher than $8.2$~mm and $8.8$ at ED and ES, respectively. From this experiment, we found that 18\% of the segmentations produced by both U-Nets, ACNN or SHG can be seen as outliers. This value goes up to 30 \% for U-Net++. For comparison purpose, the outliers rate from two series of annotations on fold 5 produced by the same expert \emph{O\textsubscript{1a}} is equal to 13\%. Even if the overall performances of the EDNs are remarkable, this confirms the interest of still improving deep learning solution to produce highly reliable segmentation results on daily clinical practice data.

\vspace*{0.2cm}

\subsubsection{Accuracy of EDNs at estimating clinical indices}

Clinical scores provided in table~\ref{tab:comparison_clinical_indices_dl_vs_others_full} show that EDNs produce results below the inter-observer scores for all the metrics. It thus appears that the evaluated EDNs are serious candidates to automatically produce trustworthy estimations of the \LVedv~and \LVesv~indices, on par with medical expertise. Concerning the \LVef, even if the results are better than the inter-observer scores, a correlation value of $0.82$ (for the best performing method) appears too low in comparison with the intra-observer value of $0.90$ to consider the automatic estimate of this index as sufficiently robust to be dependable in clinical practice. The lower \LVef~scores compared to \LVedv~and \LVesv~measures can be partially explained by the lack of temporal coherency in the tested EDN implementations. Indeed, for each patient, the ED and ES frames are viewed as two independent images, potentially generating less efficient estimation of the corresponding \LVef~measures. Numerous deep learning strategies that integrate temporal coherence such as the recurrent neural networks (the Long Short Term Memory - LSTM - model being one of the famous network of this family) has been described. The integration of such concepts into the \mbox{U-Net} formalism seems to be a solution of interest in order to make the \LVef~estimation more accurate. 

\section{Conclusions}
\label{sec:conclusions}

In this paper, we introduced the largest publicly-available and fully-annotated dataset for 2D echocardiographic assessment (to our knowledge). A dedicated Girder\footnote{https://girder.readthedocs.io} on-line platform has been setup for new result submissions at
\href{https://camus.creatis.insa-lyon.fr/challenge/}{https://camus.creatis.insa-lyon.fr/challenge/}, where the CAMUS dataset, containing 2D apical four-chamber and two-chamber view sequences acquired from 500 patients, is also available for download. Thanks to this dataset, the following new insights were underlined:
\begin{itemize}
\item Encoder-decoder networks produced highly accurate segmentation results in 2D echocardiography;
\item Among the different tested architectures, U-Net appeared to be most effective in terms of trade-off between the number of parameters and the achieved performance;
\item The reasons for the lack of improvement of the more sophisticated networks (ACNN, SHG and U-Net++) compared to U-Net was addressed with additional testings provided in the supplementary materials /multimedia tab;
\item U-Net reached a plateau in terms of its performances when training on more than 250 patients but still continued to improve, implying that though 250 patients was enough to generalize well on CAMUS, it has the potential to integrate additional variability;
\item U-Net showed impressive robustness to variability, especially to image quality. Considering the wide range of image quality involved in echocardiography, this result is another positive element to consider encoder-decoder-based techniques as a solution of choice to solve the problem of 2D echocardiographic image segmentation;
\item U-Net learned to reproduce a specific way of contouring;
\item The segmentation and clinical results (left ventricle volumes at ED and ES) of the encoder-decoder networks were all below the inter-observer scores;
\item The segmentation and clinical results of the encoder-decoder networks were close to but slightly worse than the intra-observer scores. This reveals that even if encoder-decoder networks produced remarkable results, there is still room for improvement to faithfully reproduce the manual annotation of a given expert. 
\end{itemize}

\section*{Acknowledgment}
\label{sec:Acknowledgment}

This work was performed within the framework of the LABEX PRIMES (ANR- 11-LABX-0063) of Université de Lyon, within the
program "Investissements d'Avenir" (ANR-11-IDEX-0007) operated by the French National Research Agency (ANR). The Centre for Innovative Ultrasound Solutions (CIUS) is funded by the Norwegian Research Council (project code 237887).
\bibliographystyle{IEEEtran}
\bibliography{IEEEabrv,tmi_camus_2018}

\begin{thebibliography}{10}
\providecommand{\url}[1]{#1}
\csname url@samestyle\endcsname
\providecommand{\newblock}{\relax}
\providecommand{\bibinfo}[2]{#2}
\providecommand{\BIBentrySTDinterwordspacing}{\spaceskip=0pt\relax}
\providecommand{\BIBentryALTinterwordstretchfactor}{4}
\providecommand{\BIBentryALTinterwordspacing}{\spaceskip=\fontdimen2\font plus
\BIBentryALTinterwordstretchfactor\fontdimen3\font minus
  \fontdimen4\font\relax}
\providecommand{\BIBforeignlanguage}[2]{{%
\expandafter\ifx\csname l@#1\endcsname\relax
\typeout{** WARNING: IEEEtran.bst: No hyphenation pattern has been}%
\typeout{** loaded for the language `#1'. Using the pattern for}%
\typeout{** the default language instead.}%
\else
\language=\csname l@#1\endcsname
\fi
#2}}
\providecommand{\BIBdecl}{\relax}
\BIBdecl

\bibitem{Armstrong2015}
C.~Armstrong, E.~P. Ricketts, C.~Cox, P.~Adler, A.~Arynchyn \emph{et~al.},
  ``{Quality Control and Reproducibility in M-Mode, Two-Dimensional, and
  Speckle Tracking Echocardiography Acquisition and Analysis: The CARDIA Study,
  Year 25 Examination Experience},'' \emph{{Echocardiography}}, vol.~32, no.~8,
  pp. 1233--1240, 2015.

\bibitem{Bernard2018}
O.~Bernard, A.~Lalande, C.~Zotti, F.~Cervenansky,  \emph{et~al.}, ``{Deep
  Learning Techniques for Automatic MRI Cardiac Multi-structures Segmentation
  and Diagnosis: Is the Problem Solved?}'' \emph{IEEE Transactions on Medical
  Imaging}, in press, 2018.

\bibitem{Bernard2016}
O.~Bernard, J.~G. Bosch, B.~Heyde, M.~Alessandrini, D.~Barbosa,
  S.~Camarasu-Pop, F.~Cervenansky, S.~Valette, O.~Mirea \emph{et~al.},
  ``{Standardized Evaluation System for Left Ventricular Segmentation
  Algorithms in 3D Echocardiography},'' \emph{IEEE Transactions on Medical
  Imaging}, vol.~35, no.~4, pp. 967--977, 2016.

\bibitem{Barbosa2014}
D.~Barbosa, D.~Friboulet, J.~D’hooge, and O.~Bernard, ``{Fast tracking of the
  left ventricle using global anatomical affine optical flow and local
  recursive block matching},'' in \emph{Proc. MICCAI Challenge on
  Echocardiographic Three-Dimensional Ultrasound Segmentation (CETUS), Boston,
  MIDAS Journal}, 2014, pp. 17--24.

\bibitem{Pedrosa2017}
J.~Pedrosa, S.~Queirós, O.~Bernard, J.~Engvall, T.~Edvardsen, E.~Nagel, and
  J.~D’hooge, ``{Fast and Fully Automatic Left Ventricular Segmentation and
  Tracking in Echocardiography Using Shape-Based B-Spline Explicit Active
  Surfaces},'' \emph{IEEE Transactions on Medical Imaging}, vol.~36, no.~11,
  pp. 2287--2296, 2017.

\bibitem{Noble2006}
J.~A. Noble and D.~Boukerroui, ``{Ultrasound image segmentation: a survey},''
  \emph{IEEE Transactions on Medical Imaging}, vol.~25, no.~8, pp. 987--1010,
  2006.

\bibitem{Carneiro2012}
G.~Carneiro, J.~C. Nascimento, and A.~Freitas, ``{The Segmentation of the Left
  Ventricle of the Heart From Ultrasound Data Using Deep Learning Architectures
  and Derivative-Based Search Methods},'' \emph{IEEE Transactions on Image
  Processing}, vol.~21, no.~3, pp. 968--982, 2012.

\bibitem{Leung2010}
K.~Y.~E. Leung and J.~G. Bosch, ``{Automated border detection in
  three-dimensional echocardiography: principles and promises},''
  \emph{European Journal of Echocardiography}, vol.~11, no.~2, pp. 97--108,
  2010.

\bibitem{Wang2014}
C.~Wang and O.~Smedby, ``{Model-based left ventricle segmentation in 3d
  ultrasound using phase image},'' in \emph{Proc. MICCAI Challenge on
  Echocardiographic Three-Dimensional Ultrasound Segmentation (CETUS), Boston,
  MIDAS Journal}, 2014, pp. 81--88.

\bibitem{Smistad2014}
E.~Smistad and F.~Lindseth, ``{Real-time tracking of the left ventricle in 3d
  ultrasound using kalman filter and mean value coordinates},'' in \emph{Proc.
  MICCAI Challenge on Echocardiographic Three-Dimensional Ultrasound
  Segmentation (CETUS), Boston, MIDAS Journal}, 2014, pp. 65--72.

\bibitem{Bernier2014}
M.~Bernier, P.~Jodoin, and A.~Lalande, ``{Automatized evaluation of the left
  ventricular ejection fraction from echocardiographic images using graph
  cut},'' in \emph{Proc. MICCAI Challenge on Echocardiographic
  Three-Dimensional Ultrasound Segmentation (CETUS), Boston, MIDAS Journal},
  2014, pp. 25--32.

\bibitem{Milletari2014}
F.~Milletari, M.~Yigitsoy, N.~Navab, and S.~Ahmadi, ``{Left ventricle
  segmentation in cardiac ultrasound using hough-forests with implicit shape
  and appearance priors},'' in \emph{Proc. MICCAI Challenge on
  Echocardiographic Three-Dimensional Ultrasound Segmentation (CETUS), Boston,
  MIDAS Journal}, 2014, pp. 49--56.

\bibitem{Vanstralen2014}
M.~van Stralen, A.~Haak, K.~Leung, G.~van Burken, and J.~Bosch, ``{Segmentation
  of multi-center 3d left ventricular echocardiograms by active appearance
  models},'' in \emph{Proc. MICCAI Challenge on Echocardiographic
  Three-Dimensional Ultrasound Segmentation (CETUS), Boston, MIDAS Journal},
  2014, pp. 73--80.

\bibitem{Oktay2014}
O.~Oktay, W.~Shi, K.~Keraudren, J.~Caballero, and D.~Rueckert, ``{Learning
  shape representations for multi-atlas endocardium segmentation in 3D echo
  images},'' in \emph{Proc. MICCAI Challenge on Echocardiographic
  Three-Dimensional Ultrasound Segmentation (CETUS), Boston, MIDAS Journal},
  2014, pp. 57--64.

\bibitem{Keraudren2014}
K.~Keraudren, O.~Oktay, W.~Shi, J.~Hajnal, and D.~Rueckert, ``{Endocardial 3d
  ultrasound segmentation using autocontext random forests},'' in \emph{Proc.
  MICCAI Challenge on Echocardiographic Three-Dimensional Ultrasound
  Segmentation (CETUS), Boston, MIDAS Journal}, 2014, pp. 41--48.

\bibitem{Domingos2014}
J.~Domingos, R.~Stebbing, and J.~Noble, ``{Endocardial segmentation using
  structured random forests in 3D echocardiography},'' in \emph{Proc. MICCAI
  Challenge on Echocardiographic Three-Dimensional Ultrasound Segmentation
  (CETUS), Boston, MIDAS Journal}, 2014, pp. 33--40.

\bibitem{Smistad2017}
E.~Smistad, A.~Østvik, B.~O. Haugen, and L.~Lovstakken, ``{2D left ventricle
  segmentation using deep learning},'' in \emph{2017 IEEE International
  Ultrasonics Symposium (IUS)}, 2017, pp. 1--4.

\bibitem{Ronneberger2015}
O.~Ronneberger, P.~Fischer, and T.~Brox, ``{U-Net: Convolutional Networks for
  Biomedical Image Segmentation},'' in \emph{Proc. MICCAI}, 2015, pp. 234--241.

\bibitem{Oktay2018}
O.~Oktay, E.~Ferrante, K.~Kamnitsas, M.~Heinrich, W.~Bai, J.~Caballero, S.~A.
  Cook, A.~de~Marvao, T.~Dawes, D.~P. O‘Regan, B.~Kainz, B.~Glocker, and
  D.~Rueckert, ``{Anatomically Constrained Neural Networks (ACNNs): Application
  to Cardiac Image Enhancement and Segmentation},'' \emph{IEEE Transactions on
  Medical Imaging}, vol.~37, no.~2, pp. 384--395, 2018.

\bibitem{Cicek2016}
{\"O}.~{\c{C}}i{\c{c}}ek, A.~Abdulkadir, S.~S. Lienkamp, T.~Brox, and
  O.~Ronneberger, ``{3D U-Net: Learning Dense Volumetric Segmentation from
  Sparse Annotation},'' in \emph{Medical Image Computing and Computer-Assisted
  Intervention -- MICCAI 2016}.\hskip 1em plus 0.5em minus 0.4em\relax Cham:
  Springer International Publishing, 2016, pp. 424--432.

\bibitem{Folland1979}
E.~D. Folland, A.~F. Parisi, P.~F. Moynihan, D.~R. Jones, C.~L. Feldman, and
  D.~E. Tow, ``{Assessment of left ventricular ejection fraction and volumes by
  real-time, two-dimensional echocardiography. A comparison of cineangiographic
  and radionuclide techniques},'' \emph{Circulation}, vol.~60, no.~4, pp.
  760--766, 1979.

\bibitem{Lang2015}
R.~M. Lang, L.~P. Badano, V.~Mor-Avi, J.~Afilalo \emph{et~al.},
  ``{Recommendations for Cardiac Chamber Quantification by Echocardiography in
  Adults: An Update from the American Society of Echocardiography and the
  European Association of Cardiovascular Imaging},'' \emph{Circulation},
  vol.~28, no.~1, pp. 1--39, 2015.

\bibitem{Litjens2017}
G.~Litjens, T.~Kooi, B.~E. Bejnordi, A.~A.~A. Setio, F.~Ciompi, M.~Ghafoorian,
  J.~A. van~der Laak, B.~van Ginneken, and C.~I. Sánchez, ``A survey on deep
  learning in medical image analysis,'' \emph{Med. Image Anal.}, vol.~42, pp.
  60 -- 88, 2017.

\bibitem{newell2016}
A.~Newell, K.~Yang, and J.~Deng, ``Stacked hourglass networks for human pose
  estimation,'' in \emph{Computer Vision -- ECCV 2016}, B.~Leibe, J.~Matas,
  N.~Sebe, and M.~Welling, Eds.\hskip 1em plus 0.5em minus 0.4em\relax Cham:
  Springer International Publishing, 2016, pp. 483--499.

\bibitem{zhou2018}
Z.~Zhou, M.~Siddiquee, N.~Tajbakhsh, and J.~Liang, ``Unet++: A nested u-net
  architecture for medical image segmentation,'' in \emph{in proc. of Deep
  Learning in Medical Image Analysis and Multimodal Learning for Clinical
  Decision Support}, 2018, pp. 3--11.

\bibitem{Leclerc2017}
S.~Leclerc, T.~Grenier, F.~Espinosa, and O.~Bernard, ``{A fully automatic and
  multi-structural segmentation of the left ventricle and the myocardium on
  highly heterogeneous 2D echocardiographic data},'' in \emph{IEEE
  International Ultrasonics Symposium (IUS)}, 2017.

\bibitem{dollar2015}
P.~Dollár and C.~L. Zitnick, ``{Fast Edge Detection Using Structured
  Forests},'' \emph{IEEE Transactions on Pattern Analysis and Machine
  Intelligence}, vol.~37, no.~8, pp. 1558--1570, 2015.

\bibitem{Barbosa2012}
D.~Barbosa, T.~Dietenbeck, J.~Schaerer, J.~D'hooge, D.~Friboulet, and
  O.~Bernard, ``{B-Spline Explicit Active Surfaces: An Efficient Framework for
  Real-Time 3-D Region-Based Segmentation},'' \emph{IEEE Transactions on Image
  Processing}, vol.~21, no.~1, pp. 241--251, 2012.

\bibitem{Pedrosa2016}
J.~Pedrosa, D.~Barbosa, B.~Heyde, F.~Schnell, A.~Rösner, P.~Claus, and
  J.~D’hooge, ``{Left Ventricular Myocardial Segmentation in 3-D Ultrasound
  Recordings: Effect of Different Endocardial and Epicardial Coupling
  Strategies},'' \emph{{IEEE Transactions on Ultrasonics, Ferroelectrics, and
  Frequency Control}}, vol.~64, no.~3, pp. 525--536, March 2017.

\bibitem{Barbosa2013}
D.~Barbosa, T.~Dietenbeck, B.~Heyde, H.~Houle, D.~Friboulet, J.~D’hooge, and
  O.~Bernard, ``{Fast and Fully Automatic 3-D Echocardiographic Segmentation
  Using B-Spline Explicit Active Surfaces: Feasibility Study and Validation in
  a Clinical Setting},'' \emph{{Ultrasound in Medicine \& Biology}}, vol.~39,
  no.~1, pp. 89 -- 101, 2013.

\end{thebibliography}
\setcounter{table}{2}  
\onecolumn
\clearpage

\clearpage

\begin{minipage}{.95\textwidth}
\begin{center}
\vspace*{0.6cm}
\underline{\textbf{Supplementary materials for:}}
\vspace*{0.6cm}


\textsl{\textbf{Deep Learning for Segmentation using an Open Large-Scale Dataset in 2D Echocardiography}}

\vspace*{0.6cm}

S.~Leclerc\textsuperscript{1}, E.~Smistad\textsuperscript{2}, J.~Pedrosa\textsuperscript{3}, A.~Ostvik\textsuperscript{2}, Frederic Cervenansky\textsuperscript{1}, F.~Espinosa\textsuperscript{4}, T.~Espeland\textsuperscript{2, 5}, E.A.~Rye Berg\textsuperscript{2,5}, P.M.~Jodoin\textsuperscript{6}, T.~Grenier\textsuperscript{1}, C.~Lartizien\textsuperscript{1}, J.~D'hooge\textsuperscript{3}, L.~Lovstakken\textsuperscript{2}, O.~Bernard\textsuperscript{1}, 
\vspace*{0.6cm}

\end{center}

\textsuperscript{1}~University of Lyon, CREATIS, CNRS UMR5220, Inserm U1044, INSA-Lyon, University of Lyon 1, France
\vspace*{0.1cm}

\textsuperscript{2}~Center of Innovative Solutions (CIUS), Department of Circulation and Medical Imaging, NTNU, Norwegian University of Science and Technology
\vspace*{0.1cm}

\textsuperscript{3}~Dept. of Cardiovascular Sciences, KU Leuven, Leuven, Belgium
\vspace*{0.1cm}

\textsuperscript{4}~Cardiovascular department Centre Hospitalier de Saint-Etienne
Saint-Etienne, France
\vspace*{0.1cm}

\textsuperscript{5}~Clinic of cardiology, St. Olavs Hospital, Norway
\vspace*{0.1cm}

\textsuperscript{6}~Computer Science Department, University of Sherbrooke, Sherbrooke, Canada

\vspace*{0.6cm}

Email: \textcolor{blue}{\underline{sarah.leclerc@gmx.fr}}

\vspace*{1.0cm}

\rule{\linewidth}{.5pt}

\vspace*{0.5cm}
Dear readers,
\vspace*{0.3cm}

We hope that these supplementary materials will appropriately complement our paper. This document provides the following additional information:

\vspace*{0.2cm}
\begin{enumerate}
 \item Details on the U-Net architectures and their implementation (Table I, II);
 \correction{\item Impact of the main hyperparameters (Table III) };
  \correction{\item Overlapping densities of U-Net 1 and U-Net 2 for the geometrical results};
 \correction{\item Behavior of ACNNs on 2D echocardiography };
 \correction{\item Tests on U-Net ++};
 \item Dice performance of the mono/multi/GM experiment (Figure 1);
 \item Geometrical results on fold $5$ restricted to good \& medium image quality (Table \ref{tab:comparison_segmentation_dl_vs_others_fold5});
 \item Clinical metrics computed on fold $5$ restricted to good \& medium image quality (Table \ref{tab:comparison_clinical_indices_dl_vs_others_fold5});
 \correction{\item Evolution of the number of outliers when increasing the training dataset size;}
  \item Segmentation visual examples (Figure 4 to 11);
  \item Limitations examples (Figure 12 and 13).
\end{enumerate}
\vspace*{1.0cm}

With my best regards,
\vspace*{0.3cm}

Sarah Leclerc, on behalf of the co-authors.

\end{minipage}


\clearpage

\section{Further details on U-Net 1 and U-Net 2}
\vspace*{0.5cm}
\subsection{Detailed architectures} 
\vspace*{0.3cm}

The two U-Net implementations are summarized in tables I and II. As can be seen, U-Net 1 shows a more compact architecture, with the addition of one downsampling level. In the U-Net 2 design, the number of filters per convolutional layers increases and decreases linearly from an initial kernel size of 48, which makes for a wider net. As can be infered from the small gap in performance between the two architectures, U-Net 1 appears as a more efficient architecture when considering a robustness / capacity trade-off. \correction{Please note that the kernel size value corresponds to the number of feature maps (dimensionality of the output of the layer)}.

\vspace*{0.3cm}


\scalebox{0.75}{
\begin{tabular}[t]{*{5}{c}}
\multicolumn{5}{c}{\multirow{1}{*}{TABLE I}} \\
\multicolumn{5}{c}{\multirow{2}{*}{U-NET 1 \small ARCHITECTURE}} \\
 & & & & \\

\toprule 

Level & Layer & Kernel / Pool size & Activation & Connection\\

\midrule
\multirow{3}{*}{D1} & Conv  & 32 (3,3) & ReLU &  \\
& Conv & 32 (3,3) & ReLU & * \\
& MaxPooling & (2*2) & & \\

\midrule
\multirow{3}{*}{D2} & Conv  & 32 (3,3) & ReLU &  \\
& Conv & 32 (3,3) & ReLU & ** \\
& MaxPooling & (2*2) & & \\

\midrule
\multirow{3}{*}{D3} & Conv  & 64 (3,3) & ReLU &  \\
& Conv & 64 (3,3) & ReLU & *** \\
& MaxPooling & (2*2) & & \\

\midrule
\multirow{3}{*}{D4} & Conv  & 128 (3,3) & ReLU &  \\
& Conv & 128 (3,3) & ReLU & **** \\
& MaxPooling & (2*2) & & \\

\midrule
\multirow{3}{*}{D5} & Conv  & 128 (3,3) & ReLU &  \\
& Conv & 128 (3,3) & ReLU & ***** \\
& MaxPooling & (2*2) & & \\

\midrule
\multirow{2}{*}{D6} & Conv  & 128 (3,3) & ReLU &  \\
& Conv & 128 (3,3) & ReLU & \\

\midrule
\multirow{3}{*}{U1} & UpSampling  & (2,2) & &  \\
& Conv & 128 (3,3) & ReLU & ***** \\
& Conv & 128 (3,3) & ReLU & \\

\midrule
\multirow{3}{*}{U2} & UpSampling  & (2,2) & &  \\
& Conv & 128 (3,3) & ReLU & **** \\
& Conv & 128 (3,3) & ReLU & \\

\midrule
\multirow{3}{*}{U3} & UpSampling  & (2,2) & &  \\
& Conv & 64 (3,3) & ReLU & *** \\
& Conv & 64 (3,3) & ReLU & \\

\midrule
\multirow{3}{*}{U4} & UpSampling  & (2,2) & &  \\
& Conv & 32 (3,3) & ReLU & ** \\
& Conv & 32 (3,3) & ReLU & \\

\midrule
\multirow{3}{*}{U5} & UpSampling  & (2,2) & &  \\
& Conv & 16 (3,3) & ReLU & * \\
& Conv & 16 (3,3) & ReLU & \\

\midrule
\multirow{1}{*}{Seg} & Conv  & 4 (1,1) & Softmax &  \\
\bottomrule
\label{tab:unet1a}
\end{tabular}
}
\hfill
\scalebox{0.75}{
  \begin{tabular}[t]{*{5}{c}}
\multicolumn{5}{c}{\multirow{1}{*}{TABLE II}} \\
\multicolumn{5}{c}{\multirow{2}{*}{U-NET 2 \small ARCHITECTURE}} \\
 & & & & \\
\toprule 

Level & Layer & Kernel / Pool size & Activation & Connection\\

\midrule
\multirow{5}{*}{D1} & Conv  & 48 (3,3) & &  \\
& BatchNorm & & ReLU & \\
& Conv  & 48 (3,3) & &  \\
& BatchNorm & & ReLU & * \\
& MaxPooling & (2*2) & & \\
\midrule

\multirow{5}{*}{D2} & Conv  & 96 (3,3) & &  \\
& BatchNorm & & ReLU & \\
& Conv  & 96 (3,3) & &  \\
& BatchNorm & & ReLU & ** \\
& MaxPooling & (2*2) & & \\
\midrule

\multirow{5}{*}{D3} & Conv  & 192 (3,3) & &  \\
& BatchNorm & & ReLU & \\
& Conv  & 192 (3,3) & &  \\
& BatchNorm & & ReLU & *** \\
& MaxPooling & (2*2) & & \\
\midrule

\multirow{5}{*}{D4} & Conv  & 384 (3,3) & &  \\
& BatchNorm & & ReLU & \\
& Conv  & 384 (3,3) & &  \\
& BatchNorm & & ReLU & **** \\
& MaxPooling & (2*2) & & \\
\midrule

\multirow{4}{*}{D5} & Conv  & 768 (3,3) & &  \\
& BatchNorm & & ReLU & \\
& Conv  & 768 (3,3) & &  \\
& BatchNorm & & ReLU & \\
\midrule

\multirow{3}{*}{U1} & ConvTranspose  & 384 (2,2) - s(2,2) & &  \\
& BatchNorm & & ReLU & \\
& Conv & 384 (3,3) & & **** \\
& BatchNorm & & ReLU & \\
& Conv & 384 (3,3) & & \\
& BatchNorm & & ReLU & \\
\midrule

\multirow{3}{*}{U2} & ConvTranspose  & 192 (2,2) - s(2,2) & &  \\
& BatchNorm & & ReLU & \\
& Conv & 192 (3,3) & & *** \\
& BatchNorm & & ReLU & \\
& Conv & 192 (3,3) & & \\
& BatchNorm & & ReLU & \\
\midrule

\multirow{3}{*}{U3} & ConvTranspose  & 96 (2,2) - s(2,2) & &  \\
& BatchNorm & & ReLU & \\
& Conv & 96 (3,3) & & ** \\
& BatchNorm & & ReLU & \\
& Conv & 96 (3,3) & & \\
& BatchNorm & & ReLU & \\
\midrule

\multirow{3}{*}{U4} & ConvTranspose  & 48 (2,2) - s(2,2) & &  \\
& BatchNorm & & ReLU & \\
& Conv & 48 (3,3) & & * \\
& BatchNorm & & ReLU & \\
& Conv & 48 (3,3) & & \\
& BatchNorm & & ReLU & \\
\midrule

\multirow{1}{*}{Seg} & Conv  & 4 (1,1) & Softmax &  \\

\bottomrule
  \multicolumn{2}{l}{\emph{s: height and width strides}}
\label{tab:unet2a}
\end{tabular}
}

\clearpage

\subsection{\correction{Data formatting}}
\vspace*{0.3cm}

For a fair comparison between U-Net implementations, the same data pre- and post-processing were applied. In particular \emph{i)} images were resized to $256$x$256$ pixels before performing density normalization; \emph{ii)} no data augmentation strategy was involved; \emph{iii)} padding was applied before the 3*3 convolutions; \emph{iv)} initialization of weights were sampled from Glorot uniform distribution; \emph{v)} $2$*$2$ max-pooling was used as downsampling; \emph{vi)} ReLu activations were used; \emph{vii)} segmentation was obtained from a final softmax layer; \emph{viii)} principal simply connected objects selection was performed to remove small false positive detections. The Adam method was used for stochastic optimization. When learning on the full dataset, the number of epochs was set to 30 and the model that performed best on the validation set was saved.

\correction{\subsection{From U-Net 1 to U-Net 2}}
	
\vspace*{0.3cm}

\correction{In table \ref{tab:fromunet1to2}, we evaluate how the segmentation accuracy (\LVendo~and \LVepi) obtained by U-Net 1 on the ten test sets restricted to patients having good or medium image quality ($406$ patients) on ED and ES images is affected when a single hyperparameter differing between U-Net 1 and U-Net 2 is changed. We also retrained a U-Net 1 to see how initialization can impact the results.}

\correction{As mentioned in the paper, batchNormalization is the only modification that brings an improvement on the LV that is consistently superior to what may be brought by initialization. The loss, the batch size and the upsampling scheme didn't bring consistent differences. A lower learning rate was slightly detrimental, as the number of epochs is kept the same. Surprisingly, adding more filters doesn't appear to be meaningfully beneficial, and neither does learning upsampling parameters. Though only the original number of filters was changed here, it suggests that U-Net 1 already has a sufficient number of parameters and that supplementary filters learn redundant information.}

\vspace*{0.5cm}

\begin{table*}[h]
	\renewcommand{\arraystretch}{1.0}
	\caption{Segmentation accuracy (\LVendo~and \LVepi) from U-Net 1 to U-Net 2.\hspace{\textwidth}Bold values indicate a superior value to U-Net 1.}
	\centering
	\begin{tabular}{c*{6}{c}}
		
		\toprule 
		\multicolumn{1}{c}{\multirow{6}{*}{\bf \small Methods *}} \\
		&  \multicolumn{3}{c}{\bf$\boldsymbol{\LVendo}$} & \multicolumn{3}{c}{\bf $\boldsymbol{\LVepi}$} \\
		\cmidrule(r){2-4} \cmidrule(r){5-7} 
		& \multicolumn{1}{c}{\bf \emph{D}} & \multicolumn{1}{c}{\bf \emph{d\textsubscript{m}}} & \multicolumn{1}{c}{\bf \bf \emph{d\textsubscript{H}}} & \multicolumn{1}{c}{\bf \emph{D}} & \multicolumn{1}{c}{\bf \emph{d\textsubscript{m}}} & \multicolumn{1}{c}{\bf \emph{d\textsubscript{H}}} \\
		\cmidrule(r){2-2} \cmidrule(r){3-3} \cmidrule(r){4-4} \cmidrule(r){5-5} \cmidrule(r){6-6} \cmidrule(r){7-7}
		& val. & mm & mm & val.& mm & mm \\
		
		\midrule
		
		\addlinespace[+0.7ex]
		
		\multicolumn{1}{l}{\multirow{2}{*}{U-Net~1}} & 0.920 & 1.7 & 5.6 & 0.947
		& 1.9 & 6.0 \\
		
		& \scriptsize{$\pm$0.056} & \scriptsize{$\pm$1.2}
		& \scriptsize{$\pm$3.3} & \scriptsize{$\pm$0.030}
		& \scriptsize{$\pm$1.1} & \scriptsize{$\pm$3.8} \\
		
		\addlinespace[+0.7ex]
		
		\multicolumn{1}{l}{\multirow{2}{*}{U-Net~1~bis}} & \bf 0.921 & 1.8& \bf 5.5 & \bf 0.949  & 1.9 & 6.0 \\
		
		& \scriptsize{$\pm 0.052$} & \scriptsize{$\pm 1.5$}
		& \scriptsize{$\pm 3.1$} & \scriptsize{$\pm 0.028$}
		& \scriptsize{$\pm 1.0$} & \scriptsize{$\pm 3.5$} \\
		
		\addlinespace[+0.7ex]
		
		\multicolumn{1}{l}{\multirow{2}{*}{U-Net~1~Kernel 48}} & \bf 0.923 & 1.9& \bf 5.5& \bf 0.948 & 1.9 & 6.0 \\
		
		& \scriptsize{$\pm 0.047 $} & \scriptsize{$\pm 1.0 $}
		& \scriptsize{$\pm 3.2$} & \scriptsize{$\pm 0.026$}
		& \scriptsize{$\pm 1.2 $} & \scriptsize{$\pm 3.8$} \\
		
		\addlinespace[+0.7ex]
		
		\multicolumn{1}{l}{\multirow{2}{*}{U-Net~1~Cross-Entropy loss}} & 0.916 & 1.9 & 5.7 & \bf 0.948 & 1.9 & 6.0\\
		
		& \scriptsize{$\pm 0.047$} & \scriptsize{$\pm 1.3$} 
		& \scriptsize{$\pm 3.5$} & \scriptsize{$\pm 0.030$} 
		& \scriptsize{$\pm 1.0$} & \scriptsize{$\pm 3.5$}  \\
		\addlinespace[+0.7ex]

		\multicolumn{1}{l}{\multirow{2}{*}{U-Net~1~Conv2DTranspose}} & \bf 0.921 & 1.7 & 5.6 & \bf 0.948 & 1.9 & 6.1 \\

		& \scriptsize{$\pm$0.052} & \scriptsize{$\pm$1.3}
		& \scriptsize{$\pm$3.7} & \scriptsize{$\pm$0.027}
		& \scriptsize{$\pm$1.0} & \scriptsize{$\pm$4.5} \\
		
		\multicolumn{1}{l}{\multirow{2}{*}{U-Net~1~BatchSize 10}} & \bf 0.923 & 1.7 & 5.6 & \bf 0.949 & \bf 1.8 & 6.4 \\
		
		& \scriptsize{$\pm$0.055} & \scriptsize{$\pm$1.0}
		& \scriptsize{$\pm$4.3} & \scriptsize{$\pm$0.031}
		& \scriptsize{$\pm$1.1} & \scriptsize{$\pm$5.2} \\
		
		\multicolumn{1}{l}{\multirow{2}{*}{U-Net~1~LR(1e-4)}} & 0.913 & 1.9 & 6.1 & 0.943 & 2.1 & 6.7 \\
		
		& \scriptsize{$\pm$0.055} & \scriptsize{$\pm$1.1}
		& \scriptsize{$\pm$3.9} & \scriptsize{$\pm$0.034}
		& \scriptsize{$\pm$1.2} & \scriptsize{$\pm$4.4} \\
		
		\multicolumn{1}{l}{\multirow{2}{*}{U-Net~1~BatchNorm}} & \bf 0.926 & \bf 1.6 & \bf 5.3 & \bf 0.948 & 1.9 & 6.0 \\
		
		& \scriptsize{$\pm$0.053} & \scriptsize{$\pm$1.4}
		& \scriptsize{$\pm$3.5} & \scriptsize{$\pm$0.030}
		& \scriptsize{$\pm$1.0} & \scriptsize{$\pm$3.5} \\

		\multicolumn{1}{l}{\multirow{2}{*}{U-Net~2}} & \bf 0.928 & \bf 1.6 & \bf 5.4 & \bf 0.950 & \bf 1.8 & 6.1 \\
		
		& \scriptsize{$\pm$0.054} & \scriptsize{$\pm$1.5}
		& \scriptsize{$\pm$3.7} & \scriptsize{$\pm$0.032}
		& \scriptsize{$\pm$1.0} & \scriptsize{$\pm$4.1} \\

		\multicolumn{1}{l}{\multirow{2}{*}{U-Net~2 bis}} & \bf 0.930 & \bf 1.5 & \bf 5.4 & \bf 0.951 & \bf 1.8 & \bf 5.9 \\

	& \scriptsize{$\pm$0.049} & \scriptsize{$\pm$0.9}
		& \scriptsize{$\pm$3.4} & \scriptsize{$\pm$0.024}
	& \scriptsize{$\pm$0.9} & \scriptsize{$\pm$3.3} \\

		\addlinespace[+0.7ex]
		
		\bottomrule
		\label{tab:fromunet1to2}
	\end{tabular}
\end{table*}

\vspace*{0.5cm}

\subsection{Comparison of densities for geometrical metrics}
\vspace*{0.3cm}
\correction{
In addition to the p-values obtained from Wilcoxon Signed-Rank tests, we provide visuals of the overlapping distributions of U-Net 1 and U-Net 2 for $d_m$ and $d_H$ evaluated on the endocardium and the epicardium. As can be seen, both distributions have chi-square aspects, with the production of a few outliers in spite of very good mean performances and overall good robustness.}

\begin{figure}[htbp]
    \centering
    \subfigure[$d_m$ endo / epi]{{\includegraphics[width=16cm]{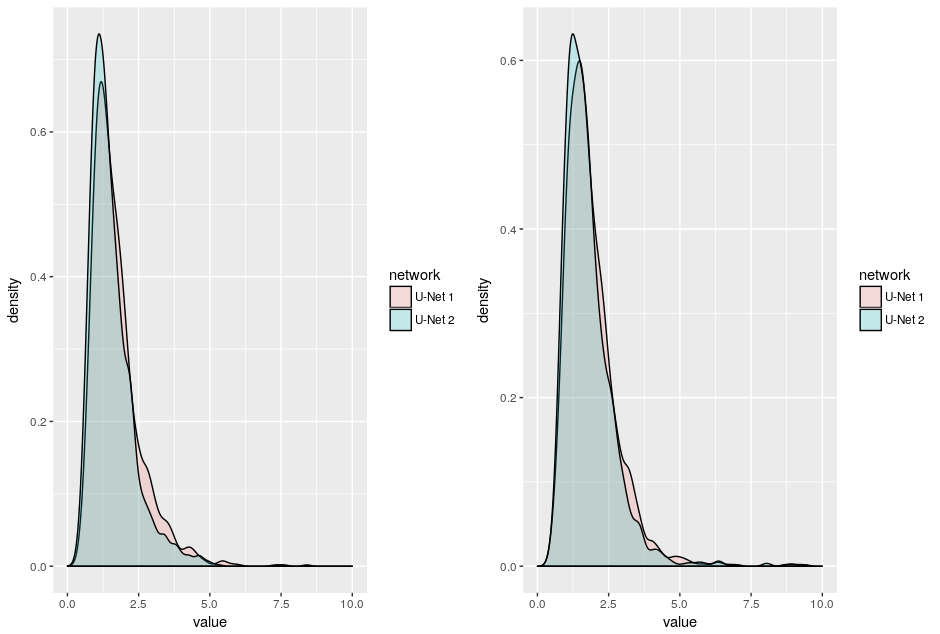}}}
    \\
    \subfigure[$d_H$ endo / epi]{{\includegraphics[width=16cm]{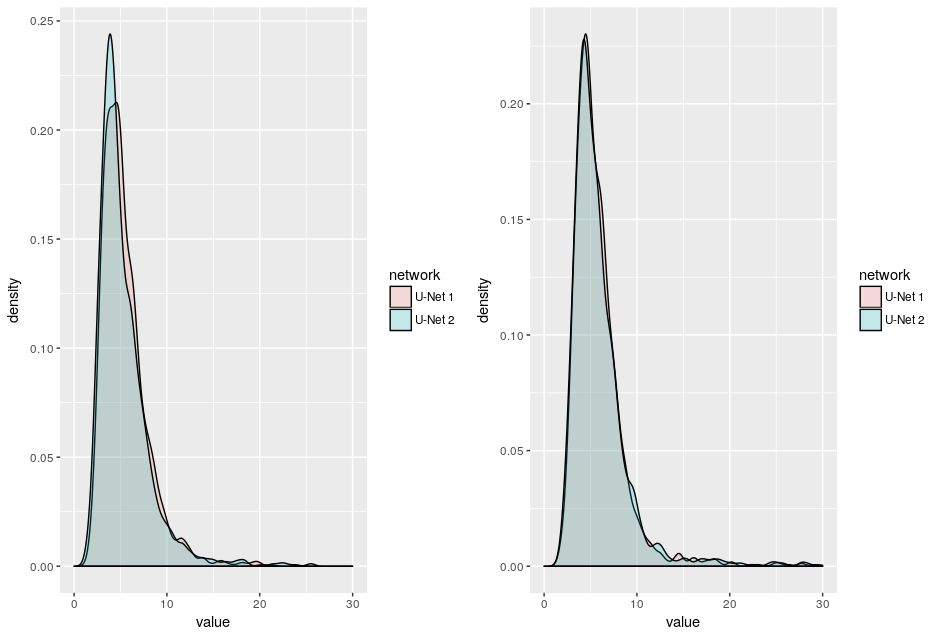}}}
    \caption{$d_m$ and $d_H$ overlapping distributions from U-Net 1 and U-Net 2}
\end{figure}
 
\clearpage

\correction{\section{Additional tests performed on ACNNs}}

\subsection{Auto-encoder implementation and training}
\vspace*{0.3cm}

\correction{The auto-encoders learn to reconstruct all segmentation masks of the training dataset from a low dimensional vector of fixed size with $32$ values. The detailed architecture is given in table IV. Convolutions are not followed by any padding so the image size is gradually reduced in the encoding part and enlarged in the decoding part. L1 regularization is applied on the code so that code coefficients remain small and of similar magnitude for all auto-encoders. ELU activations are used as they showed to lead to a better convergence.}

\correction{We train the networks during 50 epochs (long after convergence) with a batch size of 50. Cross-Entropy is used as the loss to optimize with the Adam optimizer and a learning rate of 1e-4. Train, validation and test sets are the same as the corresponding segmentation network and all ten auto-encoders achieve an accuracy superior to 92$\%$.}

\vspace*{0.3cm}

\begin{table*}[h]
	\caption{AUTO-ENCODER ARCHITECTURE}
	\begin{center}
		
	\begin{tabular}{*{4}{c}}
		\toprule 
		Level & Layer & Kernel size & Activation \\
		
		\midrule
		\multirow{6}{*}{Encoder} & Conv  & 16 (3,3) & eLU \\
		& Conv & 16 (3,3) & eLU \\
		& Conv & 32 (3,3) & eLU \\
		& Conv & 32 (3,3) & eLU \\
		& Conv & 64 (3,3) & eLU \\
		& Conv & 64 (3,3) & eLU \\
		\midrule
		\multirow{2}{*}{Code} & Conv+Flatten  & 1 (3,3) & eLU \\
		& Dense & 32 &\\
		\midrule
		\multirow{10}{*}{Decoder}& Dense & 169 & eLU \\
		& Reshape & 1 (13,13) & \\
		& ConvTranspose & 64 (4,4) & eLU \\
		& ConvTranspose & 64 (4,4) & eLU \\
		& Conv & 64 (3,3) & eLU \\
		& ConvTranspose & 32 (4,4) & eLU \\
		& Conv & 32 (3,3) & eLU \\
		& ConvTranspose & 16 (4,4) & eLU \\
		& Conv & 16 (3,3) & eLU \\
		& ConvTranspose & 16 (4,4) & eLU \\
		\midrule
		Seg & Conv & 4 (3,3) & Softmax \\
		\bottomrule
	\end{tabular}
\end{center}
\label{tab:ae}
\end{table*}

\subsection{Impact of the regularization loss}

\vspace*{0.3cm}

\correction{The regularization loss corresponds to the mean squared error of the code coefficients. As its values do not scale with the ones provided by the segmentation loss, a multiplying factor is applied. In our study, it is set at $10^4$ so that the two losses have close initial values. In Table \ref{tab:acnns}, we compare the segmentation accuracy reached by our ACNN implementation for two different values of the regularization weight : the chosen one, and one 100 times stronger. From this table, it can be seen that the increase of the regularization weight affects the quality of the results, especially for the dH metric.}

\begin{table*}[htbp]
	\renewcommand{\arraystretch}{1.0}
	\caption{Segmentation accuracy for ACNN architecture with different shape regularization strength}
	\centering
	\begin{tabular}{c*{6}{c}}
		
		\toprule 
		\multicolumn{1}{c}{\multirow{6}{*}{\bf \small Methods *}} \\
		&  \multicolumn{3}{c}{\bf$\boldsymbol{\LVendo}$} & \multicolumn{3}{c}{\bf $\boldsymbol{\LVepi}$} \\
		\cmidrule(r){2-4} \cmidrule(r){5-7} 
		& \multicolumn{1}{c}{\bf \emph{D}} & \multicolumn{1}{c}{\bf \emph{d\textsubscript{m}}} & \multicolumn{1}{c}{\bf \bf \emph{d\textsubscript{H}}} & \multicolumn{1}{c}{\bf \emph{D}} & \multicolumn{1}{c}{\bf \emph{d\textsubscript{m}}} & \multicolumn{1}{c}{\bf \emph{d\textsubscript{H}}} \\
		\cmidrule(r){2-2} \cmidrule(r){3-3} \cmidrule(r){4-4} \cmidrule(r){5-5} \cmidrule(r){6-6} \cmidrule(r){7-7}
		& val. & mm & mm & val.& mm & mm \\
		
		\midrule
		
		\addlinespace[+0.7ex]
		
		\multicolumn{1}{l}{\multirow{2}{*}{ACNN - $\lambda=10^4$}} &\bf 0.918 & \bf 1.8 & \bf 5.9 & \bf 0.946
		& \bf 1.9 & \bf 6.4 \\
		
		& \scriptsize{$\pm$0.050} & \scriptsize{$\pm$1.0}
		& \scriptsize{$\pm$3.5} & \scriptsize{$\pm$0.030}
		& \scriptsize{$\pm$1.1} & \scriptsize{$\pm$4.2} \\
		
		\addlinespace[+0.7ex]
		
		\multicolumn{1}{l}{\multirow{2}{*}{ACNN - $\lambda = 10^6$}} & 0.912 & 1.9& 6.2 & 0.947  & \bf 1.9 & 6.9 \\
		
		& \scriptsize{$\pm 0.054$} & \scriptsize{$\pm 1.1$}
		& \scriptsize{$\pm 4.0$} & \scriptsize{$\pm 0.030$}
		& \scriptsize{$\pm 1.1$} & \scriptsize{$\pm 5.8$} \\
		
		\addlinespace[+0.7ex]
		
		\bottomrule
	\end{tabular}
\label{tab:acnns}
\end{table*}

\subsection{Influence of the training set size}
\vspace*{0.3cm}
\correction{To study the influence of the training set of the auto-encoder and segmentation network of ACNNs, we first trained two auto-encoders using the fold 6 as validation set and fold 5 as test set : SR400 was trained on 400 patients while SR15 was trained on 15 patients from the same training set. }

\correction{We then trained five ACNN networks (using U-Net 1 for the segmentation part of the network) with the same validation and test set. The $50p\_no\_SR$ model involved no shape regularization and was trained on 50 patients (200 images). The $5p\_no\_SR$ model was trained on a smaller set of only $5$ patients (20 images). Models $50p\_with\_SR15$ and $50p\_with\_SR400$ were trained on the same training set as $50p\_no\_SR$ while optimizing the ACNN shape regularization auxiliary loss returned respectively from the $SR15$ and $SR400$ auto-encoders. Finally, the $5p\_SR400$ model was trained on the same training set as the $5p\_no\_SR$ model, with the shape regularization from $SR400$.}

\vspace*{0.5cm}
\begin{figure}[htbp]
	\centering
	{\includegraphics[scale=0.6]{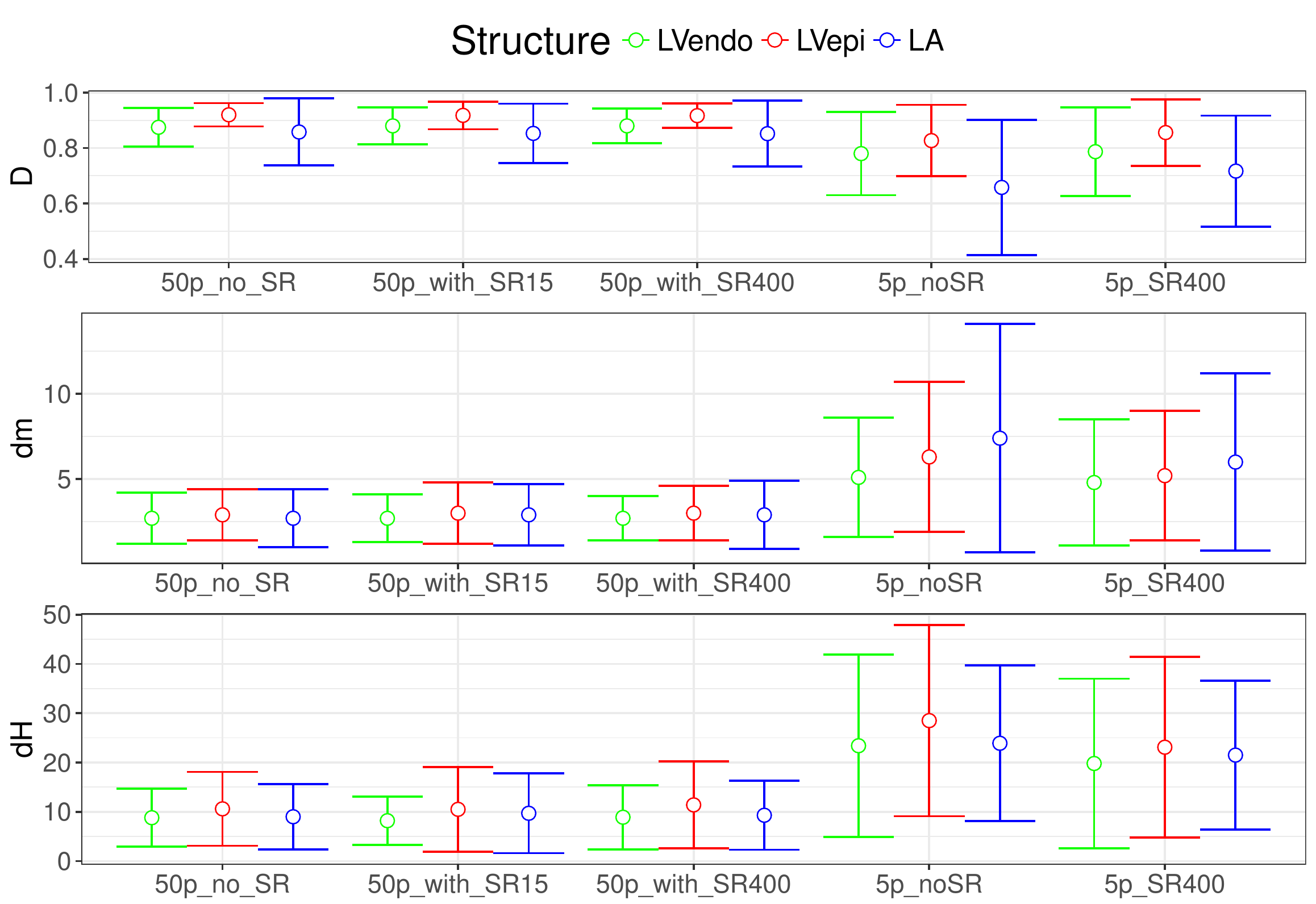}}
	\caption{Geometric performance illustrated by standard error bars around the mean values for the 5 segmentation networks. }%
	\label{fig:acnn}
\end{figure}

\vspace*{0.5cm}

Fig.\ref{fig:acnn} represent the geometrical results of all five models. It can be observed that : \emph{i)} Shape regularization does not improve results for an ACNN trained on 50 patients but it does significantly improve the results for a smaller training set of 5 patients; \emph{ii)} Using for the ACNN an auto-encoder learnt from 15 patients (accuracy of $84\%$) produced close results compared to the one learnt on 400 patients.
\\

\correction{These results support the idea that shape regularization will be helpful on datasets for which the training dataset size is not sufficient to learn the annotated shapes' complexity. For our specific task of 2D echocardiography, it appears that the necessary number of cases is low, inferior to 50 patients, probably because the shape variability is also low, allowing an auto-encoder to roughly infer it from a few number of cases. In the paper experiments, we trained encoder-decoder networks on 400 patients, which explains why the ACNN do not show any significant improvement compared to U-Net 1.}

\clearpage

\correction{\section{Tests on U-Net ++}}
\vspace*{0.3cm}
\correction{We provide in table \ref{tab:unetnest} a few results obtained from the U-Net ++ architecture, where we studied the influence of averaging the last two segmentation maps and changing the baseline with the succession of layers from U-net  1. Deep Supervision was always applied and no dropout was performed, as it was always detrimental.}

\begin{table*}[htbp]
	\renewcommand{\arraystretch}{1.0}
	\caption{Segmentation accuracy for ACNN architecture with different shape regularization strength}
	\centering
	\begin{tabular}{c*{6}{c}}
		
		\toprule 
		\multicolumn{1}{c}{\multirow{6}{*}{\bf \small Methods *}} \\
		&  \multicolumn{3}{c}{\bf$\boldsymbol{\LVendo}$} & \multicolumn{3}{c}{\bf $\boldsymbol{\LVepi}$} \\
		\cmidrule(r){2-4} \cmidrule(r){5-7} 
		& \multicolumn{1}{c}{\bf \emph{D}} & \multicolumn{1}{c}{\bf \emph{d\textsubscript{m}}} & \multicolumn{1}{c}{\bf \bf \emph{d\textsubscript{H}}} & \multicolumn{1}{c}{\bf \emph{D}} & \multicolumn{1}{c}{\bf \emph{d\textsubscript{m}}} & \multicolumn{1}{c}{\bf \emph{d\textsubscript{H}}} \\
		\cmidrule(r){2-2} \cmidrule(r){3-3} \cmidrule(r){4-4} \cmidrule(r){5-5} \cmidrule(r){6-6} \cmidrule(r){7-7}
		& val. & mm & mm & val.& mm & mm \\
		
		\midrule
		
		\addlinespace[+0.7ex]
		\multicolumn{1}{l}{\multirow{2}{*}{U-Net ++ - Mono output}} & 0.909 & 2.0 & 6.6 & 0.941
		& 2.2 & \bf 7.1 \\
		
		& \scriptsize{$\pm$0.063} & \scriptsize{$\pm$1.6}
		& \scriptsize{$\pm$4.8} & \scriptsize{$\pm$0.033}
		& \scriptsize{$\pm$1.2} & \scriptsize{$\pm$5.0} \\
		
		\addlinespace[+0.7ex]
		
		\multicolumn{1}{l}{\multirow{2}{*}{U-net ++ - 2 outputs}} & 0.905 & 2.1& 7.3 & 0.936  & 2.3 & 8.6 \\
		
		& \scriptsize{$\pm 0.070$} & \scriptsize{$\pm 4.3$}
		& \scriptsize{$\pm 7.3$} & \scriptsize{$\pm 0.040$}
		& \scriptsize{$\pm 1.4$} & \scriptsize{$\pm 6.5$} \\
		
		\addlinespace[+0.7ex]

		\multicolumn{1}{l}{\multirow{2}{*}{U-net ++ - U-Net 1}} & \bf 0.915 & \bf 1.8& \bf 6.4 & \bf 0.942  & \bf 2.1 & \bf 7.1 \\

	& \scriptsize{$\pm 0.055$} & \scriptsize{$\pm 1.0$}
	& \scriptsize{$\pm 4.0$} & \scriptsize{$\pm 0.030$}
	& \scriptsize{$\pm 1.1$} & \scriptsize{$\pm 4.7$} \\

\addlinespace[+0.7ex]
		
		\bottomrule
	\end{tabular}
	\label{tab:unetnest}
\end{table*}

\correction{From table \ref{tab:unetnest}, one can observe that averaging the feature maps of the intermediate output (Unet++ - 2 outputs model) worsened the results (especially the standard deviation), showing the uselessness of this operation on our dataset. The best U-Net ++ implementation corresponds to the one keeping the last ouput as final result and replacing the original layers by the design of the U-Net 1 architecture which has been already optimized on the CAMUS dataset. This model was used in all the experiments given in our paper. }

\clearpage

\section{Geometric and clinical scores supplements}
\vspace*{0.5cm}
\subsection{Deep learning behavior: Dice box plots}
\vspace*{0.3cm}

\begin{figure}[htbp]
    \centering
     {\includegraphics[scale=0.8]{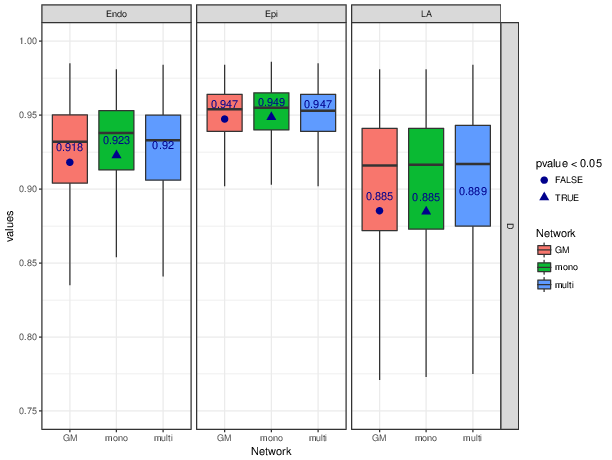}}
    \caption{Dice box plots results of the U-Net 1 method for three different schemes (GM for the learning from good \& medium image quality, mono for the learning of one structure, multi for the learning of the three structures at the same time). Values in blue displayed under the boxes correspond to mean values computed from each set of measurements. p-values are based on the Wilcoxon signed rank computed with the multi-structures strategy as reference.}%
    \label{fig:dice}
\end{figure}

Complementary to the analysis made in section V-B, we provide in Fig.\ref{fig:dice} the box plots related to the Dice metric. \correction{From the derived boxplot, one can see that, unrelated to the structure, the mono and multi-structures approaches produced close results though statistically different. Concerning the effect of poor image quality, the boxplots show that the two different strategies (learning from all the dataset or from the dataset restricted to good and medium quality images) produced results whose differences are statistically unsignificant. From these observations, the same conclusions as the ones made in the paper can be infered: i) learning the segmentation of one structure (e.g. LV Endo ) in the context of the others (e.g. LV Epi \& LA) does not improve significantly the Dice results; ii) the $19\%$ (94 patients) of poor image quality do not bring additional information.} Outliers are more numerous on the LA, which can be explained by a greater variability in shape and unoptimized acquisitions for this structure. 
\vspace*{0.5cm}

\subsection{Geometric and clinical scores computed on fold $5$}
\vspace*{0.3cm}

In addition to table III and IV given in the article, we provide in table \ref{tab:comparison_segmentation_dl_vs_others_fold5} and \ref{tab:comparison_clinical_indices_dl_vs_others_fold5} of this supplementary material the segmentation accuracy (\LVendo~and \LVepi) and the clinical scores (\LVedv, \LVesv~and \LVef) of the \correction{$8$ evaluated methods} on fold $5$ restricted to patients having good \& medium image quality ($40$ patients). As the results given in the paper, one can see that the \correction{EDNs} implementations show \emph{i)} the best segmentation performances on all metrics and structures and \emph{ii)} the best correlation scores with low biases and standard deviations.

\begin{table*}[tbp]
\renewcommand{\arraystretch}{1.0}
  \caption{Segmentation accuracy (\LVendo~and \LVepi) of $8$ evaluated methods on fold $5$ restricted to good \& medium image quality ($40$ patients). The values in bold refer to the best performance for each measure. p-values are based on the Wilcoxon signed rank between the U-Net 1 and 2 methods for each evaluation metric.}
  \centering
  \begin{tabular}{c*{6}{c}|*{6}{c}}

 \toprule 
 \multicolumn{1}{c}{\multirow{6}{*}{\bf \small Methods *}} &  \multicolumn{6}{c}{\bf \small ED} & \multicolumn{6}{c}{\bf \small ES} \\
 \cmidrule(r){2-7} \cmidrule(r){8-13}
 &  \multicolumn{3}{c}{\bf$\boldsymbol{\LVendo}$} & \multicolumn{3}{c}{\bf $\boldsymbol{\LVepi}$} & 
 \multicolumn{3}{c}{\bf $\boldsymbol{\LVendo}$} & \multicolumn{3}{c}{\bf $\boldsymbol{\LVepi}$} \\
 \cmidrule(r){2-4} \cmidrule(r){5-7} \cmidrule(r){8-10} \cmidrule(r){11-13}
 & \multicolumn{1}{c}{\bf \emph{D}} & \multicolumn{1}{c}{\bf \emph{d\textsubscript{m}}} & \multicolumn{1}{c}{\bf \bf \emph{d\textsubscript{H}}} & \multicolumn{1}{c}{\bf \emph{D}} & \multicolumn{1}{c}{\bf \emph{d\textsubscript{m}}} & \multicolumn{1}{c}{\bf \emph{d\textsubscript{H}}} & \multicolumn{1}{c}{\bf \emph{D}} & \multicolumn{1}{c}{\bf \emph{d\textsubscript{m}}} & \multicolumn{1}{c}{\bf \emph{d\textsubscript{H}}} & \multicolumn{1}{c}{\bf \emph{D}} & \multicolumn{1}{c}{\bf \emph{d\textsubscript{m}}} & \multicolumn{1}{c}{\bf \emph{d\textsubscript{H}}} \\
\cmidrule(r){2-2} \cmidrule(r){3-3} \cmidrule(r){4-4} \cmidrule(r){5-5} \cmidrule(r){6-6} \cmidrule(r){7-7} \cmidrule(r){8-8} \cmidrule(r){9-9} \cmidrule(r){10-10} \cmidrule(r){11-11} \cmidrule(r){12-12} \cmidrule(r){13-13}
 & val. & mm & mm & val.& mm & mm & val. & mm & mm & val. & mm & mm \\
 
\midrule

\multicolumn{1}{l}{\emph{O\textsubscript{1a}} vs \emph{O\textsubscript{2}}} & 0.919 & 2.2 & 6.0 & 0.913
& 3.5 & 8.0 & 0.873 & 2.7 & 6.6 & 0.890
& 3.9 & 8.6 \\

\multicolumn{1}{l}{(inter-obs)} & \scriptsize{$\pm$0.033} & \scriptsize{$\pm$0.9}
& \scriptsize{$\pm$2.0} & \scriptsize{$\pm$0.037}
& \scriptsize{$\pm$1.7} & \scriptsize{$\pm$2.9}
& \scriptsize{$\pm$0.060} & \scriptsize{$\pm$1.2}
& \scriptsize{$\pm$2.4} & \scriptsize{$\pm$0.047}
& \scriptsize{$\pm$1.8} & \scriptsize{$\pm$3.3} \\

\addlinespace[+0.7ex]

\multicolumn{1}{l}{\emph{O\textsubscript{1a}} vs \emph{O\textsubscript{3}}} & 0.886 & 3.3 & 8.2 & 0.943
& 2.3 & 6.5 & 0.823 & 4.0 & 8.8 & 0.931
& 2.4 & 6.4 \\

\multicolumn{1}{l}{(inter-obs)} & \scriptsize{$\pm$0.050} & \scriptsize{$\pm$1.5}
& \scriptsize{$\pm$2.5} & \scriptsize{$\pm$0.018}
& \scriptsize{$\pm$0.8} & \scriptsize{$\pm$2.6}
& \scriptsize{$\pm$0.091} & \scriptsize{$\pm$2.0}
& \scriptsize{$\pm$3.5} & \scriptsize{$\pm$0.025}
& \scriptsize{$\pm$1.0} & \scriptsize{$\pm$2.4} \\

\addlinespace[+0.7ex]

\multicolumn{1}{l}{\emph{O\textsubscript{2}} vs \emph{O\textsubscript{3}}} & 0.921 & 2.3 & 6.3 & 0.922
& 3.0 & 7.4 & 0.888 & 2.6 & 6.9 & 0.885
& 3.9 & 8.4 \\

\multicolumn{1}{l}{(inter-obs)} & \scriptsize{$\pm$0.037} & \scriptsize{$\pm$1.2}
& \scriptsize{$\pm$2.5} & \scriptsize{$\pm$0.036}
& \scriptsize{$\pm$1.5} & \scriptsize{$\pm$3.0}
& \scriptsize{$\pm$0.058} & \scriptsize{$\pm$1.3}
& \scriptsize{$\pm$2.9} & \scriptsize{$\pm$0.054}
& \scriptsize{$\pm$1.9} & \scriptsize{$\pm$2.8} \\

\addlinespace[+0.7ex]

\midrule

\addlinespace[+0.7ex]

		\multicolumn{1}{l}{\multirow{2}{*}{SRF}} & 0.905 & 2.4 & 9.5 & 0.920
& 3.0 & 10.8 & 0.854 & 3.0 & 10.7 & 0.900
& 3.3 & 13.4 \\

& \scriptsize{$\pm$0.053} & \scriptsize{$\pm$1.3}
& \scriptsize{$\pm$6.3} & \scriptsize{$\pm$0.053}
& \scriptsize{$\pm$1.8} & \scriptsize{$\pm$7.0}
& \scriptsize{$\pm$0.099} & \scriptsize{$\pm$1.8}
& \scriptsize{$\pm$8.6} & \scriptsize{$\pm$0.062}
& \scriptsize{$\pm$1.8} & \scriptsize{$\pm$9.4} \\

\addlinespace[+0.7ex]

\multicolumn{1}{l}{\multirow{2}{*}{BEASM fully}} & 0.882 & 3.2 & 8.5 & 0.897
& 3.9 & 9.9 & 0.832 & 3.6 & 9.3 & 0.883
& 4.1 & 10.4 \\

& \scriptsize{$\pm$0.065} & \scriptsize{$\pm$2.1}
& \scriptsize{$\pm$5.0} & \scriptsize{$\pm$0.052}
& \scriptsize{$\pm$2.3} & \scriptsize{$\pm$5.0}
& \scriptsize{$\pm$0.089} & \scriptsize{$\pm$2.2}
& \scriptsize{$\pm$5.4} & \scriptsize{$\pm$0.051}
& \scriptsize{$\pm$2.2} & \scriptsize{$\pm$4.9} \\

\addlinespace[+0.7ex]

\multicolumn{1}{l}{\multirow{2}{*}{BEASM semi}} & 0.922 & 2.1 & 5.6 & 0.916
& 3.2 & 7.8 & 0.868 & 2.8 & 7.1 & 0.904
& 3.3 & 8.7 \\

& \scriptsize{$\pm$0.031} & \scriptsize{$\pm$0.8}
& \scriptsize{$\pm$2.0} & \scriptsize{$\pm$0.040}
& \scriptsize{$\pm$1.6} & \scriptsize{$\pm$2.8}
& \scriptsize{$\pm$0.057} & \scriptsize{$\pm$1.2}
& \scriptsize{$\pm$3.1} & \scriptsize{$\pm$0.036}
& \scriptsize{$\pm$1.4} & \scriptsize{$\pm$3.0} \\

\addlinespace[+0.7ex]

\midrule

\addlinespace[+0.7ex]

\multicolumn{1}{l}{\multirow{2}{*}{U-Net~1}} & 0.941 & 1.5 & 5.0 & \bf 0.954
& \bf 1.7 & \bf 5.2 & 0.917 & 1.6 & 5.0 & 0.947
& \bf 1.8 & \bf \bf 5.4 \\

& \scriptsize{$\pm$0.021} & \scriptsize{$\pm$0.5}
& \scriptsize{$\pm$1.4} & \scriptsize{$\pm$0.018}
& \scriptsize{$\pm$0.7} & \scriptsize{$\pm$1.9}
& \scriptsize{$\pm$0.037} & \scriptsize{$\pm$0.6}
& \scriptsize{$\pm$2.1} & \scriptsize{$\pm$0.020}
& \scriptsize{$\pm$0.7} & \scriptsize{$\pm$2.1} \\

\addlinespace[+0.7ex]

\multicolumn{1}{l}{\multirow{2}{*}{U-Net~2}} & \bf 0.945 & \bf 1.4 & \bf 4.9 & \bf 0.954 & 1.8 & 5.7 & \bf 0.927 & \bf 1.3 & 4.9 & 0.946
& 1.9 & 6.0 \\

& \scriptsize{$\pm$0.021} & \scriptsize{$\pm$0.5}
& \scriptsize{$\pm$1.9} & \scriptsize{$\pm$0.017}
& \scriptsize{$\pm$0.7} & \scriptsize{$\pm$3.1}
& \scriptsize{$\pm$0.039} & \scriptsize{$\pm$0.6}
& \scriptsize{$\pm$2.5} & \scriptsize{$\pm$0.023}
& \scriptsize{$\pm$0.8} & \scriptsize{$\pm$3.4} \\

\addlinespace[+0.7ex]

\multicolumn{1}{l}{\multirow{2}{*}{ACNN}} & 0.942 & \bf 1.4& 5.0&  \bf 0.954 & \bf 1.7 & 5.5 & 0.917 & 1.6 & 5.0 & \bf 0.948
& \bf 1.8 & 5.9\\

 & \scriptsize{$\pm 0.021$} & \scriptsize{$\pm 0.5$}
& \scriptsize{$\pm 1.5$} & \scriptsize{$\pm 0.018$}
& \scriptsize{$\pm 0.7$} & \scriptsize{$\pm 2.3$}
& \scriptsize{$\pm 0.040$} & \scriptsize{$\pm 0.6$}
& \scriptsize{$\pm 2.0$} & \scriptsize{$\pm 0.022$}
& \scriptsize{$\pm 0.8$} & \scriptsize{$\pm 3.1$} \\

\multicolumn{1}{l}{\multirow{2}{*}{SHG}} & 0.941 & \bf 1.4& 5.1&  0.952& 1.9 & 5.4& 0.918& 1.5& \bf 4.8& 0.944
& 2.0& 5.6\\

& \scriptsize{$\pm 0.024$} & \scriptsize{$\pm 0.6$}
& \scriptsize{$\pm 2.0$} & \scriptsize{$\pm 0.019$}
& \scriptsize{$\pm 0.8$} & \scriptsize{$\pm 2.8$}
& \scriptsize{$\pm 0.04$} & \scriptsize{$\pm 0.6$}
& \scriptsize{$\pm 1.7$} & \scriptsize{$\pm 0.025$}
& \scriptsize{$\pm 0.9$} & \scriptsize{$\pm 2.0$} \\

\multicolumn{1}{l}{\multirow{2}{*}{U-Net~++}} & 0.934 & 1.6 & 5.4 & 0.947 & 2.0 & 6.6 & 0.916 & 1.5 & 5.3 & 0.939 & 2.1 & 7.7\\

 & \scriptsize{$\pm 0.025$} & \scriptsize{$\pm 0.6$} 
 & \scriptsize{$\pm 1.9$} & \scriptsize{$\pm 0.023$} 
 & \scriptsize{$\pm 0.9$} & \scriptsize{$\pm 4.3$} 
 & \scriptsize{$\pm 0.041$} & \scriptsize{$\pm 0.7$}
& \scriptsize{$\pm 2.4$} & \scriptsize{$\pm 0.027$}
& \scriptsize{$\pm 1.1$} & \scriptsize{$\pm 6.8$} \\

\addlinespace[+0.7ex]

\midrule

\multicolumn{1}{l}{p-values} & \scriptsize{$\approx0.053$} & \scriptsize{$\approx0.054$} & \scriptsize{$\approx0.245$}& \scriptsize{$\approx0.710$} & \scriptsize{$\approx0.520$} & \scriptsize{$\approx0.692$} & \scriptsize{$\ll 0.05$} & \scriptsize{$\ll 0.05$} & \scriptsize{$\approx0.517$} &  \scriptsize{$\approx0.926$} & \scriptsize{$\approx0.675$} & \scriptsize{$\approx0.0988$}  \\

\bottomrule
  \label{tab:comparison_segmentation_dl_vs_others_fold5}
  \end{tabular}
\end{table*}

\begin{table*}[tbp]
\renewcommand{\arraystretch}{1.0}
  \caption{Clinical metrics of the 8 evaluated methods on fold $5$ restricted to patients having good \& medium image quality ($40$ patients in total). The values in bold correspond to the best performance for each measure. p-values are based on the Wilcoxon signed rank between the U-Net 1 and 2 methods for the \LVendo, \LVepi~and \LVef.}
  \centering
  \begin{tabular}{c*{3}{c}|*{3}{c}|*{3}{c}}

 \toprule 
 \multicolumn{1}{c}{\multirow{5}{*}{\bf \small Methods *}} &  \multicolumn{3}{c}{\bf \small \emph{LV\textsubscript{EDV}}} & \multicolumn{3}{c}{\bf \small \emph{LV\textsubscript{ESV}}} & \multicolumn{3}{c}{\bf \small \emph{LV\textsubscript{EF}}}\\
 \cmidrule(r){2-4} \cmidrule(r){5-7} \cmidrule(r){8-10}
 & \multicolumn{1}{c}{\bf \emph{corr}} & \multicolumn{1}{c}{\bf \emph{bias$\pm\sigma$}} & \multicolumn{1}{c}{\bf \emph{mae}} & \multicolumn{1}{c}{\bf \emph{corr}} & \multicolumn{1}{c}{\bf \emph{bias$\pm\sigma$}} & \multicolumn{1}{c}{\bf \emph{mae}} & \multicolumn{1}{c}{\bf \emph{corr}} & \multicolumn{1}{c}{\bf \emph{bias$\pm\sigma$}} & \multicolumn{1}{c}{\bf \emph{mae}} \\
\cmidrule(r){2-2} \cmidrule(r){3-3} \cmidrule(r){4-4} \cmidrule(r){5-5} \cmidrule(r){6-6} \cmidrule(r){7-7} \cmidrule(r){8-8} \cmidrule(r){9-9} \cmidrule(r){10-10}
 & val. & ml & ml & val.& ml & ml & val. & \% & \% \\
 
\midrule

\multicolumn{1}{l}{\emph{O\textsubscript{1a}} vs \emph{O\textsubscript{2}} (inter-obs)} & 0.940 & 18.7$\pm$12.9& 18.7 & 0.956 & 18.9$\pm$9.3 & 18.9 & 0.801 & -9.1$\pm$8.1 & 10.0 \\

\multicolumn{1}{l}{\emph{O\textsubscript{1a}} vs \emph{O\textsubscript{3}} (inter-obs)} & 0.895 & 39.0$\pm$18.8& 39.0 & 0.860 & 35.9$\pm$17.1 & 35.9 & 0.646 & -12.6$\pm$10.0 & 13.4 \\

\multicolumn{1}{l}{\emph{O\textsubscript{2}} vs \emph{O\textsubscript{3}} (inter-obs)} & 0.926 & -20.3$\pm$15.6& 21.0 & 0.916 & -17.0$\pm$13.5 & 17.7 & 0.569 & 3.5$\pm$11.0 & 8.5 \\

\midrule

\multicolumn{1}{l}{SRF} & 0.843 & 5.3$\pm$18.6& 13.9 & 0.845 & 12.4$\pm$15.6 & 15.9 & 0.603 & -11.9$\pm$10.8 & 13.1 \\

\multicolumn{1}{l}{BEASM fully} & 0.809 & 19.1$\pm$23.3& 21.3 & 0.791 & 23.2$\pm$20.6 & 24.0 & 0.776 & -12.1$\pm$8.2 & 12.6 \\

\multicolumn{1}{l}{BEASM semi} & 0.913 & 12.0$\pm$15.4& 15.3 & 0.875 & 16.3$\pm$15.1 & 18.0 & 0.853 & -10.2$\pm$6.7 & 10.5 \\

\midrule

\multicolumn{1}{l}{U-Net~1} & 0.973 & -7.7$\pm$8.3& 8.7 & 0.945 & -3.1$\pm$10.1 & 7.6 & 0.820 & -2.6$\pm$7.5 & 5.9 \\

\multicolumn{1}{l}{U-Net~2} & \bf 0.977 & -4.0 $\pm$7.3& 6.6&  \bf 0.976 & -0.9$\pm$ 7.0 & \bf 5.2  & \bf 0.928 & -2.3$\pm$4.8 & \bf 4.0 \\

\multicolumn{1}{l}{ACNN} & 0.970 &-3.3$\pm $ 8.4 & \bf6.3 & 0.943 & 0.8$\pm 9.9$ & 7.0 & 0.851 & -3.9$\pm$6.8 & 6.5\\

\multicolumn{1}{l}{SHG} & 0.967 & -3.1$\pm$8.5 & 6.6 & 0.947 & 0.7$\pm$9.6 & 7.0 &  0.842 & -3.9$\pm$6.9 & 6.3\\

\multicolumn{1}{l}{U-Net~++} & 0.971 & -11.5$\pm$ 9.0& 12.0 &  0.963& -5.7$\pm$ 9.5 & 7.9&  0.872 & -2.0$\pm$ 6.3& 4.7 \\

\midrule

\multicolumn{1}{l}{p-values} & \multicolumn{3}{c}{\scriptsize{$< 0.05$}} & \multicolumn{3}{c}{\scriptsize{$\ll 0.05$}} & \multicolumn{3}{c}{\scriptsize{$\ll 0.05$}} \\

\bottomrule
 
 \multicolumn{10}{l}{} \\
 \multicolumn{10}{l}{*~\emph{LV\textsubscript{EDV}: End diastolic left ventricular volume; LV\textsubscript{ESV}: End systolic left ventricular volume;}} \\
 \multicolumn{10}{l}{~~\emph{corr: Pearson correlation coefficient; mae: mean absolute error. \vspace{-0.4cm}}} \\ 
 
\label{tab:comparison_clinical_indices_dl_vs_others_fold5}
  \end{tabular}
\end{table*}

\clearpage

\subsection{Geometric scores computed on the poor quality images ($94$ patients)}
\vspace*{0.3cm}

In addition to table III given in the article, we provide in table \ref{tab:comparison_segmentation_dl_vs_others_poor_quality_images} of this supplementary material the segmentation accuracy (\LVendo~and \LVepi) of \correction{the $8$ evaluated methods} on the ten test sets restricted to patients having poor image quality ($94$ patients). As the results given in the paper, one can see that the \correction{EDNs} architectures show the best segmentation performances on all metrics and structures \correction{except for the $d_H$ metric of the \LVendo~ at ED}.

\vspace*{0.5cm}

\begin{table*}[h]
\renewcommand{\arraystretch}{1.0}
  \caption{Segmentation accuracy (\LVendo~and \LVepi) of $8$ evaluated methods on the ten test datasets restricted to patients having poor image quality ($94$ patients in total). The values in bold refer to the best performance for each measure. p-values are based on the Wilcoxon signed rank between the U-Net methods for each evaluation metric.}
  \centering
  \begin{tabular}{c*{6}{c}|*{6}{c}}

 \toprule 
 \multicolumn{1}{c}{\multirow{6}{*}{\bf \small Methods *}} &  \multicolumn{6}{c}{\bf \small ED} & \multicolumn{6}{c}{\bf \small ES} \\
 \cmidrule(r){2-7} \cmidrule(r){8-13}
 &  \multicolumn{3}{c}{\bf$\boldsymbol{\LVendo}$} & \multicolumn{3}{c}{\bf $\boldsymbol{\LVepi}$} & 
 \multicolumn{3}{c}{\bf $\boldsymbol{\LVendo}$} & \multicolumn{3}{c}{\bf $\boldsymbol{\LVepi}$} \\
 \cmidrule(r){2-4} \cmidrule(r){5-7} \cmidrule(r){8-10} \cmidrule(r){11-13}
 & \multicolumn{1}{c}{\bf \emph{D}} & \multicolumn{1}{c}{\bf \emph{d\textsubscript{m}}} & \multicolumn{1}{c}{\bf \bf \emph{d\textsubscript{H}}} & \multicolumn{1}{c}{\bf \emph{D}} & \multicolumn{1}{c}{\bf \emph{d\textsubscript{m}}} & \multicolumn{1}{c}{\bf \emph{d\textsubscript{H}}} & \multicolumn{1}{c}{\bf \emph{D}} & \multicolumn{1}{c}{\bf \emph{d\textsubscript{m}}} & \multicolumn{1}{c}{\bf \emph{d\textsubscript{H}}} & \multicolumn{1}{c}{\bf \emph{D}} & \multicolumn{1}{c}{\bf \emph{d\textsubscript{m}}} & \multicolumn{1}{c}{\bf \emph{d\textsubscript{H}}} \\
\cmidrule(r){2-2} \cmidrule(r){3-3} \cmidrule(r){4-4} \cmidrule(r){5-5} \cmidrule(r){6-6} \cmidrule(r){7-7} \cmidrule(r){8-8} \cmidrule(r){9-9} \cmidrule(r){10-10} \cmidrule(r){11-11} \cmidrule(r){12-12} \cmidrule(r){13-13}
 & val. & mm & mm & val.& mm & mm & val. & mm & mm & val. & mm & mm \\

\midrule

\addlinespace[+0.7ex]

\multicolumn{1}{l}{\multirow{2}{*}{SRF}} & 0.869 & 3.6 & 14.3 & 0.891
& 4.2 & 15.9 & 0.801 & 4.6 & 17.0 & 0.852
& 4.9 & 18.0 \\

 & \scriptsize{$\pm$0.062} & \scriptsize{$\pm$2.0}
& \scriptsize{$\pm$9.4} & \scriptsize{$\pm$0.063}
& \scriptsize{$\pm$2.4} & \scriptsize{$\pm$8.5}
& \scriptsize{$\pm$0.123} & \scriptsize{$\pm$3.5}
& \scriptsize{$\pm$13.2} & \scriptsize{$\pm$0.112}
& \scriptsize{$\pm$3.1} & \scriptsize{$\pm$12.1} \\

\addlinespace[+0.7ex]

\multicolumn{1}{l}{\multirow{2}{*}{BEAS fully}} & 0.857 & 4.1 & 10.5 & 0.888
& 4.5 & 11.9 & 0.801 & 4.7 & 12.3 & 0.873
& 4.7 & 12.4 \\

 & \scriptsize{$\pm$0.083} & \scriptsize{$\pm$2.6}
& \scriptsize{$\pm$6.3} & \scriptsize{$\pm$0.058}
& \scriptsize{$\pm$2.6} & \scriptsize{$\pm$6.2}
& \scriptsize{$\pm$0.102} & \scriptsize{$\pm$2.7}
& \scriptsize{$\pm$6.6} & \scriptsize{$\pm$0.063}
& \scriptsize{$\pm$2.6} & \scriptsize{$\pm$6.2} \\

\addlinespace[+0.7ex]

\multicolumn{1}{l}{\multirow{2}{*}{BEAS semi}} & 0.915 & 2.4 & \bf 6.4 & 0.914
& 3.4 & 8.5 & 0.859 & 3.3 & 8.3 & 0.900
& 3.6 & 9.5 \\

 & \scriptsize{$\pm$0.039} & \scriptsize{$\pm$1.2}
& \scriptsize{$\pm$2.7} & \scriptsize{$\pm$0.035}
& \scriptsize{$\pm$1.5} & \scriptsize{$\pm$2.9}
& \scriptsize{$\pm$0.063} & \scriptsize{$\pm$1.6}
& \scriptsize{$\pm$3.6} & \scriptsize{$\pm$0.039}
& \scriptsize{$\pm$1.6} & \scriptsize{$\pm$3.4} \\

\addlinespace[+0.7ex]

\midrule

\addlinespace[+0.7ex]

\multicolumn{1}{l}{\multirow{2}{*}{U-Net~1}} & \bf 0.921 & \bf 2.1 & 6.5 & 0.945
& 2.2 & \bf 6.8 & 0.893 & 2.2 & 6.8 & 0.935
& 2.4 & \bf 7.2 \\

& \scriptsize{$\pm$0.037} & \scriptsize{$\pm$1.0}
& \scriptsize{$\pm$3.0} & \scriptsize{$\pm$0.021}
& \scriptsize{$\pm$1.0} & \scriptsize{$\pm$3.0}
& \scriptsize{$\pm$0.059} & \scriptsize{$\pm$1.2}
& \scriptsize{$\pm$4.2} & \scriptsize{$\pm$0.031}
& \scriptsize{$\pm$1.3} & \scriptsize{$\pm$4.7} \\

\addlinespace[+0.7ex]

\multicolumn{1}{l}{\multirow{2}{*}{U-Net~2}} & \bf 0.921 & \bf 2.1 & 6.8 &\bf 0.947 & \bf 2.1 & 7.2 & \bf 0.898 & \bf 2.1 & \bf 6.6 & \bf 0.937
& \bf 2.2 & 7.6 \\

& \scriptsize{$\pm$0.037} & \scriptsize{$\pm$1.0}
& \scriptsize{$\pm$3.3} & \scriptsize{$\pm$0.023}
& \scriptsize{$\pm$1.0} & \scriptsize{$\pm$3.8}
& \scriptsize{$\pm$0.057} & \scriptsize{$\pm$1.2}
& \scriptsize{$\pm$3.6} & \scriptsize{$\pm$0.032}
& \scriptsize{$\pm$1.2} & \scriptsize{$\pm$4.8} \\

\addlinespace[+0.7ex]

\multicolumn{1}{l}{\multirow{2}{*}{ACNN}} & 0.915 & 2.2& 7.2 & 0.939  & 2.4& 8.0 & 0.885 & 2.4& 7.3 & 0.929 & 2.5 & 8.0\\

 & \scriptsize{$\pm 0.041$} & \scriptsize{$\pm 1.1$}
& \scriptsize{$\pm 3.7$} & \scriptsize{$\pm 0.027$}
& \scriptsize{$\pm 1.2$} & \scriptsize{$\pm 5.5$}
& \scriptsize{$\pm0.059 $} & \scriptsize{$\pm 1.2$}
& \scriptsize{$\pm 4.3 $} & \scriptsize{$\pm 0.034$}
& \scriptsize{$\pm 1.3$} & \scriptsize{$\pm 4.7$} \\

\addlinespace[+0.7ex]

\multicolumn{1}{l}{\multirow{2}{*}{SHG}} & 0.919 & \bf 2.1& 6.8& 0.943 & 2.3 & 7.1 & 0.889 & 2.3 & 7.0 & 0.934
& 2.4& 7.3\\

 & \scriptsize{$\pm 0.041 $} & \scriptsize{$\pm 1.2 $}
& \scriptsize{$\pm 3.9$} & \scriptsize{$\pm 0.026$}
& \scriptsize{$\pm 1.2 $} & \scriptsize{$\pm 4.0$}
& \scriptsize{$\pm 0.057$} & \scriptsize{$\pm 1.2$}
& \scriptsize{$\pm 4.0$} & \scriptsize{$\pm 0.032$}
& \scriptsize{$\pm 1.3$} & \scriptsize{$\pm 4.6$} \\

\addlinespace[+0.7ex]

\multicolumn{1}{l}{\multirow{2}{*}{U-Net~++}} & 0.910 & 2.3 & 7.9 & 0.938 & 2.5 & 8.8& 0.886 & 2.3 &  7.9 &  0.924 & 2.7 & 9.6\\

 & \scriptsize{$\pm 0.041$} & \scriptsize{$\pm 1.1$} 
 & \scriptsize{$\pm 3.8$} & \scriptsize{$\pm 0.028$} 
 & \scriptsize{$\pm 1.2$} & \scriptsize{$\pm 5.1$} 
 & \scriptsize{$\pm  0.084$} & \scriptsize{$\pm 1.4 $}
& \scriptsize{$\pm$ 5.5} & \scriptsize{$\pm0.036$}
& \scriptsize{$\pm 1.4$} & \scriptsize{$\pm$ 7.2} \\

\addlinespace[+0.7ex]

\midrule

\multicolumn{1}{l}{p-values} & \scriptsize{$\approx0.939$} & \scriptsize{$\approx0.656$}& \scriptsize{$\approx0.409$} & \scriptsize{$\approx0.054$} & \scriptsize{$\approx0.088$} & \scriptsize{$\approx0.541$} & \scriptsize{$\approx0.194$} & \scriptsize{$\approx0.177$} & \scriptsize{$\approx0.708$} & \scriptsize{$\approx0.061$} & \scriptsize{$\approx0.059$} & \scriptsize{$\approx0.278$}  \\
\bottomrule
  \label{tab:comparison_segmentation_dl_vs_others_poor_quality_images}
  \end{tabular}
\end{table*}

\vspace*{0.3cm}

\correction{From these results, one can see that the performances of all the algorithms (except the SRF) remain close to their corresponding results on good and medium image quality (table III in the article). This suggests that the image quality criteria does not influence so much the quality of the results of automatic solutions. This observation is especially true for encoder-decoder networks since they show less performance drop compared to the non-deep learning techniques. P-values indicate that on this test set of 94 patients, the differences between U-Net 1 and U-Net 2 predictions are not significant.}

\clearpage
\subsection{Bland Altman plots of the \LVef~scores}
\begin{figure}[htbp]
    \centering
       \subfigure[O\textsubscript{2} vs O\textsubscript{1a}]{{\includegraphics[width=7cm]{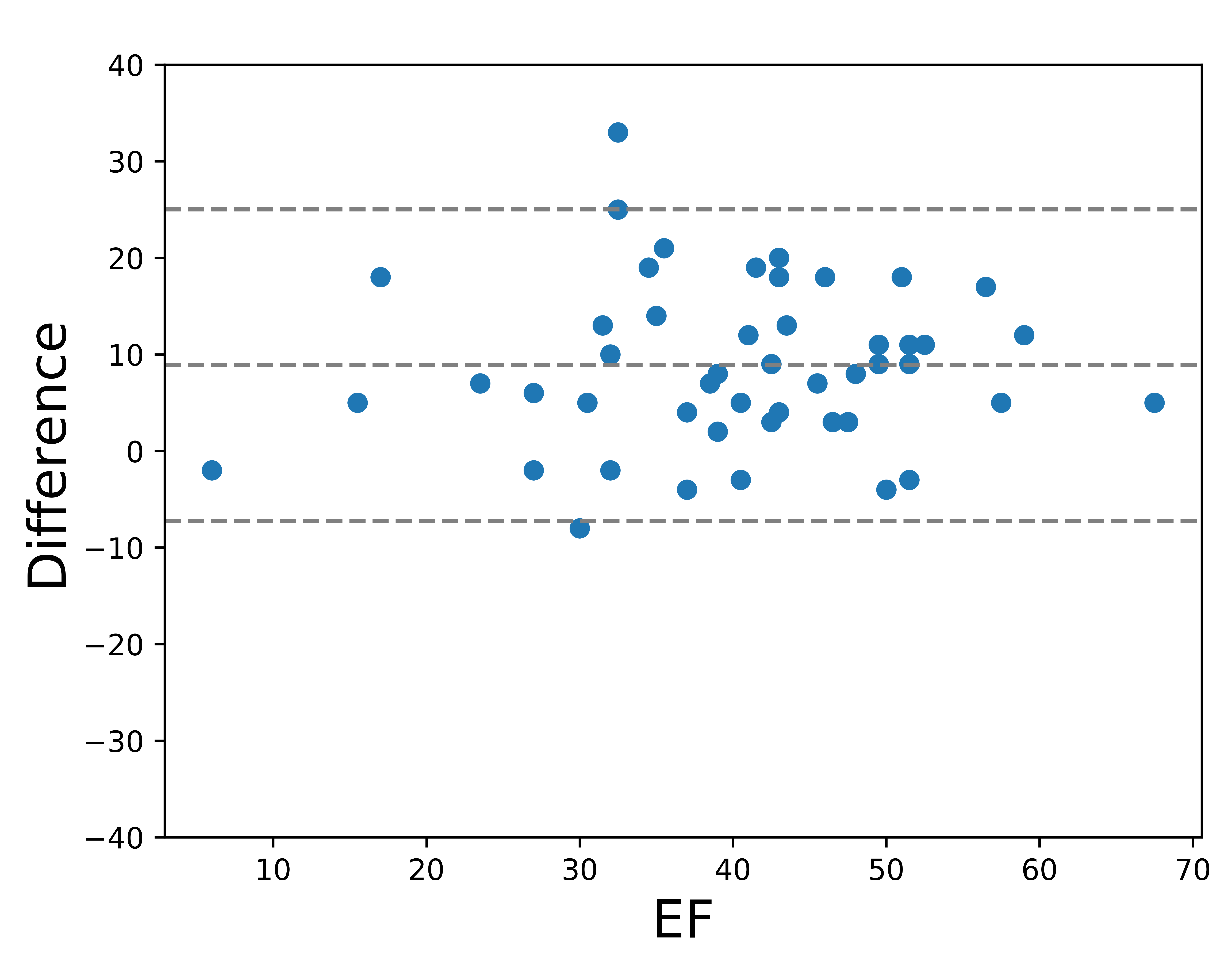}}}
    \qquad
    \subfigure[O\textsubscript{3} vs O\textsubscript{1a}]{{\includegraphics[width=7cm]{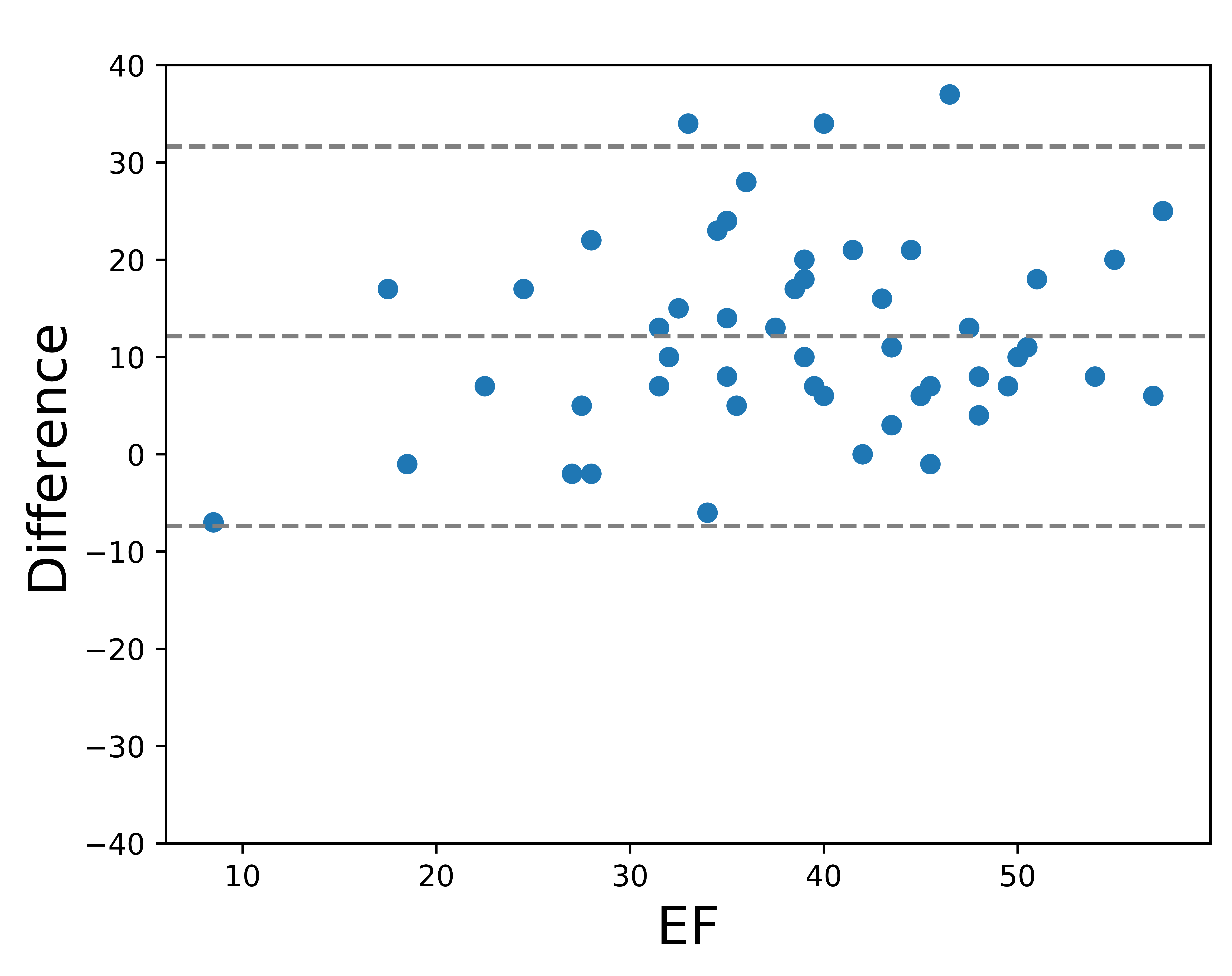}}}
    \\
    \subfigure[O\textsubscript{3} vs O\textsubscript{2}]{{\includegraphics[width=7cm]{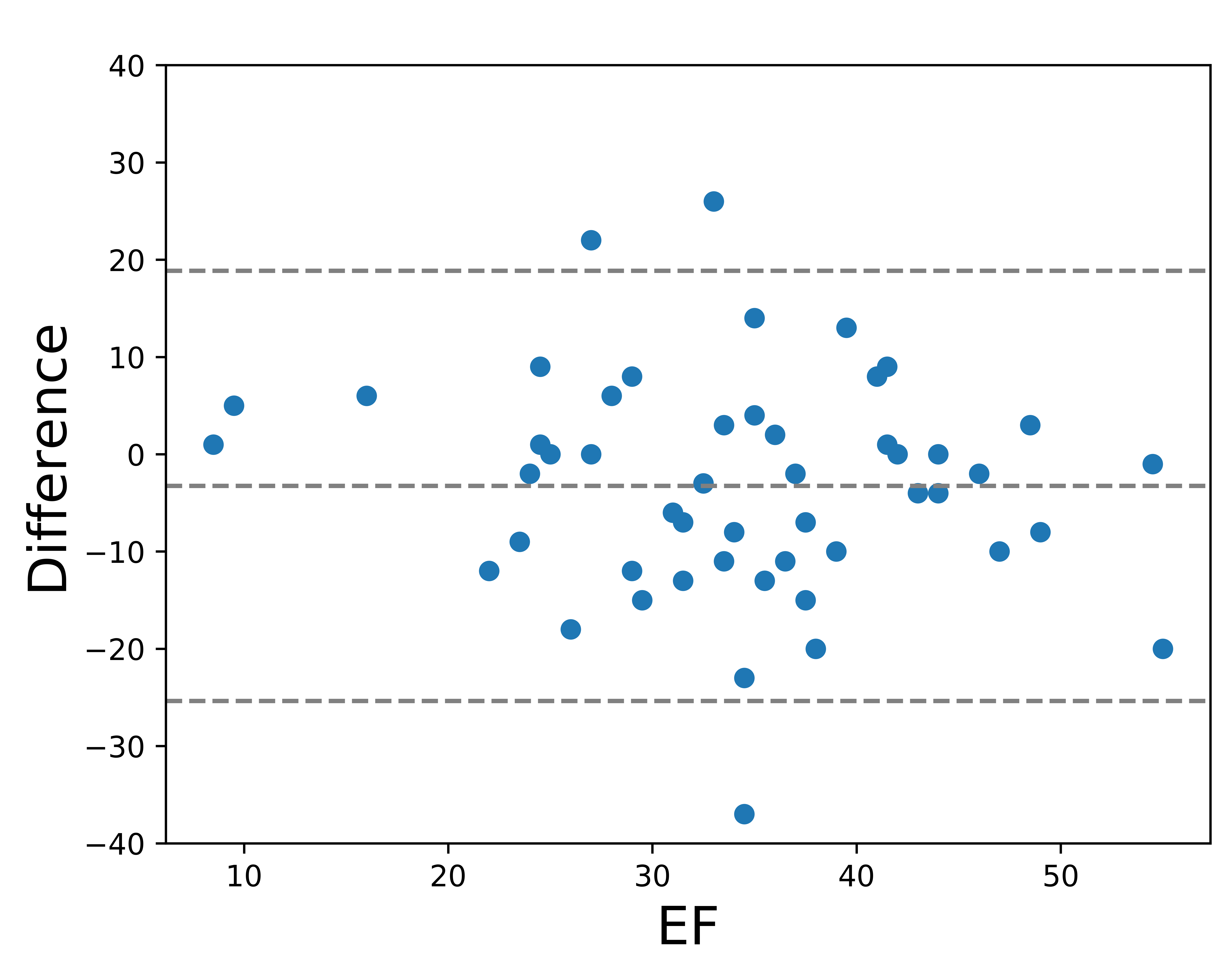}}}
\qquad
\subfigure[O\textsubscript{1b} vs O\textsubscript{1a}]{{\includegraphics[width=7cm]{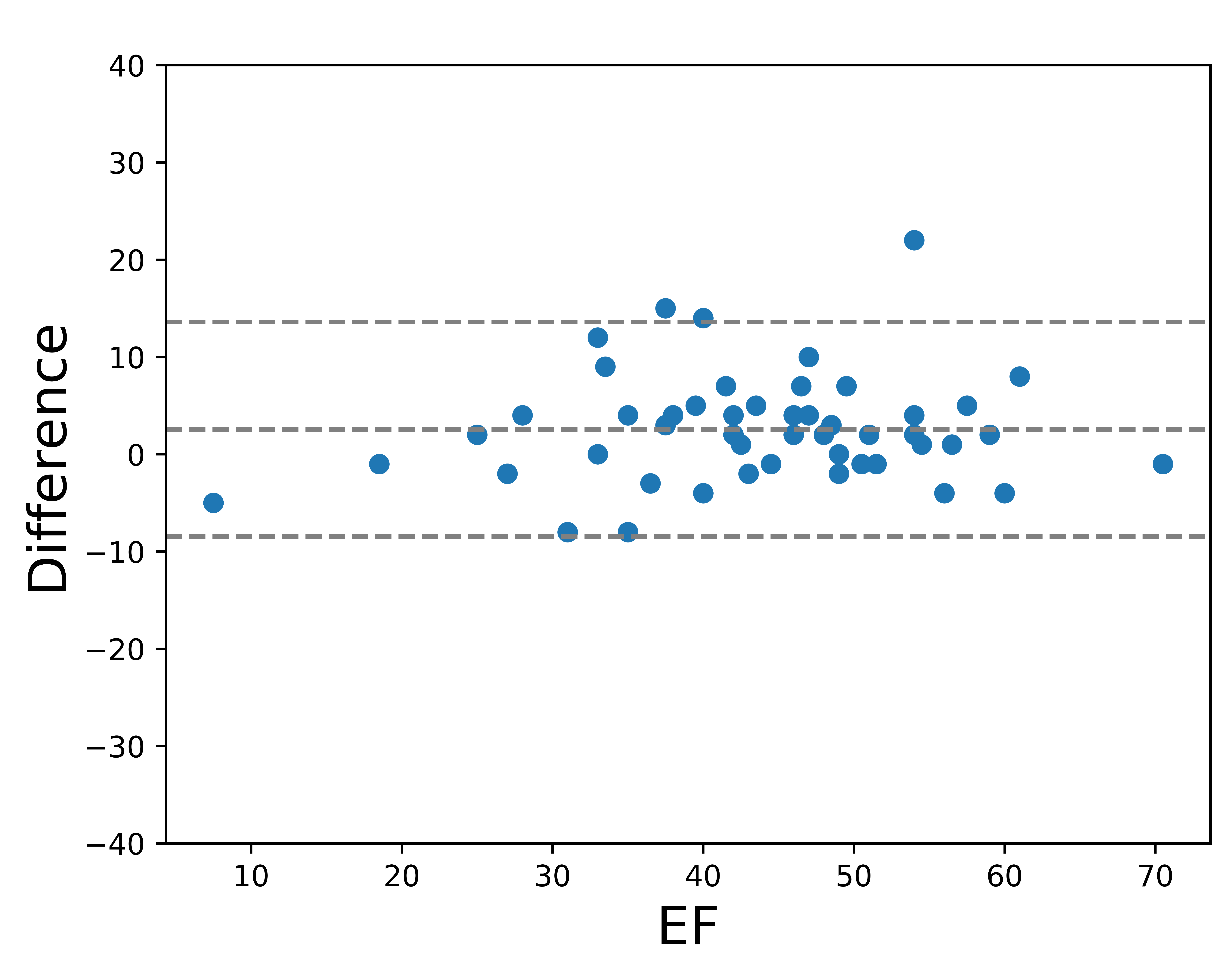}}}
\\    
    \caption{Bland Altman plots of the \LVef~scores computed between the cardiologists from fold $5$}
    \label{fig:bland_altman_plots_cardiologists}
\end{figure}

\begin{figure}[htbp]
    \centering
    \subfigure[U-Net 1 vs O\textsubscript{1a}]{{\includegraphics[width=6.8cm]{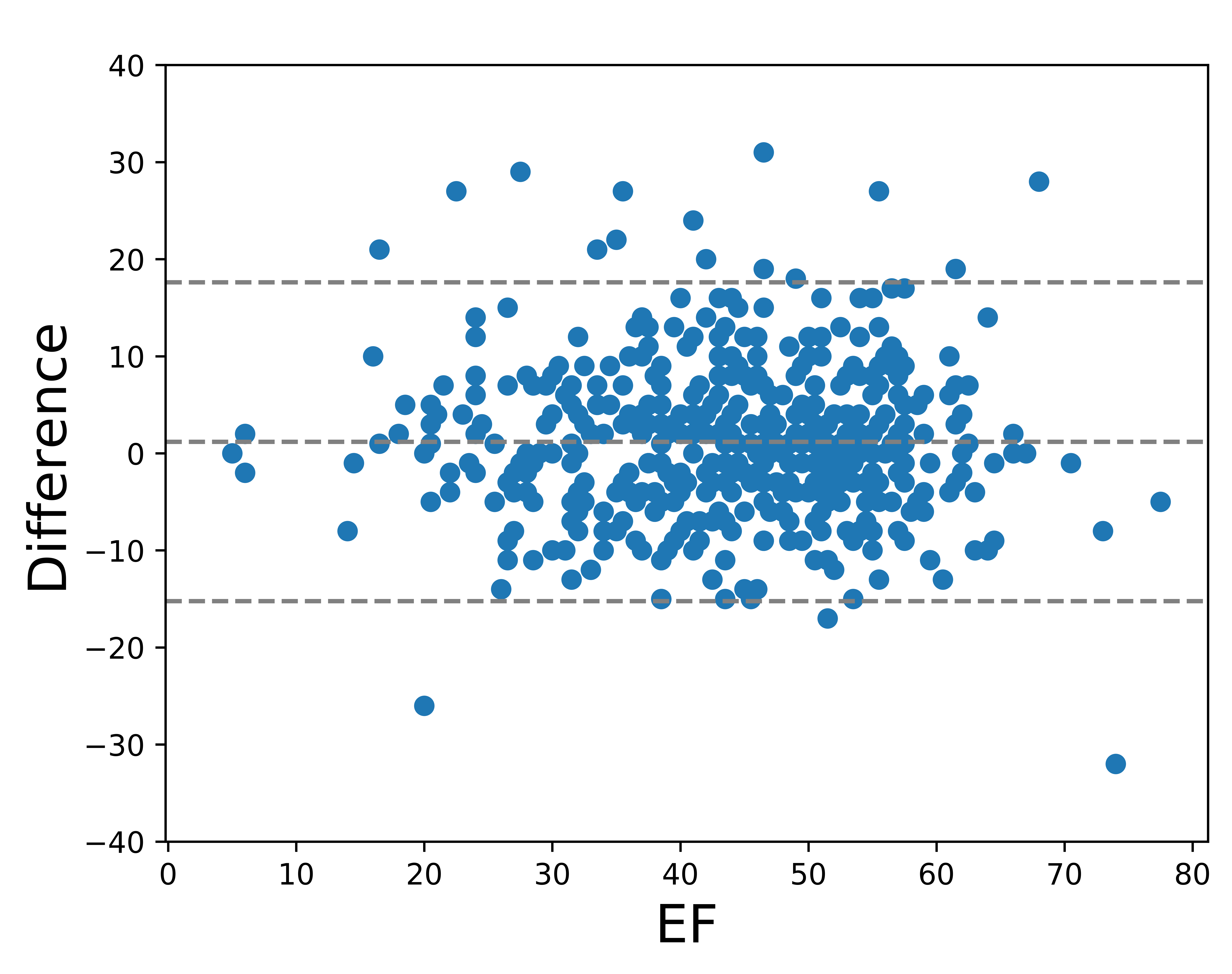}}}
    \qquad
    \subfigure[U-Net 2 vs O\textsubscript{1a}]{{\includegraphics[width=6.8cm]{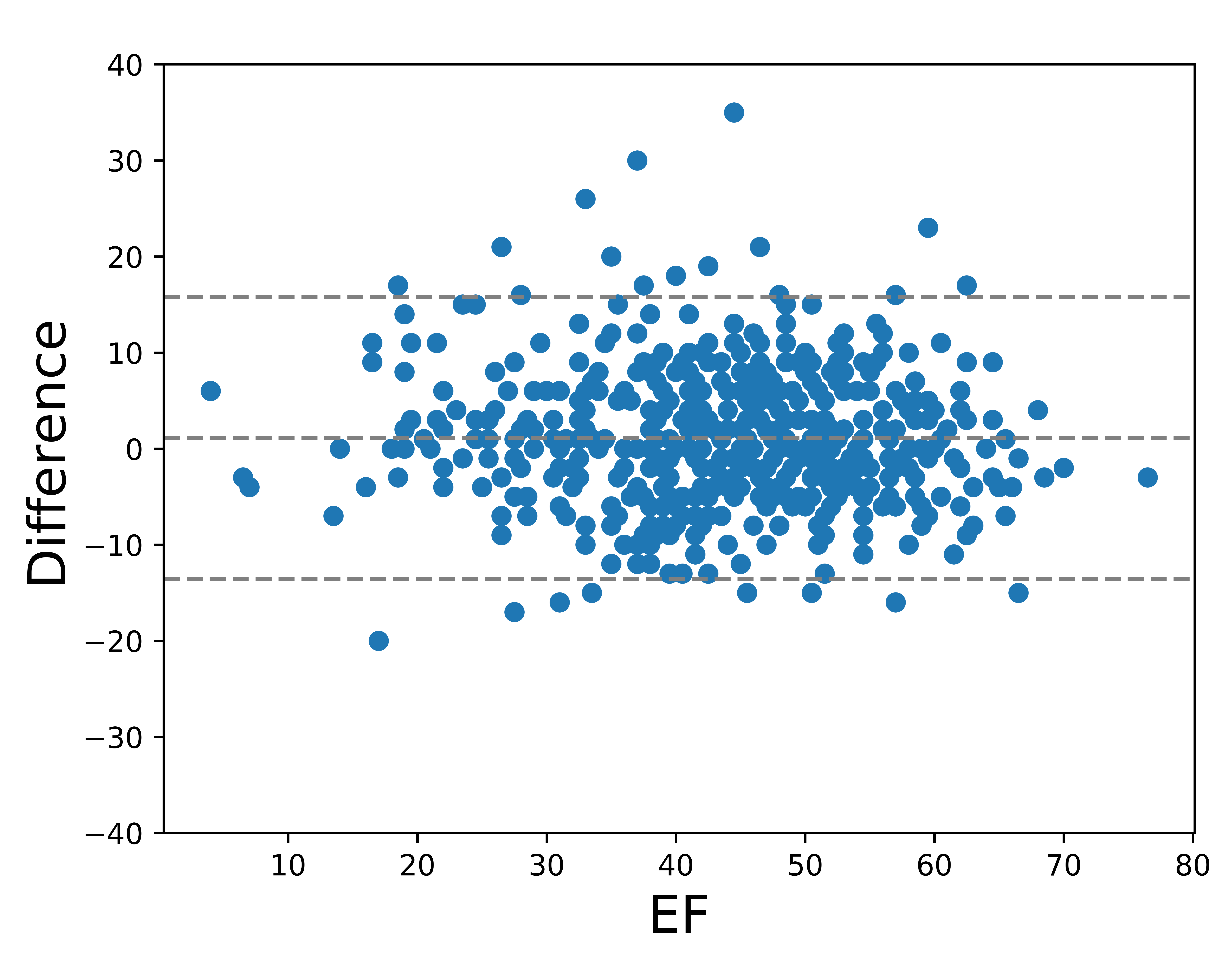}}}
    \\
    \subfigure[ACNN vs O\textsubscript{1a}]{{\includegraphics[width=6.8cm]{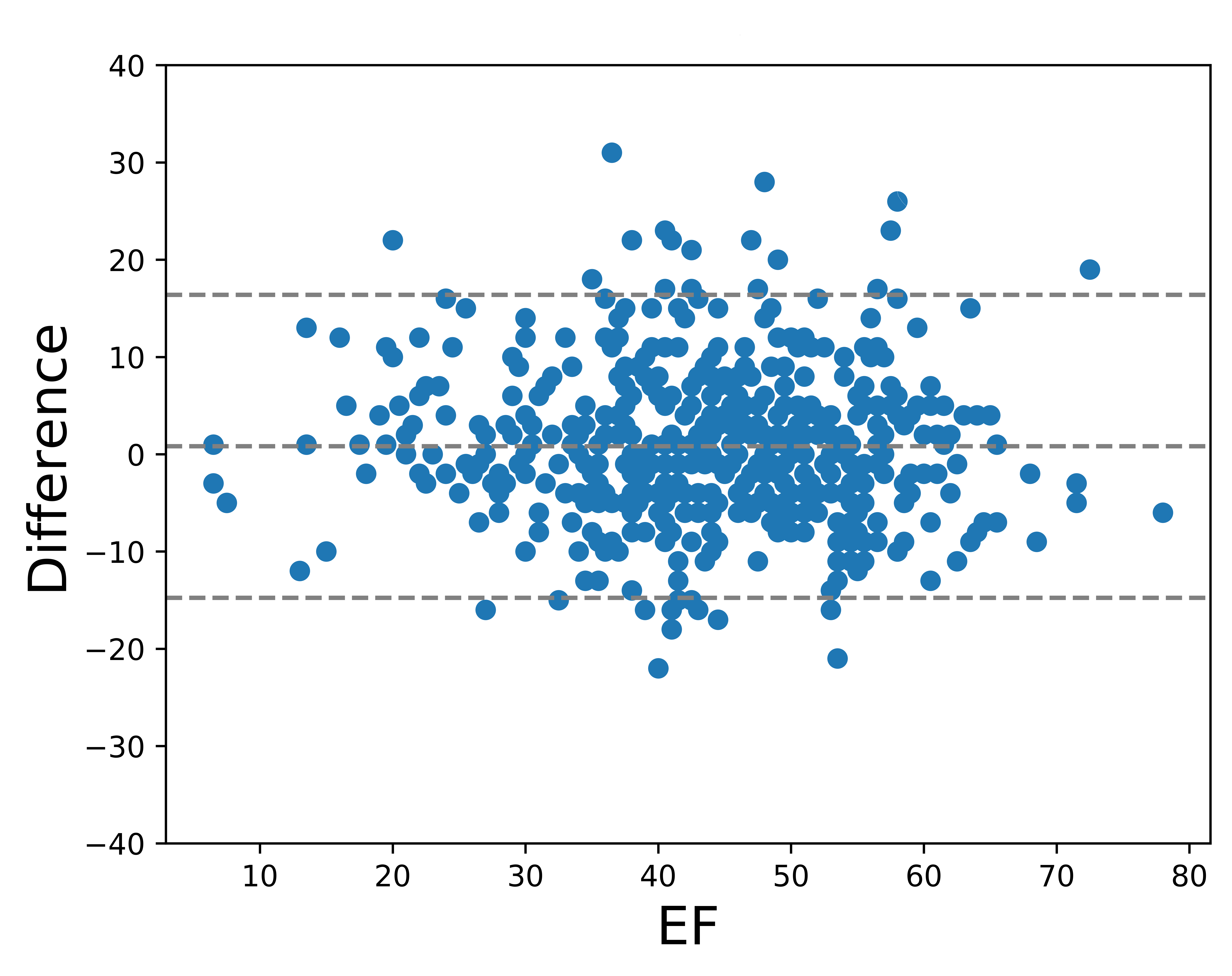}}}
	\qquad
	\subfigure[SHG vs 		O\textsubscript{1a}]{{\includegraphics[width=6.8cm]{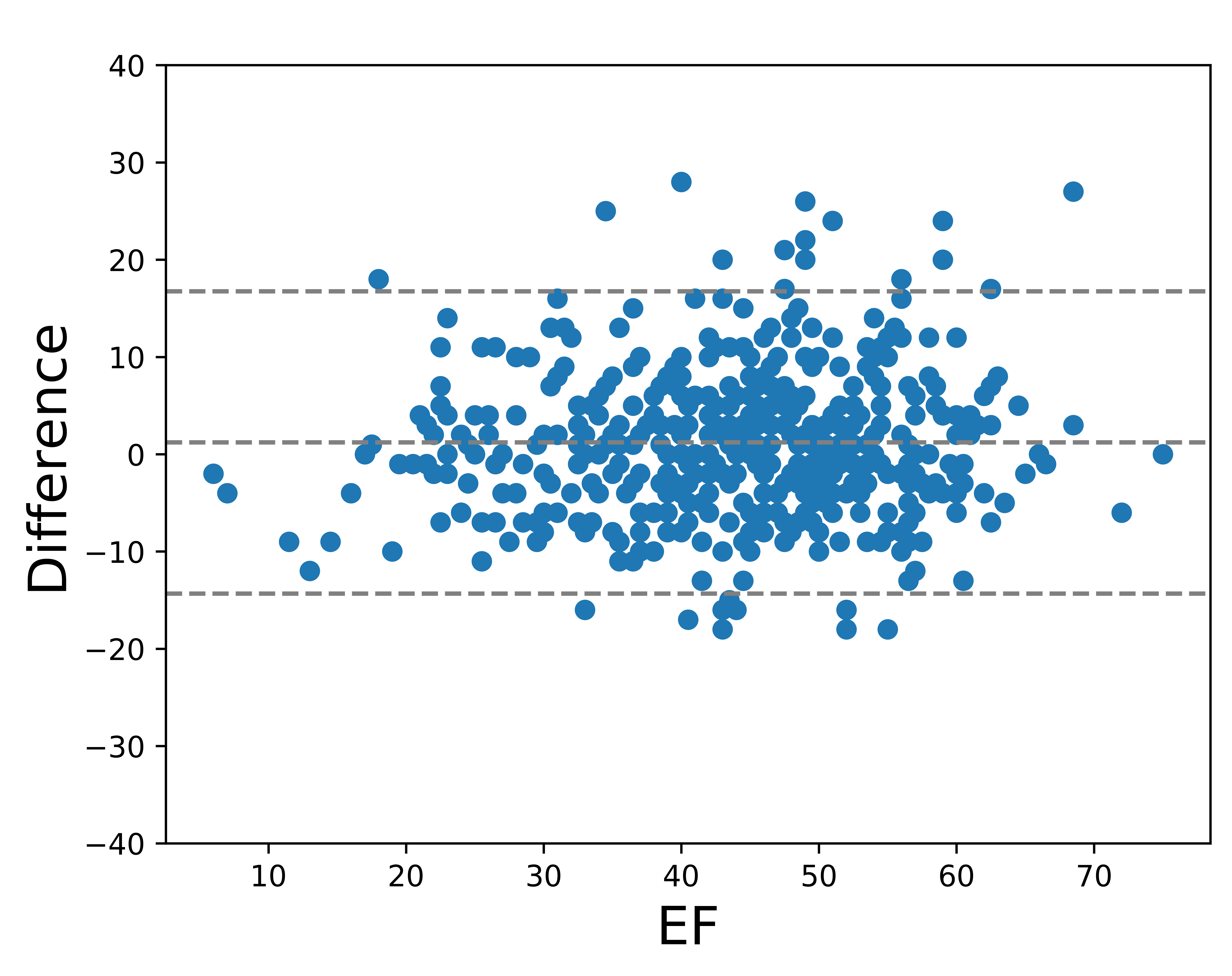}}}
	\\    
	\subfigure[U-Net ++ vs 		O\textsubscript{1a}]{{\includegraphics[width=6.8cm]{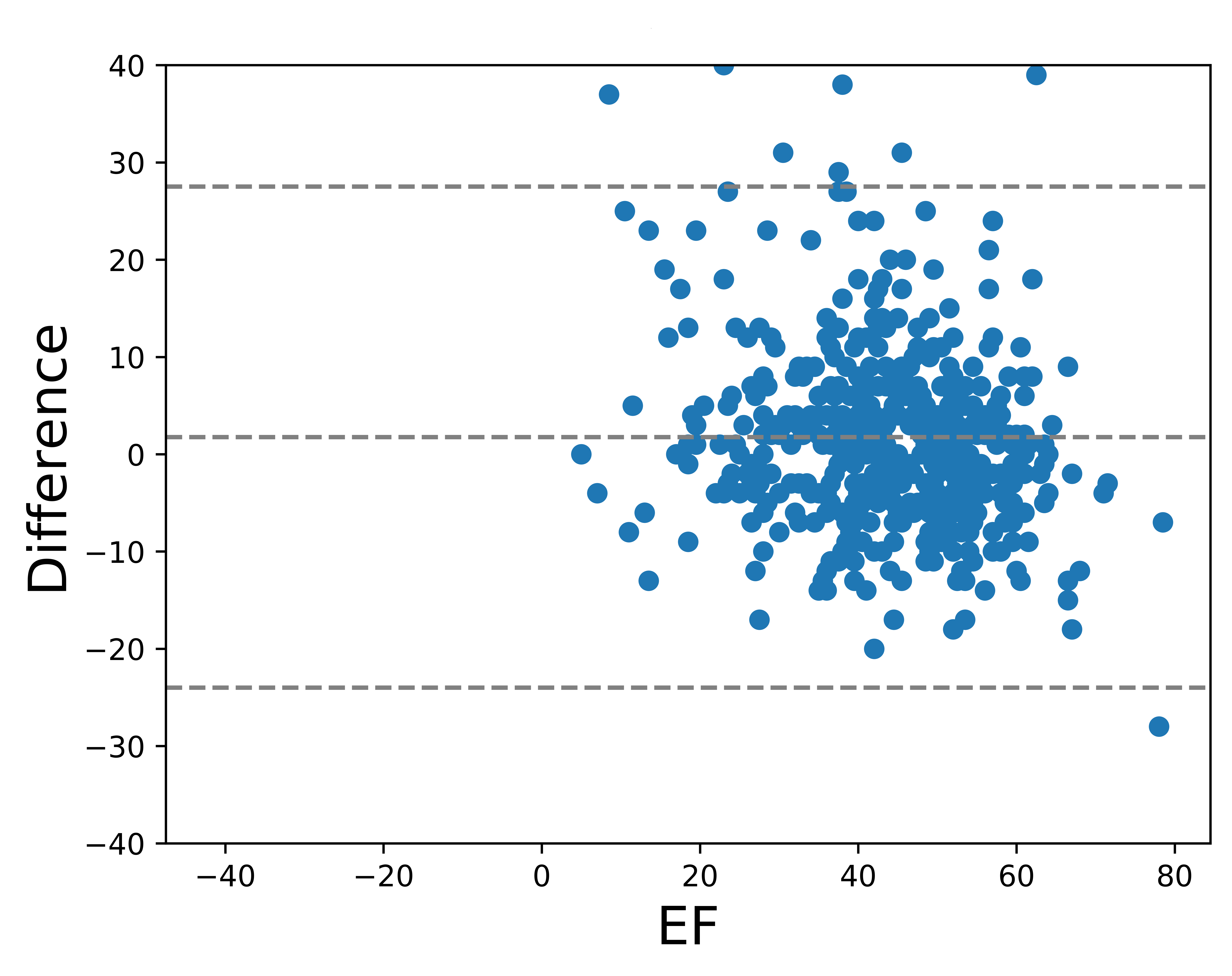}}}\qquad	
	\subfigure[SRF vs O\textsubscript{1a}]{{\includegraphics[width=6.8cm]{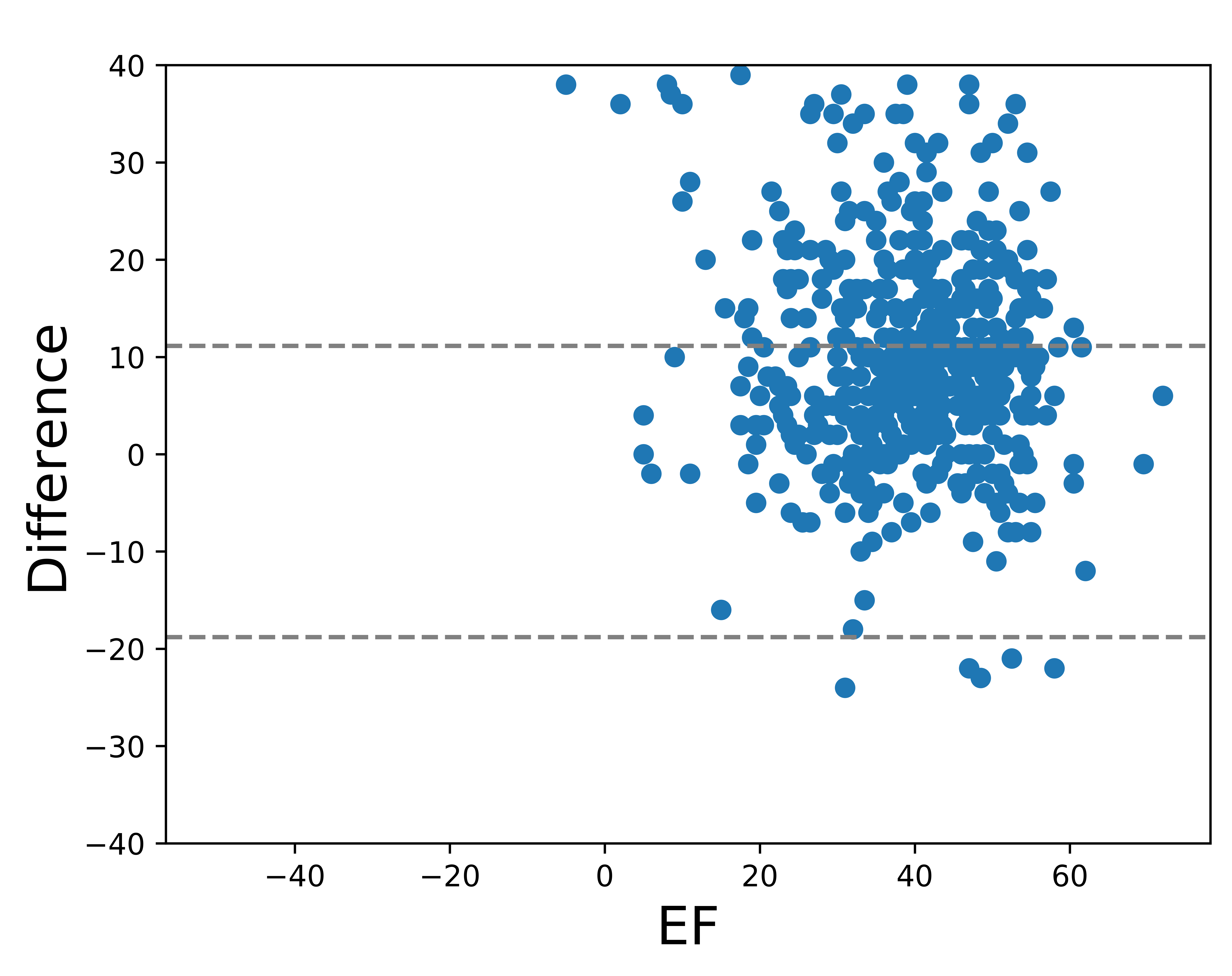}}}
	\\
    \subfigure[BEASM-auto vs O\textsubscript{1a}]{{\includegraphics[width=6.8cm]{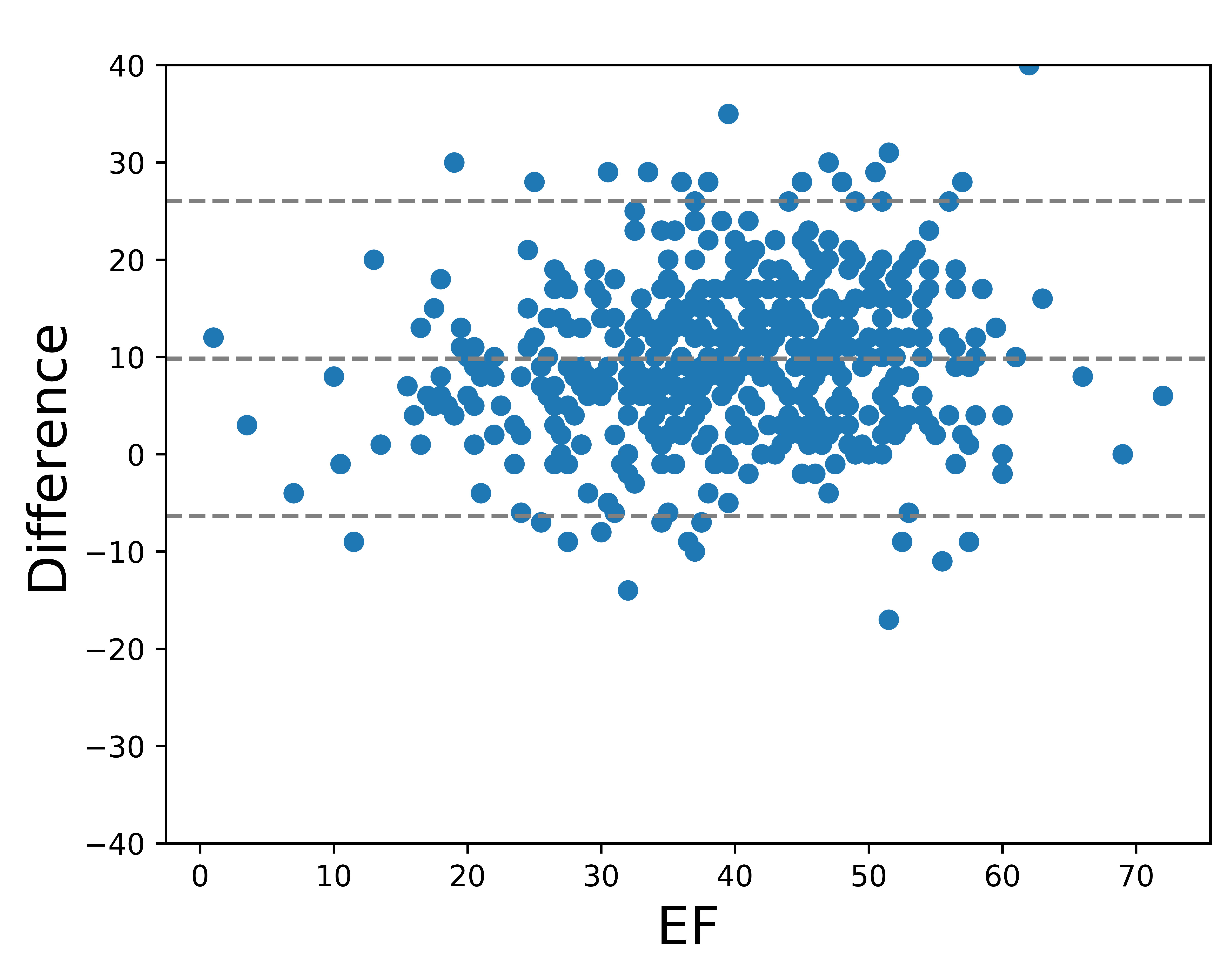}}}
    \qquad
    \subfigure[BEASM-semi vs O\textsubscript{1a}]{{\includegraphics[width=6.8cm]{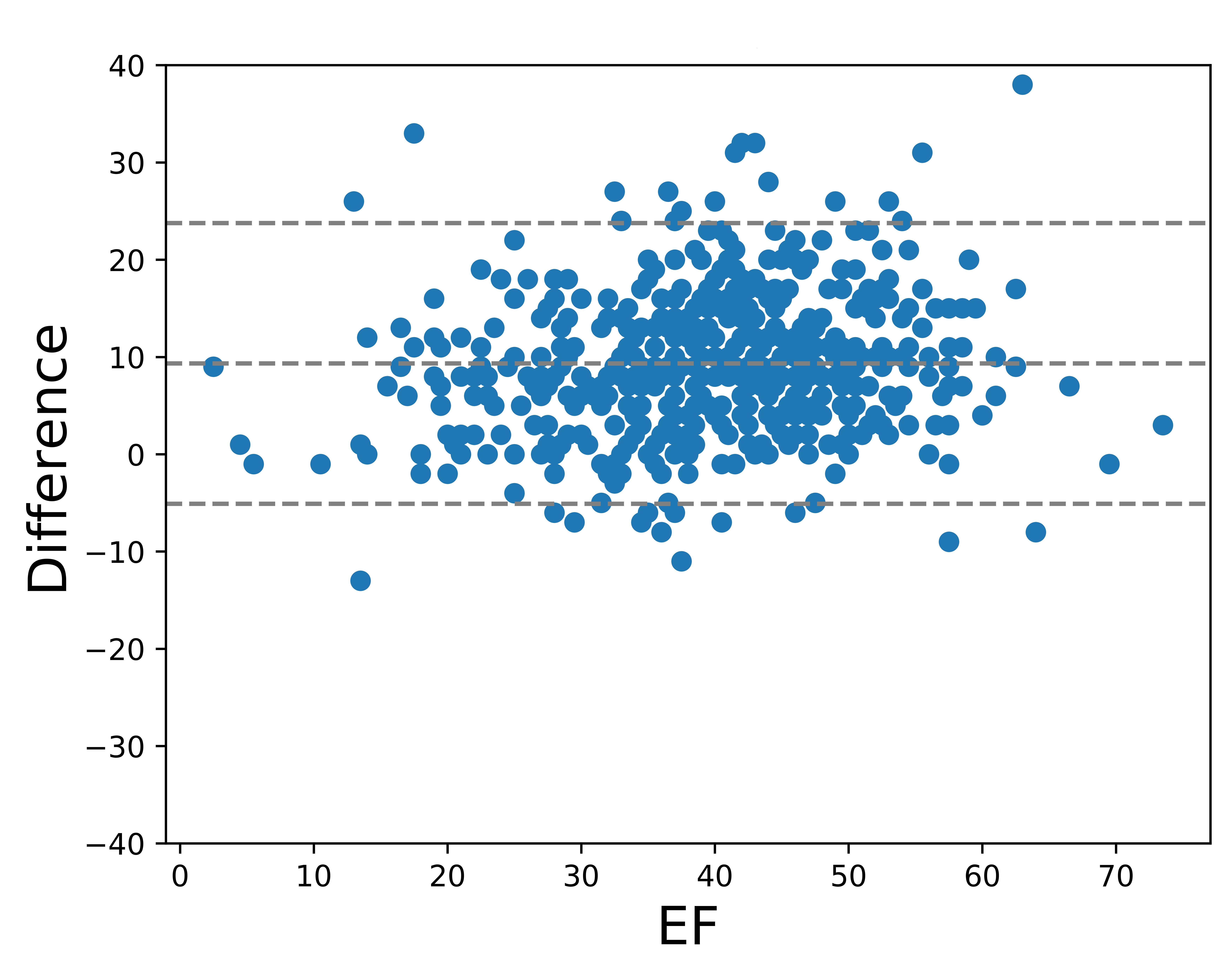}}}
    \caption{Bland Altman plots of the \LVef~scores computed for the $8$ evaluated methods from the full dataset. Mean difference and $95\%$ confidence interval are represented with dotted horizontal lines.}
    \label{fig:bland_altman_plots_methods}
\end{figure}

\vspace*{0.3cm}
As complements to the results given in table IV of the article, we provide in Fig.~\ref{fig:bland_altman_plots_cardiologists} and \ref{fig:bland_altman_plots_methods} the Bland Altman plots of the \LVef~measurements \correction{between the three cardiologists (inter-variability) and the two set of annotations from the first cardiologist (intra-variability), as well as for the $8$ evaluated methods. For a good visualization and an easy comparison, results are shown for errors between -40 and 40. }

\vspace*{0.6cm}

\clearpage
\correction{\section{Additional insights on outliers}}
\vspace*{0.5cm}

\correction{\subsection{Influence of the size of the training dataset}}
\vspace*{0.3cm}

\begin{figure}[htbp]
	\centering
	{\includegraphics[scale=0.5]{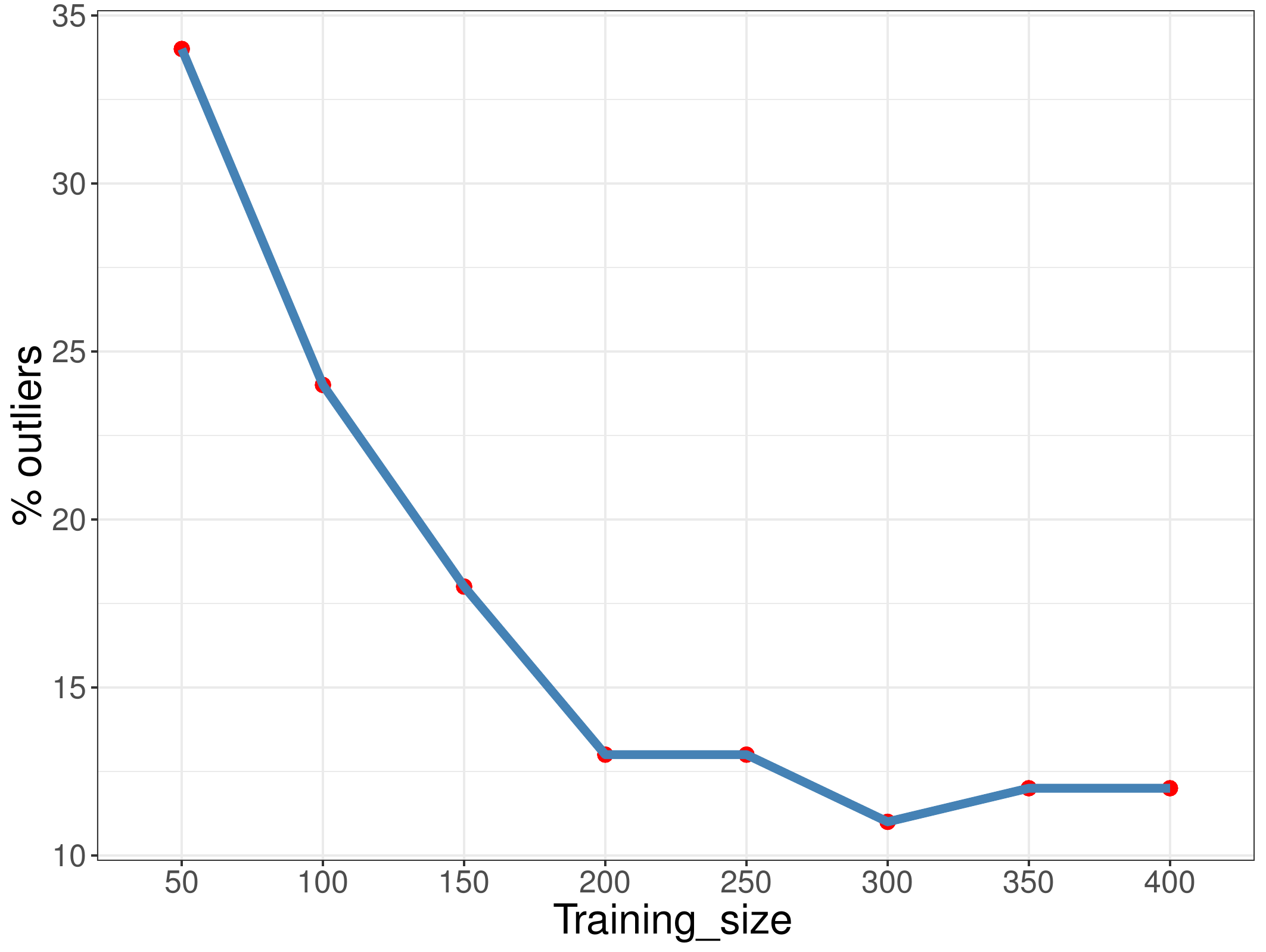}}
	\caption{Number of outliers (in \% of the CAMUS dataset) produced on fold 5 restricted to patients having good or medium image quality for different training set size.}%
	\label{fig:out_inc}
\end{figure}

\correction{In complement to Section VI of the article, Fig. \ref{fig:out_inc} show the evolution of the number of outliers derived from the U-Net 1 model on fold 5 when gradually increasing the training dataset, considering only images of good and medium image quality : }

\vspace*{0.5cm}

\correction{Interestingly, while the improvement between 50 to 200 patients is quite pronounced (e.g. of the outliers rate from $34\%$ to $13\%$), one can observe a change in the evolution of the performance of the U-Net 1 method for more than 200 patients. Indeed, from this value on, the outliers rate seems to stabilize around 12-13$\%$. These results confirm the observations made from Figure 3 of the article, i.e. the U-Net 1 method needs a training dataset of at least 200 patients to reach highly competitive results, which can slightly be further improved with an even larger training dataset.}

\vspace*{0.3cm}
\correction{\subsection{Comparison between methods}}
\vspace*{0.3cm}

\correction{As complement, we provide in table \ref{tab:outliers} the outliers rate produced by each of the 8 methods as well as by each expert (using cardiologist $O1_a$ as reference) on fold 5, for patients having good \& medium image quality (40 patients).}

\begin{table}[h]
\renewcommand{\arraystretch}{1.4}
\caption{Outliers rate of the 8 methods compared to Observer $O_{1a}$ on the whole dataset and to the cardiologists on the fifth fold, restricted to patients having good \& medium image quality (respectively 406 and 40 patients in total).}
\centering
\begin{tabular}{{c}*{12}{c}}
	
	\toprule
	
	\multicolumn{1}{c}{\multirow{2}{*}{\bf \small Outliers}} & SRF & BEASM-fully & BEASM-semi & U-Net 1 & U-Net 2 & ACNN & SHG & U-net ++ & $O_2$ & $O_3$ & $O_2 | O_3$ & $O_{1b}$ \\
	\addlinespace[+0.5ex]
	\cline{2-13}
	\addlinespace[+0.5ex]
	& 69\% & 79\% & 59\% & 18\% & 18\%& 22\%& 18\%& 30\%& 56\%& 58\% & 46\% & 13\% \\
	
	\bottomrule 
	
	\label{tab:outliers}
\end{tabular}
\end{table}

%
%

\section{Visual assessment of the obtained segmentation results}
\vspace*{0.5cm}

\begin{figure}[htbp]
    \centering
    \subfigure[$2$CH-ED: $d_m$-endo = 1.0 , $d_m$-epi = 2.2 , $d_H$-endo = 4.2 , $d_H$-epi = 7.1 mm.]{{\includegraphics[width=5cm]{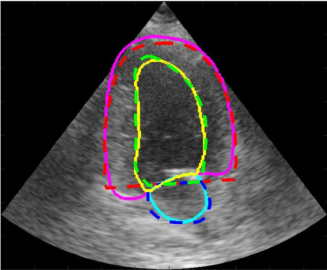}}}
    \qquad
    \subfigure[$2$CH-ES: $d_m$-endo = 1.3 , $d_m$-epi = 2.8 , $d_H$-endo = 7.8 , $d_H$-epi = 8.3 mm.]{{\includegraphics[width=5cm]{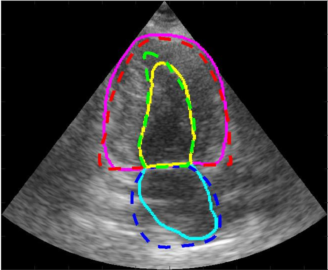}}}
    \\
    \subfigure[$4$CH-ED: $d_m$-endo = 1.7 , $d_m$-epi = 1.7 , $d_H$-endo = 6.3 , $d_H$-epi = 4.0 mm.]{{\includegraphics[width=5cm]{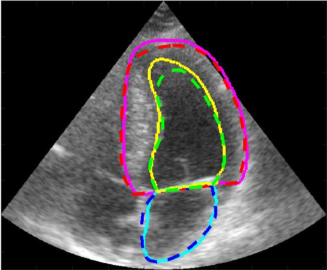}}}
    \qquad
    \subfigure[$4$CH-ES: $d_m$-endo = 1.5 , $d_m$-epi = 1.5 , $d_H$-endo = 5.9 , $d_H$-epi = 3.7 mm.]{{\includegraphics[width=5cm]{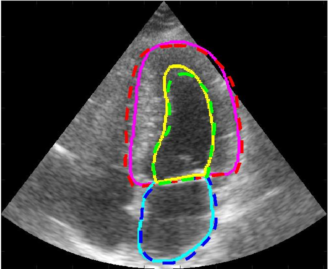}}}
    \caption{Segmentation results obtained by the \mbox{U-Net 1} architecture on Patient 252 (image defined as medium quality). Ground-truth contours are dotted and prediction contours are drawn in full line.}
    \label{fig:unet1_patient252}
\end{figure}

Fig.~\ref{fig:unet1_patient252} provides the segmentation results obtained from U-Net1 on a patient with medium quality. This case has been chosen since it reflect on at least one image the median scores obtained by the \mbox{U-Net 1} algorithm on the full dataset for the endocardium and epicardium: $d_m$-endo = 1.6 mm, $d_m$-epi = 1.7 mm.

\vspace*{1.0cm}

\begin{figure}[htbp]
    \centering
    \subfigure[$2$CH-ED: $d_m$-endo = 1.4 , $d_m$-epi = 1.5 , $d_H$-endo = 4.4 , $d_H$-epi = 4.8 mm.]{{\includegraphics[width=5cm]{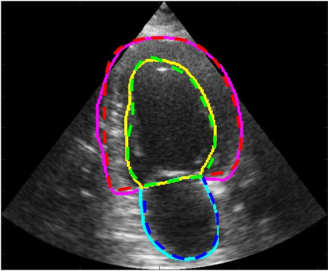}}}
    \qquad
    \subfigure[$2$CH-ES: $d_m$-endo = 1.6 , $d_m$-epi = 2.4 , $d_H$-endo = 5.1 , $d_H$-epi = 7.7 mm.]{{\includegraphics[width=5cm]{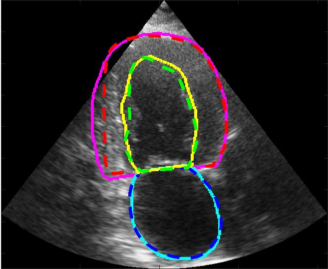}}}
    \\
    \subfigure[$4$CH-ED: $d_m$-endo = 1.0 , $d_m$-epi = 2.2 , $d_H$-endo = 5.9 , $d_H$-epi = 4.5 mm.]{{\includegraphics[width=5cm]{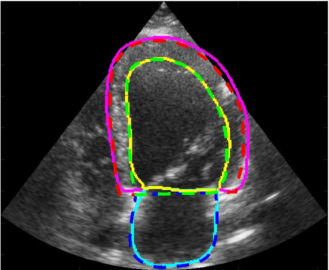}}}
    \qquad
    \subfigure[$4$CH-ES: $d_m$-endo = 1.0 , $d_m$-epi = 1.7 , $d_H$-endo = 3.8 , $d_H$-epi = 4.5 mm.]{{\includegraphics[width=5cm]{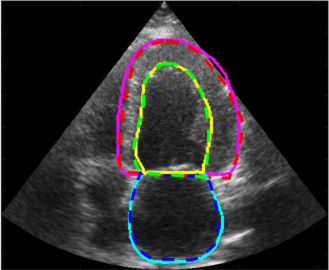}}}
    \caption{Segmentation results obtained by the \mbox{U-Net 1} architecture on Patient 27 (image defined as good quality). Ground-truth contours are dotted and prediction contours are drawn in full line. }
    \label{fig:unet1_patient27}
\end{figure}

\begin{figure}[htbp]
    \centering
    \subfigure[$2$CH-ED : $d_m$-endo = 1.1 , $d_m$-epi = 1.4 , $d_H$-endo = 4.0 , $d_H$-epi = 4.0 mm.]{{\includegraphics[width=5 cm]{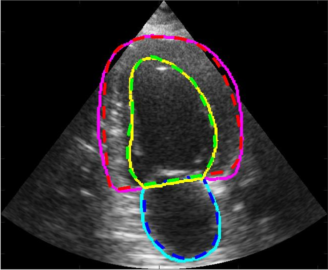}}}
    \qquad
    \subfigure[$2$CH-ES: $d_m$-endo = 1.4 , $d_m$-epi = 1.2 , $d_H$-endo = 4.4 , $d_H$-epi = 4.33 mm.]{{\includegraphics[width=5 cm]{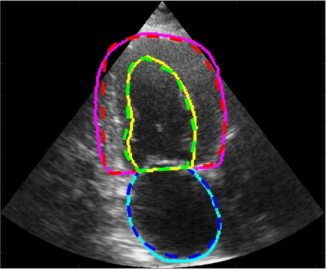}}}
    \\
    \subfigure[$4$CH-ED: $d_m$-endo = 1.6 , $d_m$-epi = 2.0 , $d_H$-endo = 6.8 , $d_H$-epi = 4.6 mm.]{{\includegraphics[width=5 cm]{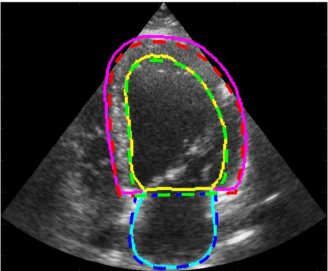}}}
    \qquad
    \subfigure[$4$CH-ES: $d_m$-endo = 1.6 , $d_m$-epi = 1.5 , $d_H$-endo = 4.7 , $d_H$-epi = 4.3 mm.]{{\includegraphics[width=5 cm]{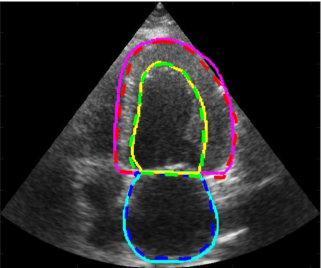}}}
    \caption{Segmentation results obtained by the \mbox{U-Net 2} architecture on Patient 27.}
    \label{fig:unet2_patient27}
\end{figure}

\begin{figure}[htbp]
	\centering
	\subfigure[$2$CH-ED : $d_m$-endo = 0.7 , $d_m$-epi = 1.1 , $d_H$-endo = 2.7 , $d_H$-epi = 3.3 mm.]{{\includegraphics[width=5cm]{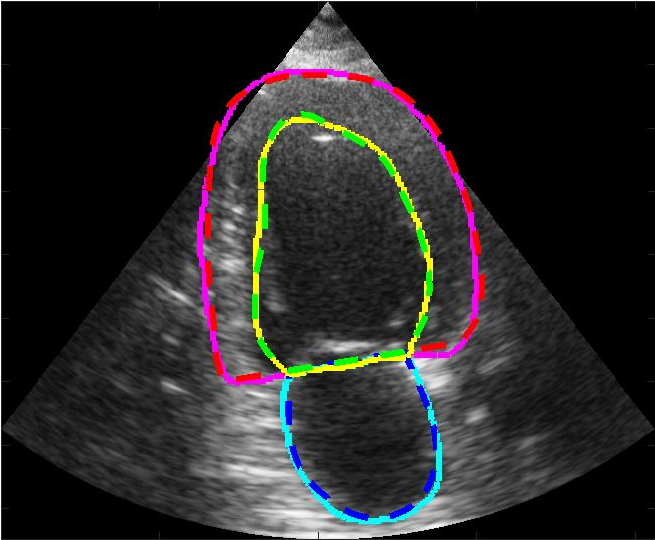}}}
	\qquad
	\subfigure[$2$CH-ES: $d_m$-endo = 1.2 , $d_m$-epi = 1.5 , $d_H$-endo = 4.0 , $d_H$-epi = 4.0 mm.]{{\includegraphics[width=5 cm]{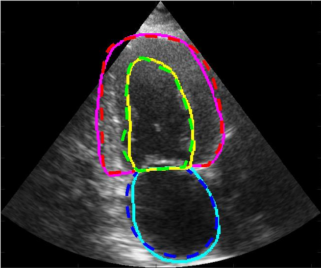}}}
	\\
	\subfigure[$4$CH-ED: $d_m$-endo = 1.2 , $d_m$-epi = 1.7 , $d_H$-endo = 5.9 , $d_H$-epi = 5.1 mm.]{{\includegraphics[width=5 cm]{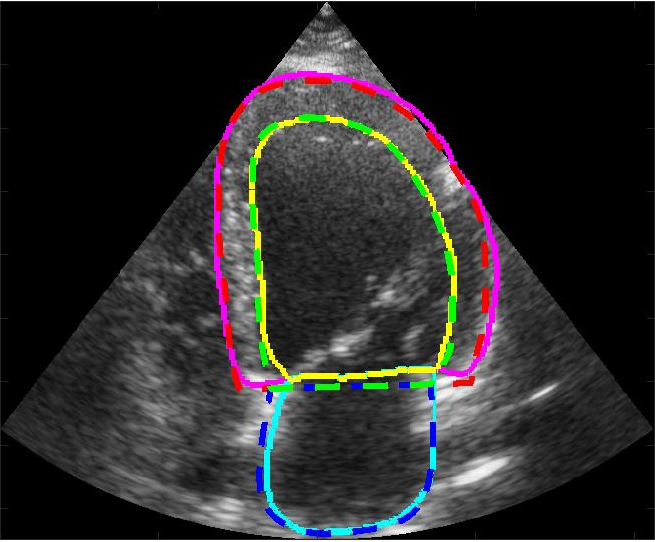}}}
	\qquad
	\subfigure[$4$CH-ES: $d_m$-endo = 1.3 , $d_m$-epi = 1.3 , $d_H$-endo = 3.7 , $d_H$-epi = 3.7 mm.]{{\includegraphics[width=5 cm]{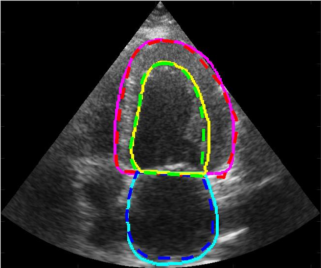}}}
	\caption{Segmentation results obtained by the \mbox{U-Net 1 ACNN} architecture on Patient 27.}
	\label{fig:unetacnn_patient27}
\end{figure}

\begin{figure}[htbp]
	\centering
	\subfigure[$2$CH-ED : $d_m$-endo = 1.1 , $d_m$-epi = 1.1 , $d_H$-endo = 4.0 , $d_H$-epi = 3.1 mm.]{{\includegraphics[width=5 cm]{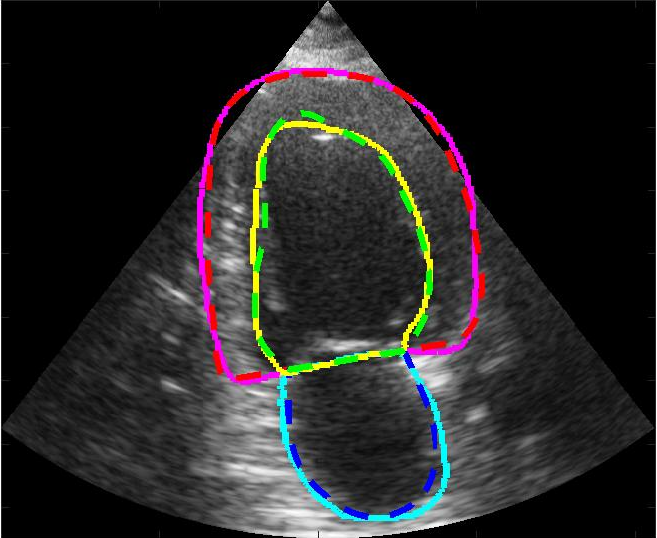}}}
	\qquad
	\subfigure[$2$CH-ES : $d_m$-endo = 1.2 , $d_m$-epi = 1.4 , $d_H$-endo = 4.1 , $d_H$-epi = 4.7 mm.]{{\includegraphics[width=5 cm]{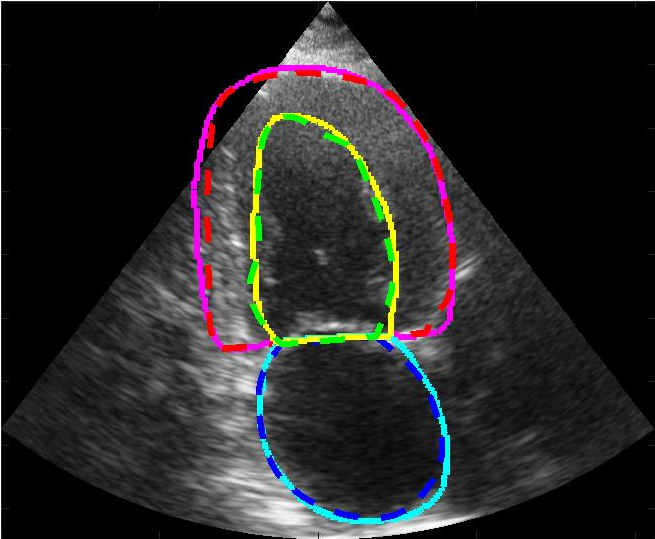}}}
	\\
	\subfigure[$4$CH-ED : $d_m$-endo = 1.1 , $d_m$-epi = 6.3 , $d_H$-endo = 2.7 , $d_H$-epi = 5.1 mm.]{{\includegraphics[width=5 cm]{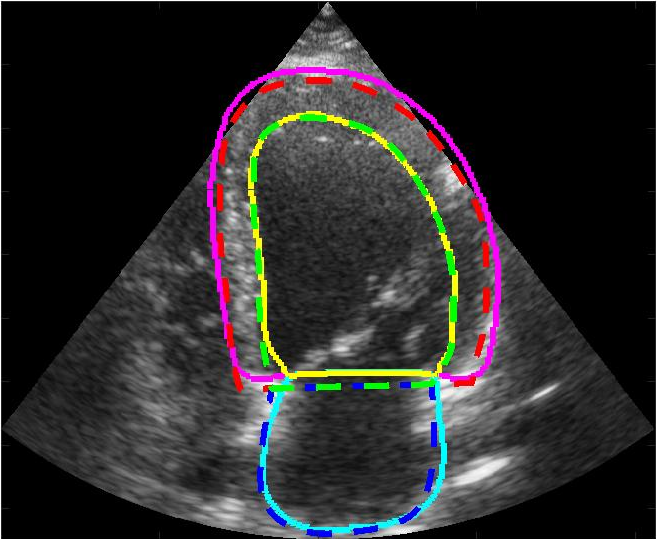}}}
	\qquad
	\subfigure[$4$CH-ES : $d_m$-endo = 1.5 , $d_m$-epi = 5.5 , $d_H$-endo = 1.6 , $d_H$-epi = 4.1 mm.]{{\includegraphics[width=5 cm]{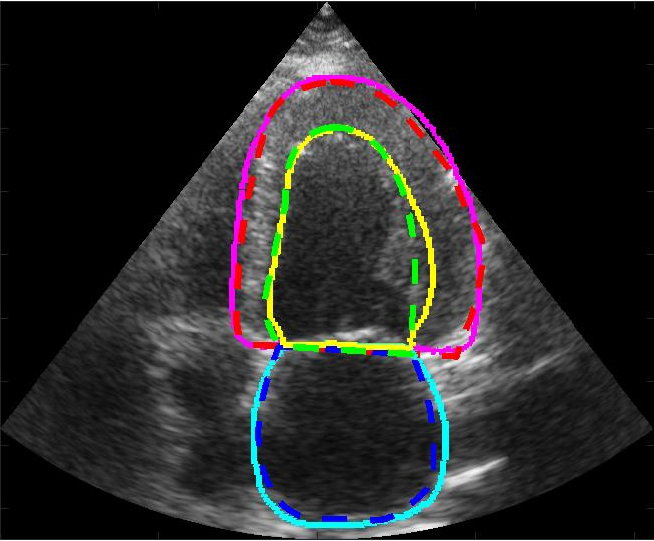}}}
	\caption{Segmentation results obtained by the \mbox{U-Net 1 SHG} architecture on Patient 27.}
	\label{fig:unetshg_patient27}
\end{figure}

\begin{figure}[htbp]
	\centering
	\subfigure[$2$CH-ED : $d_m$-endo = 1.3 , $d_m$-epi = 1.7 , $d_H$-endo = 3.7 , $d_H$-epi = 4.6 mm.]{{\includegraphics[width=5 cm]{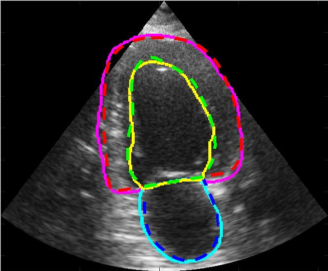}}}
	\qquad
	\subfigure[$2$CH-ES : $d_m$-endo = 1.8 , $d_m$-epi = 1.7 , $d_H$-endo = 4.9 , $d_H$-epi = 5.2 mm.]{{\includegraphics[width=5 cm]{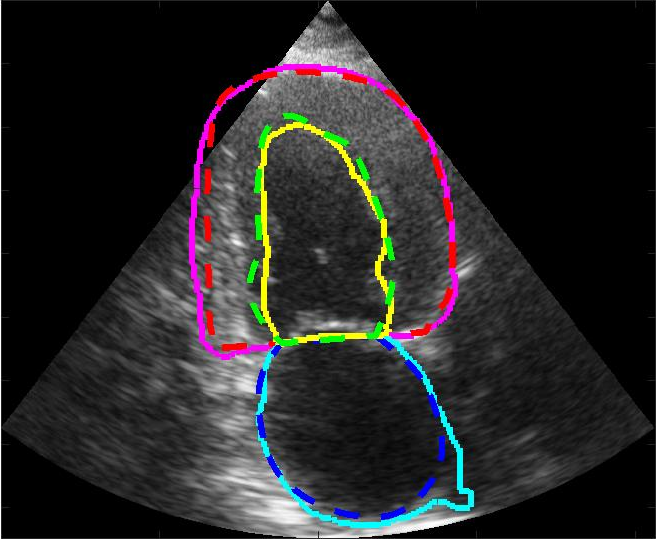}}}
	\\
	\subfigure[$4$CH-ED : $d_m$-endo = 2.5 , $d_m$-epi = 2.2 , $d_H$-endo = 6.3 , $d_H$-epi = 4.7 mm.]{{\includegraphics[width=5 cm]{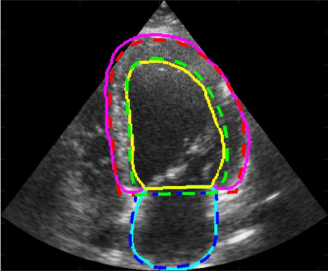}}}
	\qquad
	\subfigure[$4$CH-ES : $d_m$-endo = 1.4 , $d_m$-epi = 1.9 , $d_H$-endo = 4.1 , $d_H$-epi = 4.8 mm.]{{\includegraphics[width=5 cm]{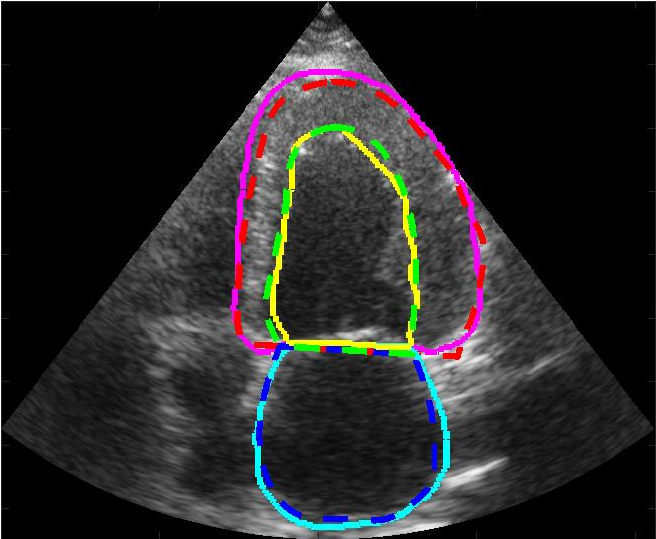}}}
	\caption{Segmentation results obtained by the \mbox{U-Net 1 ++} architecture on Patient 27.}
	\label{fig:unetn_patient27}
\end{figure}

\begin{figure}[htbp]
    \centering
    \subfigure[$2$CH-ED: $d_m$-endo = 2.7 , $d_m$-epi = 1.8 , $d_H$-endo = 8.5 , $d_H$-epi = 8.1 mm.]{{\includegraphics[width=5cm]{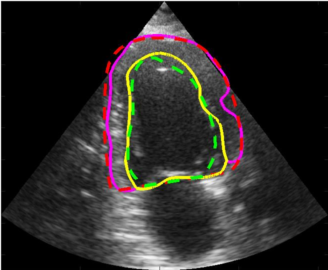}}}
    \qquad
    \subfigure[$2$CH-ES: $d_m$-endo = 3.8 , $d_m$-epi = 2.6 , $d_H$-endo = 14.0 , $d_H$-epi = 7.0 mm.]{{\includegraphics[width=5cm]{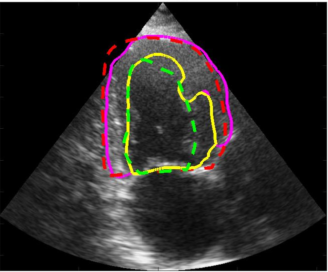}}}
    \\
    \subfigure[$4$CH-ED: $d_m$-endo = 2.3 , $d_m$-epi = 2.3 , $d_H$-endo = 11.1 , $d_H$-epi = 14.2 mm.]{{\includegraphics[width=5cm]{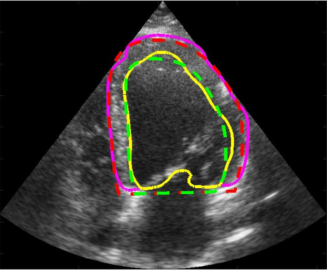}}}
    \qquad
    \subfigure[$4$CH-ES: $d_m$-endo = 3.2 , $d_m$-epi = 1.7 , $d_H$-endo = 6.8 , $d_H$-epi = 4.8 mm.]{{\includegraphics[width=5cm]{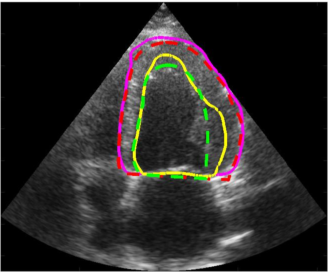}}}
    \caption{Segmentation results obtained by the SRF method on Patient 27.}
    \label{fig:srf_patient27}
\end{figure}

\begin{figure}[htbp]
    \centering
    \subfigure[$2$CH-ED: $d_m$-endo = 3.0 , $d_m$-epi = 3.6 , $d_H$-endo = 7.2 , $d_H$-epi = 7.7 mm.]{{\includegraphics[width=5cm]{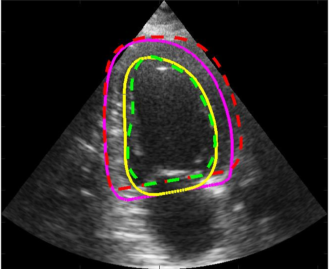}}}
    \qquad
    \subfigure[$2$CH-ES: $d_m$-endo = 2.9 , $d_m$-epi = 2.8 , $d_H$-endo = 4.7, $d_H$-epi = 5.5 mm.]{{\includegraphics[width=5cm]{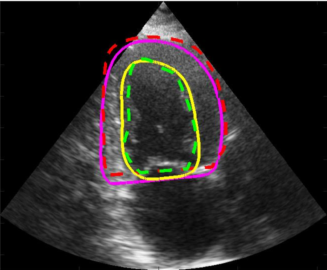}}}
    \\
    \subfigure[$4$CH-ED: $d_m$-endo = 2.4 , $d_m$-epi = 1.7 , $d_H$-endo = 6.2 , $d_H$-epi = 6.6 mm.]{{\includegraphics[width=5cm]{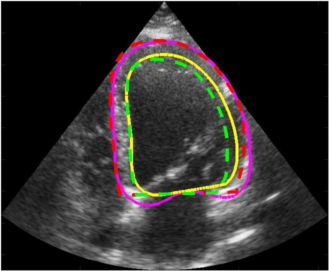}}}
    \qquad
    \subfigure[$4$CH-ES: $d_m$-endo = 5.0 , $d_m$-epi = 2.3 , $d_H$-endo = 14.2 , $d_H$-epi = 6.5 mm.]{{\includegraphics[width=5cm]{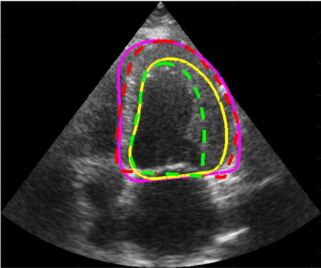}}}
    \caption{Segmentation results obtained by the BEASM method with automatic initialisation on Patient 27. }
    \label{fig:beasm_auto_patient27}
\end{figure}

\begin{figure}[htbp]
    \centering
    \subfigure[$2$CH-ED: $d_m$-endo = 1.7 , $d_m$-epi = 4.9 , $d_H$-endo = 4.1 , $d_H$-epi = 9.6 mm.]{{\includegraphics[width=5cm]{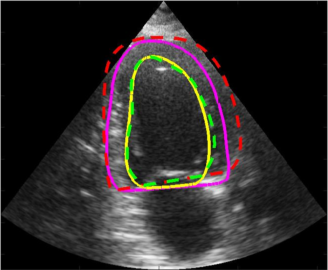}}}
    \qquad
    \subfigure[$2$CH-ES: $d_m$-endo = 1.5 , $d_m$-epi = 3.8 , $d_H$-endo = 8.5 , $d_H$-epi = 8.5 mm.]{{\includegraphics[width=5cm]{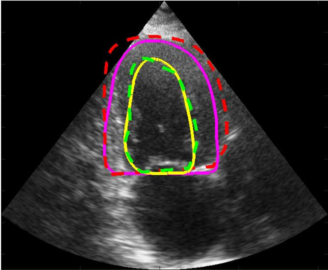}}}
    \\
    \subfigure[$4$CH-ED: $d_m$-endo = 2.1 , $d_m$-epi = 3.2 , $d_H$-endo = 6.2 , $d_H$-epi = 8.2 mm.]{{\includegraphics[width=5cm]{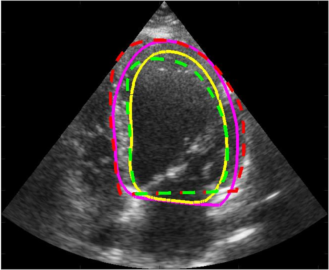}}}
    \qquad
    \subfigure[$4$CH-ES: $d_m$-endo = 3.4 , $d_m$-epi = 2.8 , $d_H$-endo = 3.5 , $d_H$-epi = 7.7 mm.]{{\includegraphics[width=5cm]{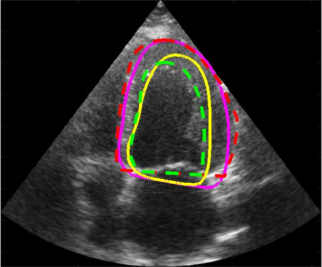}}}
    \caption{Segmentation results obtained by the BEASM method with semi-automatic initialisation on Patient 27.}
    \label{fig:beasmsemi_patient27}
\end{figure}

\begin{figure}[htbp]
    \centering
    \subfigure[$2$CH-ED: $d_m$-endo = 3.1 , $d_m$-epi = 4.2 , $d_H$-endo = 5.9 , $d_H$-epi = 9.0 mm.]{{\includegraphics[width=5cm]{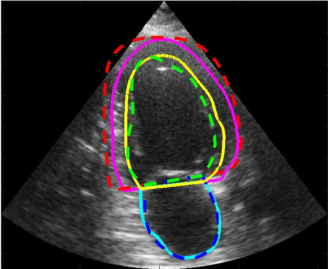}}}
    \qquad
    \subfigure[$2$CH-ES: $d_m$-endo = 3.3 , $d_m$-epi = 5.5 , $d_H$-endo = 8.1 , $d_H$-epi = 9.5 mm.]{{\includegraphics[width=5cm]{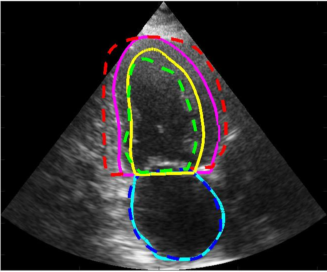}}}
    \\
    \subfigure[$4$CH-ED: $d_m$-endo = 2.0 , $d_m$-epi = 1.4, $d_H$-endo = 4.8 , $d_H$-epi = 3.4 mm.]{{\includegraphics[width=5cm]{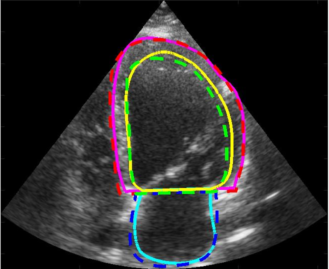}}}
    \qquad
    \subfigure[$4$CH-ES: $d_m$-endo = 2.4 , $d_m$-epi = 1.9 , $d_H$-endo = 6.1 , $d_H$-epi = 9.0 mm.]{{\includegraphics[width=5cm]{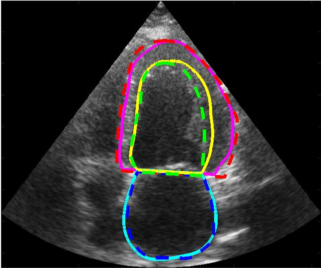}}}
    \caption{Segmentation results obtained by the cardiologist 2 on Patient 27. }
    \label{fig:cardio2_patient27}
\end{figure}

\begin{figure}[htbp]
    \centering
    \subfigure[$2$CH-ED: $d_m$-endo = 4.5 , $d_m$-epi = 4.1 , $d_H$-endo = 9.7 , $d_H$-epi = 4.1 mm.]{{\includegraphics[width=5cm]{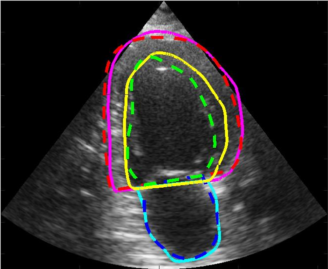}}}
    \qquad
    \subfigure[$2$CH-ES: $d_m$-endo = 5.9 , $d_m$-epi = 4.5 , $d_H$-endo = 13.2 , $d_H$-epi = 4.5 mm.]{{\includegraphics[width=5cm]{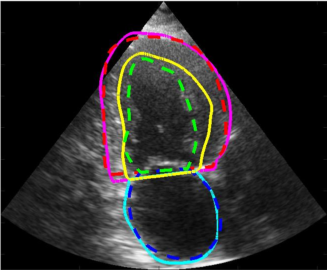}}}
    \\
    \subfigure[$4$CH-ED: $d_m$-endo = 1.7 , $d_m$-epi = 4.0 , $d_H$-endo = 6.8 , $d_H$-epi = 4.0 mm.]{{\includegraphics[width=5cm]{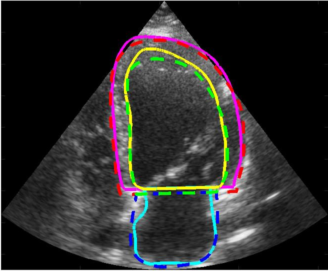}}}
    \qquad
    \subfigure[$4$CH-ES: $d_m$-endo = 2.7 , $d_m$-epi = 4.4 , $d_H$-endo = 4.9 , $d_H$-epi = 4.4 mm.]{{\includegraphics[width=5cm]{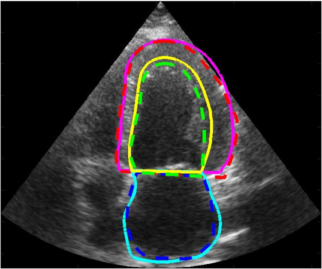}}}
    \caption{Segmentation results obtained by the cardiologist 3 on Patient 27.}
    \label{fig:cardio3_patient27}
\end{figure}

To allow visual assessment of the segmentation performance of the different methods implemented in our paper, we provide in Fig.~\ref{fig:unet1_patient27} to \ref{fig:cardio3_patient27} the segmentations results obtained by each of the presented methods and the cardiologists on a given patient with a good image quality.

\vspace*{0.3cm}


\vspace*{0.6cm}

\vspace*{0.6cm}

\vspace*{0.6cm}

\vspace*{0.6cm}

\vspace*{0.6cm}

\vspace*{0.6cm}

\vspace*{0.6cm}

\clearpage

\subsection{Unsolved cases and limitations}
\vspace*{0.3cm}

\begin{figure}[htbp]
    \centering
    \subfigure[$2$CH-ES: $d_m$-endo = 2.5, $d_m$-epi = 3.9 , $d_H$-endo = 12.6 , $d_H$-epi = 13.3 mm.]{{\includegraphics[width=5cm]{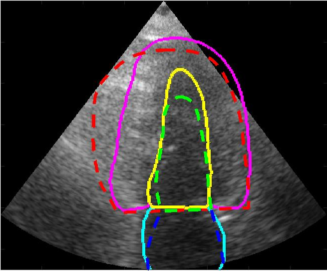}}}
    \qquad
    \subfigure[$4$CH-ES: $d_m$-endo = 3.0 , $d_m$-epi = 2.4 , $d_H$-endo = 15.7 , $d_H$-epi = 10.1 mm.]{{\includegraphics[width=5cm]{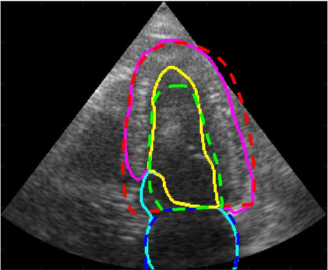}}}
    \caption{Anatomically plausible and inconceivable outliers.}
    \label{fig-art1}
\end{figure}

In this section, we show outliers predictions for the U-Net 1 algorithm. Outliers were deemed to constitute $18\%$ of the dataset. Among them, most (about $90\%$) could be associated to anatomically plausible LV and myocardium shapes, but on the last $10\%$ ($1.8\%$ of the full dataset), the provided segmentation can not be assimilated to a heart shape (Figure \ref{fig-art1}). 

\vspace*{0.5cm}

\begin{figure}[htbp]
    \centering
    \subfigure[$2$CH-ED: $d_m$-endo = 7.3, $d_m$-epi = 6.1 , $d_H$-endo = 23.2 , $d_H$-epi = 16.5 mm.]{{\includegraphics[width=5cm, height=4cm]{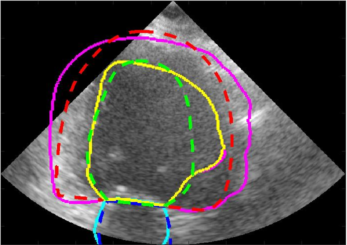}}}
    \qquad
    \subfigure[$2$CH-ES: $d_m$-endo = 3.8, $d_m$-epi = 5.7 , $d_H$-endo = 11.1 , $d_H$-epi = 12.0 mm.]{{\includegraphics[width=5cm, height=4cm]{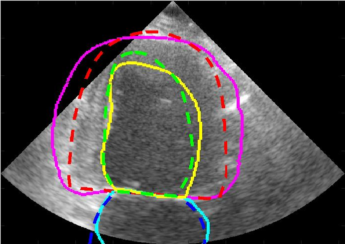}}}
    \caption{Outliers that may benefit from temporal coherency on the LV shape.}
    \label{fig-art2}
\end{figure}

Performing error analysis on outlier images reveal that peculiar context of acquisitions can mislead the network: 
\begin{itemize}
\item non-frequent zoom and probe tilt 
\item artifacts (shadowed zones, reverberation, fuzzy textures) as can be observed for the epicardium on Figure \ref{fig-art2} 
\item locally/globally extremely low contrast, which is patient-dependent (endocardium in Figure \ref{fig-art2}). 
\end{itemize}

\vspace*{0.5cm}

This suggests that artificial data augmentation recreating the conditions in which the network will be applied may be a good lead to ensure good robustness to context variations.  

\vspace*{0.3cm}

\vspace*{0.3cm}

\end{document}